\documentclass{aa}  

\usepackage{graphicx}
\usepackage{txfonts}
\usepackage{calligra}
\usepackage[T1]{fontenc}
\usepackage[mathcal]{eucal}
\usepackage{hyperref}
\newcommand{\Msun}{${\rm M}_{\odot}$\ }

\begin{document}

   \title{Microlensing and the type Ia supernova iPTF16geu.}


   \author{J.M. Diego
         \inst{1}\fnmsep\thanks{jdiego@ifca.unican.es}
          \and
         G. Bernstein\inst{2}
          \and
          W. Chen\inst{3}
           \and
         A. Goobar\inst{4}
         \and
         J.P. Johansson\inst{4}
          \and
          P.L. Kelly\inst{3}
          \and
          E. M\"ortsell\inst{4}
          \and
         J.W. Nightingale\inst{5}
          }

   \institute{Instituto de F\'isica de Cantabria (CSIC-UC). Avda. Los Castros s/n. 39005 Santander, Spain
         \and
         Department of Physics and Astronomy, University of Pennsylvania, 209 S. 33rd St, Philadelphia, PA 19104, USA
         \and
         School of Physics and Astronomy, University of Minnesota, 116 Church Street SE, Minneapolis, MN 55455, USA
         \and
         The Oskar Klein Centre, Department of Physics, Stockholm University, SE-106 91 Stockholm, Sweden
        \and
          Centre for Extragalactic Astronomy, Department of Physics, Durham University, South Road, Durham DH1 3LE, UK
             }


 \abstract{
     The observed magnifications and light curves of the quadruply-imaged iPTF16geu supernova (SN) offers a unique opportunity to study a lens system with a variety of independent constraints. The four observed positions can be used to constrain the macrolens model. The magnifications and light curves at the four SN positions are more useful to constrain microlensing models. We define the macrolens model as a combination of a baryonic component that traces the observed light distribution, and a dark matter halo component. We constrain the macrolens model using the positional constraints given by the 4 observed images, and compare it with the best model obtained when magnification constraints are included. We find that the magnification can not be explained by a macrolens model alone, and that contributions from substructures such as microlenses are needed to explain the observed magnifications. We consider microlens models based on the inferred stellar mass from the baryonic component of the macrolens model, and use the observed magnification and light curves to constrain the contribution from microlenses. We compute the likelihood of a variety of macro+micro lens models where we vary the dark matter halo, baryonic component, and microlens configurations. We use information about the position, magnification and, for the first time, the lightcurves of the four observed SN images. We combine macrolens and microlens models in order to reproduce the observations; the four SN positions, magnifications, and lack of fluctuations in the light curves. After marginalizing over the model parameters, we find that larger stellar surface mass densities are preferred. This result suggests that the mass of the baryonic component is dominated by its stellar component. We conclude that microlensing from the baryonic component suffices to explain the observed flux ratios and light curves. 
  }

   \maketitle
%

\section{Introduction}




In 2016, SN iPTF16geu became the first confirmed multiply lensed type Ia SN \citep{Goobar2017}. This type of SNe are of particular interest, not only for their possibilities as cosmological tools, but also to study gravitational lenses in greater detail. The standard candle nature of type Ia SN allows to estimate the magnification of the underlying model at the positions of the lensed SNe images. The large  magnification factors from the macromodel can be exploited to study the SN in greater detail \citep{Johansson2021}. The magnification estimates can also be used to improve the lens model, or to reveal discrepancies that could be due to substructures in the lens plane, or along the line of sight. One of the most common types of substructures invoked to explain discrepancies between predicted and observed magnifications (of small background objects, such as quasars) is microlensing by stars or remnants in the lens plane. As these type of microlenses are ubiquitous in the lens plane, the probability that the Einstein radius of one of these microlenses intersects the line of sight to the background lensed object is not negligible.  For small sources like SNe, and for a  relatively low number density of microlenses, and/or small to moderate magnification factors, the probability of intersecting a microcaustic is small during the first days after explosion \citep{Suyu2020}. On the other hand, if the number density of stars at the position of the lensed image is sufficiently high, or the total magnification is large enough, intersecting a microlens is unavoidable \citep{Diego2019}. 

Earlier work has studied iPTF16geu in detail. 
In the initial discovery paper by \cite{Goobar2017}, the authors present an initial spectral analysis and light curves for the 4 images, together with the redshifts of the lens ($z=0.216$), and SN ($z=0.409$). The velocity dispersion of the lens is also estimated to be $\sigma _v = 163^{41}_{-27}$ km s$^{-1}$. Keck and HST images reveal 4 images of the same SN in an almost perfect symmetric configuration, forming a ring with radius $\approx 0.3"$. One of the images is significantly brighter than the other three. An initial estimation of the total magnification rendered $\mu \approx 53$ (equivalent to a 4.3 magnitude boost), although with relatively large uncertainty ($\approx 0.4$ magnitudes). After fitting the four images with an isothermal ellipsoid lens model, the authors estimated a mass within the estimated critical curve of ${\rm M} = (1.69 \pm 0.06) 10^{10}$ \Msun. Based on the anomalous flux ratio, the authors point to possible microlensing or millilensing effects. 
\cite{More2017} presents lens models that fits the 4 SN positions. All lens models predict significantly less magnification, especially for the brightest image where the model predictions range between 5.2 and 8.2,  a factor at least 4 times discrepant with the estimated magnification of image 1. The authors suggest also that microlensing could be responsible for this discrepancy. This conclusion is challenged by \cite{Yahalomi2017} that argue that the discrepancy is too large to be explained by microlensing. 
Accurate magnifications and time delays are presented in \cite{Dhawan2020} and the analysis of the background source and lens based on spectroscopic information is presented in \cite{Johansson2021}. In \cite{Dhawan2020}, the authors estimate the magnification of the four individual images and correct for reddening.They  updated the estimate of the magnification to $67.8^{+2.6}_{-2.9}$. 
They also measure the time delays from the light curves and find a relative delay of order 1 day, confirming earlier expectations. 
In  \cite{Johansson2021}, the authors measure the expansion velocity based on the Si{\small II} line, from which one can estimate the size of the photosphere as a function of time. They conclude that iPTF16geu can be classified as a high-velocity SN with a velocity of 11950 $\pm 140$ km s$^{-1}$ at the peak emission, and velocity gradient -110.3 $\pm 10.0$ km s$^{-1}$. The authors provide also an improved estimate for the velocity dispersion of the lens galaxy, $\sigma _v = 129 \pm 4$ km s$^{-1}$.
The same team presents a new lens model in \cite{Mortsell2020}, where they find consistent results with earlier work, maintaining the discrepancy between the observed and model predicted magnifications for the brightest image (and to a lesser degree also in the other images). However, their lens model predicts higher magnifications for the brightest image, but still a factor $\approx 2$ below the observed value. After comparing with realizations of the microlensing effect, the authors conclude that microlensing may be responsible for the discrepant magnification. In particular, they find that the probability to obtain the observed flux in all images is 12\% when accounting for microlensing effects and for a halo model with slope $\alpha=1.2$ (equivalent to $\eta=3-\alpha=1.8$ in \cite{Mortsell2020}, see section 3.2 below for  definition of $\alpha$). 
These results are derived, however, under the simplifying assumption that all microlenses have the same mass (0.3 \Msun). The fraction of stars is estimated from the light in the i and z bands, and adopting a mass-to-light ratio. This work also constrains the Hubble constant from the estimated time delays, although the small time delays relative to its uncertainties (and possible systematic effects in the time delay estimation from microlensing) prevents the authors from establishing strong limits on $H_o$. 
In a recent work \citep{Williams2020}, a two component lens model is adopted to fit the position of the 4 images. This model has the advantage of allowing for asymmetries by introducing a relative shift between the two components. Interestingly, the predicted magnification for the most discordant image 1 agree better with the observed value.  
Regarding the lens models in earlier work, one common feature is that all previous lens models are based on simple analytical models. To the best of our knowledge, no attempt has been made to include the galaxy itself, which shows a bulge and an extended halo around it, as part of the lens model. In this work we try to remedy this situation by incorporating the baryonic component in the lens model. Adding the baryon component has the added benefit that we can estimate directly the surface mass density of microlenses (assuming most of the baryon mass is in the form of stars and remnants). 

In this work we also pay special attention to the role of microlenses \citep[as it has been done in earlier work, for instance in][]{Mortsell2020,Weisenbach2021}.  Since the redshifts of both, the lens and background source, are relatively small, the critical curve appears at relatively short distances from the centre of the lens. At this distance, the surface mass density of stars is high and microlenses are ubiquitous around the critical curve. In the case of iPTF16geu, where 4 images form near the critical curves, and the background source is smaller or comparable in size to the size of the microcaustics, the role of these stars can not be ignored. For typical QSOs, where microlensing effects have been studied in detail, the high redshifts of the QSOs results in critical curves forming at typically 5--10 kpc from the centre of the lens. At these distances, dark matter usually dominates over the stellar component in terms of surface mass density in the lens plane. For the particular case of iPTF16geu, the critical curve forms at an exceptionally small radius of $\approx 1$ kpc from the centre of the galaxy. At these very small radii, the contribution to the projected mass from stars is expected to be comparable, if not higher than, the contribution from the smooth component (dark matter). One then expects microlensing to play a very significant role in this system. In the context of QSO microlensing, an exception can be found in OGLE Q2237+0305, for which the lens galaxy is at very low redshift ($z=0.0394$), and the critical curves form very close to the centre of the lens galaxy at $\approx 1$ kpc \citep{Wyithe2000,Anguita2008}. In this case, the surface mass density of microlenses is also relatively high. In addition, the large surface mass density of microlenses expected for the images of iPTF16geu is amplified by the magnification factor. At the large magnification factors, $\mu$, estimated for iPTF16geu, an area $A$ in the image plane gets mapped into a much smaller area $A/\mu$ in the source plane, resulting in overlapping microcaustics in the source plane \citep[see for instance][]{Kayser1986,Paczynski1986,Wambsganss1990,Wyithe2001,Kochanek2004,Diego2018,Diego2019}. 

 The microlenses can be stars and remnants from the galaxy but also more exotic microlenses such as the hypothesized primordial black holes (or PBH). 
 Although numerous studies show that PBH can not account for all dark matter, in the mass regime of a few tens of solar masses, PBH can still account for a few percent of the total mass budget. Even a small fraction of dark matter in the form of PBH could account for the rate of binary black hole merger observed by LIGO/Virgo \citep[see for instance][]{Carr2017,Liu2018,Chen2018,Liu2019,Raidal2019}. Whether or not LIGO/Virgo is observing a population of PBH with masses around 30 \Msun is still an open question. Microlensing can constrain the fraction of PBH in the universe \citep[see for instance][]{Diego2018,Oguri2018}. Strongly lensed type Ia SNe offer an alternative way of constraining the abundance of PBHs through their microlensing signatures. The direct estimation of the magnification, combined with accurate measurements of the light curve can be used to reduce degeneracies in the macro+micro model. Earlier work has found sufficient evidence about the presence of microlenses and their influence in the observed magnification. In this work we focus on these microlenses. We exploit the first observations of a lensed type Ia SN to extract information on the population of the microlenses lurking around the lens galaxy. 

The structure of this paper is as follows. In section \ref{sec_data} we present a brief summary of the iPTF16geu data used in this work. Section \ref{sec_macro} describes the macrolens model used to fit the observed positions (and magnification) of iPTF16geu. Section \ref{Sect_ModelSelection} discusses the model selection and shows the best models for the cases where only the position of the four SNe images are used to constrain the lens model, and the case where both position and magnification information is used. The anomalous flux of image 1 (and to a lesser degree the other images) and the role of microlenses is discussed in section \ref{sec_micro}. Finally, we discuss our results in section \ref{sec_discus} and present our conclusions in section \ref{sec_concl}. 

Throughout the paper we use different definitions and parameters, which for the shake of clarity are summarized here.  
We refer to the smooth lens model of the galaxy as the macromodel or macrolens model. The galaxy creates critical curves in the lens plane and caustics in the source plane. The magnification predicted by the macromodel is referred to as macromodel magnification, or simply $\mu_{macro}$. Microlenses are assumed to be stars and remnants in the lens plane and they form microcaustics in the source plane. A given position in the source plane can produce multiple images in the lens plane. In particular, iPTF16geu consists of 4 multiply lensed images, which we refer to as macroimages. Two of these images have positive parity (minima points in the time delay surface), and two have negative parity (saddle points in the time delay surface). A fifth image (maximum in the time delay surface) is expected in most elliptical lens models but not observed (likely due to the very compact nature of the lens at its centre, see for instance the discussion in \cite{Mortsell2020}).  The lens is at redshift $z_l=0.2163$, and the background source at redshift $z_s=0.409$. At the redshift of the lens, 1 arcsecond corresponds to 3.615 kpc, which becomes 5.609 kpc at the redshift of the source. We adopt the same Planck flat cosmological model as in \cite{Mortsell2020} ($h=0.678$, $\Omega_m=0.308$), for which the angular diameter distances to the lens, to the source, and between the lens and the source are $D_d=745.8$ Mpc, $D_s=1157$ Mpc, and $D_{ds}=513.2$ Mpc respectively. 

\section{Observations of iPTF16geu}\label{sec_data}
SN iPTF16geu was discovered in 2016. HST/WFC3 observations of iPTF16geu were carried out through the ToO program (14862 PI:Goobar), 10 visits for high-spatial imaging, from Oct 20, 2016, through Nov 26, 2016, covering in the optical filters F390W,F475W, F625W, F814W, and with worse angular resolution in the IR filters F105W, F110W, and F160W. Reference images for the same filters were acquired using 3 orbits in 2018, i.e., after the supernova had faded. (GO 15276, PI: Goobar). The SN lightcurves, after galaxy subtraction, were presented in \cite{Dhawan2020}. Grism spectroscopy was also obtained while the SN was active, as discussed in \cite{Johansson2021}.  
The data was used to study the lensed SN, model the lens, infer relative time delays, and magnifications at the 4 positions of the lensed SN \citep{Dhawan2020,Mortsell2020,Johansson2021}.

   \begin{figure} 
   \centering
   \includegraphics[width=9.0cm]{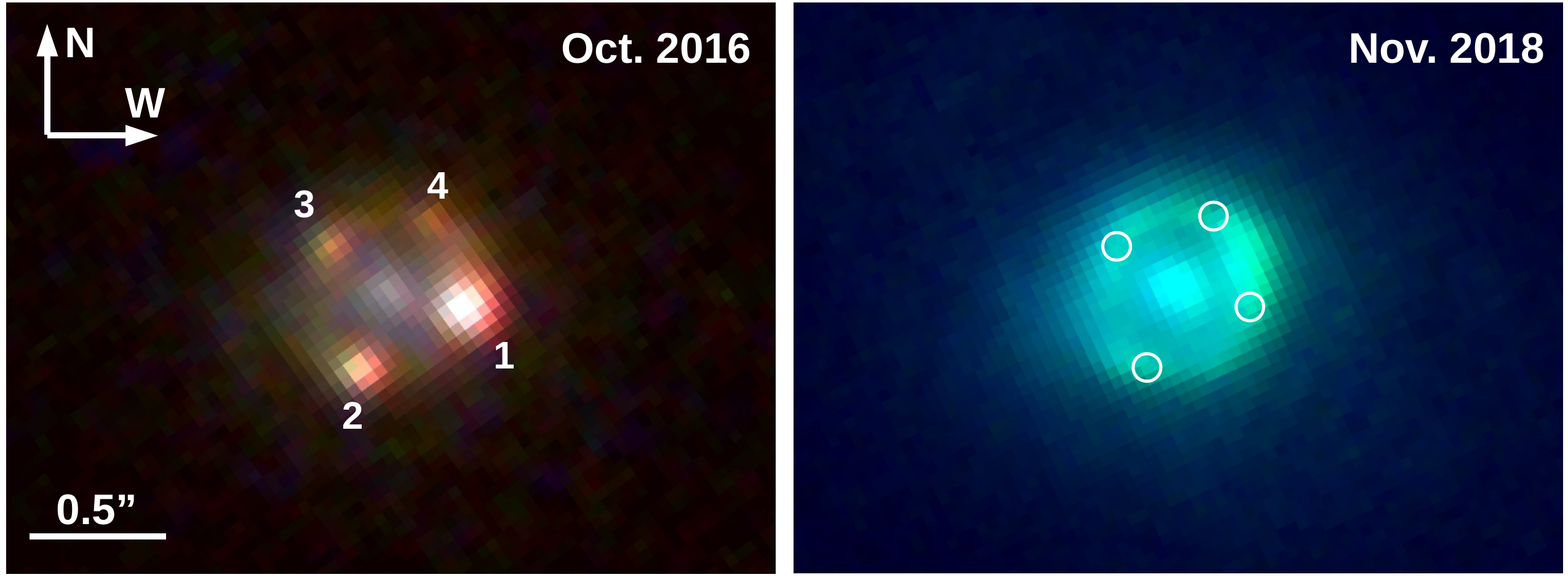}
      \caption{The left panel shows the 4 SN images of iPTF16geu on HST images taken in October 2016 (P.I Ariel Goobar). The red, green, and blue bands correspond to the F814W, F625W and F475W filters respectively. The right panel shows the same lensing system but $\approx$ 2 years later in F814W, and F625W bands. The SN is not visible anymore but the host galaxy can be seen more clearly, forming a nearly perfect Einstein ring around the centre of the lens.
              }
         \label{Fig_Data}
   \end{figure}

\begin{figure} 
   \centering
   \includegraphics[width=9.0cm]{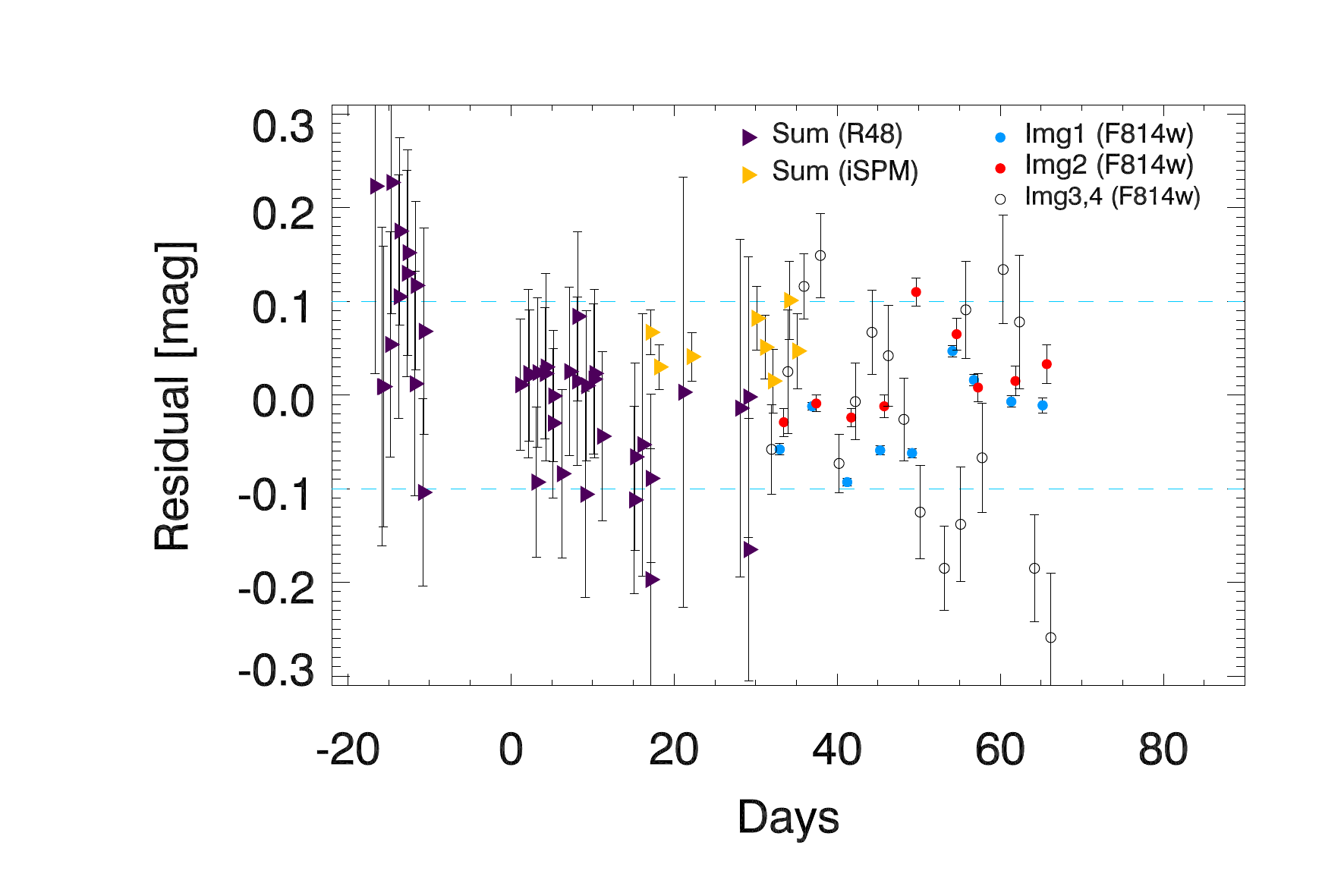}
      \caption{Residual between the observed flux and the light-curve model in \cite{Goobar2017,Dhawan2020,Johansson2021,Mortsell2020}. The x-axis shows the days since the maximum of the light curve. Only a subset of the entire data set is shown. HST data (F814W and F625W) is available only $\approx 30$ days after the maximum. Prior to this date, the observations can not resolve the 4 individual images so the residual corresponds to the sum of all four images. The dashed line marks the adopted $\pm 0.1$ magnitude upper limit, for the allowed range of variability in the light curves.
              }
         \label{Fig_Photometry}
   \end{figure}

\begin{figure} 
   \centering
   \includegraphics[width=9.0cm]{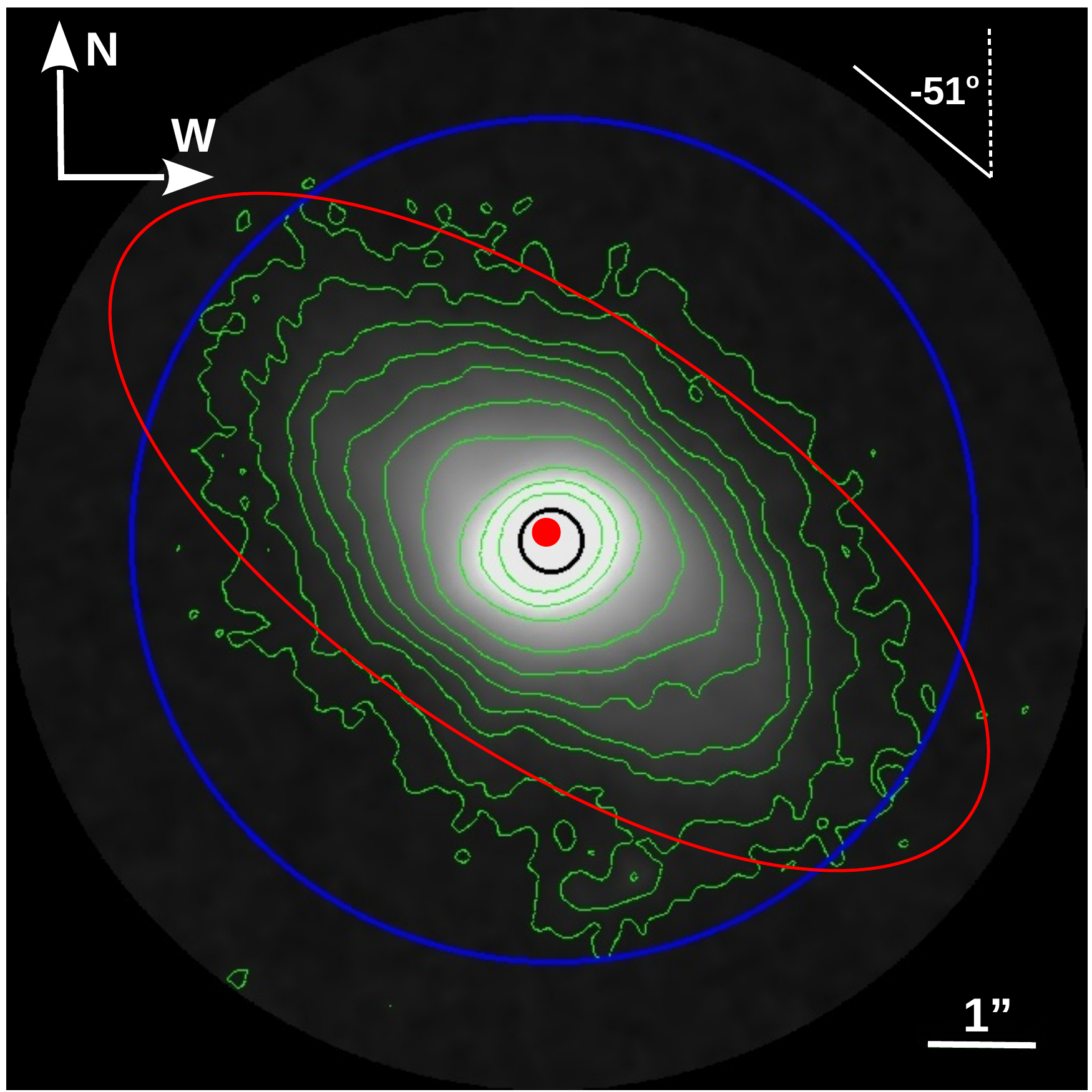}
      \caption{Contours and grey image show the lens in the F160W band. The grey image shows the square root of the flux, to better show the details at large radii. The large blue circle has a radius of 4 arcsec and marks the extent of the of the visible baryons while the small black circle has a radius of 0.3 arcsec, roughly the radius of the Einstein Ring. The galaxy is at an angle of $\approx -51^{\circ}$ with respect to the North. The red ellipse represents an elliptical dark matter halo. It is inclined by an angle of $\approx-56^{\circ}$ and has an ellipticity of 0.4. The red dot near the centre of the galaxy marks the position of the DM halo, that is slightly shifted in the NE direction by 0.23 kpc with respect to the galaxy. This shift is discussed in section~\ref{Sect_ModelSelection}. 
              }
         \label{Fig_Galaxy}
   \end{figure}

In this work we rely on the high-resolution HST images. Figure~\ref{Fig_Data} shows the 4 SN images (left panel) as observed in October 2016 with Hubble, while figure~\ref{Fig_Photometry} shows the residual flux after a SN light curve is fitted to the observations. Since the time delay between images is estimated to be of order 1 day \citep{Dhawan2020}, the difference in flux observed in the image is mostly due to differences in the underlying magnification. The images from 2016 reveal a nearly perfect Einstein ring formed by the host galaxy of the SN with a quad configuration typical of elliptical potentials. The brightness of the SN images hides some of the details of the host galaxy.  The deep images taken in 2018 (P.I Ariel Goobar) reveal more detail in the host galaxy, since the SN is no longer visible. In particular, a bright region on the NE sector between images 1--4 can be clearly appreciated, presumably a bright feature in the host galaxy that is crossing a caustic from the macromodel in the source plane. Also, the flux in the host galaxy has two minima between images 1--2 and 2--3, and a less prominent one between images 3--4. These minima can be better appreciated in Figure 1, panel e) of \cite{Dhawan2020}.   
The deeper images allow also to get a better picture of the lens. Figure \ref{Fig_Galaxy} shows a larger area around the lens. The contours tracing the light of the lens clearly indicates an extended halo stretching in the NE-SW direction, and up to $\approx 13$ kpc form the centre. Note how at radii $\approx 1"$, the orientation of the contour is nearly orthogonal to the orientation of the contours at larger radii. This is an important point, since this misalignment introduces a feature in the gravitational potential that can not be captured by classic analytical elliptical potentials. 

In earlier work \citep{Goobar2017,More2017,Mortsell2020}, it was found that the best lens models are elliptical potentials with orientations similar to that of the galaxy in large scales (and close to the red ellipse shown in the figure). This is not surprising as one expects the baryons and dark matter halo to orient almost in the same direction, especially in isolated galaxies like this one, where the closest visible galaxy is a small one found at $\approx 55$ kpc away (in projection). The alignment found in earlier work could also be indicating that the baryonic component plays a significant role in defining the lensing potential. This is also not surprising as rotational curves of galaxies indicate that the potential in the central part of galaxies is dominated by the baryonic component (and in particular the stellar part, with the gas component playing a more important role in the outer parts). For the particular case of iPTF16geu, this is even more true than for more ordinary lenses, since the relative proximity of the background SN to the lens  results in the lensed images to form much closer to the centre of the lens, where baryons are expected to contribute more significantly (in terms of projected mass). Our lens model incorporates the baryons up to the visible boundaries of the lens galaxy ($\approx 14.5$ kpc from its centre). Even though the galaxy is much larger than the size of the Einstein ring (see Figure~\ref{Fig_Galaxy}), it is important to include the galaxy shape up to the largest radii in order to capture the external shear contribution from these more distant regions. Figure~\ref{Fig_Galaxy} shows how the galaxy extends up to a radius of $\approx 4"=14.5$ kpc, while the Einstein radius can be approximated by a circle of radius $\approx 03"=1.1 kpc$. The relatively small size of the Einstein radius compared with the size of the galaxy makes this system even more appealing since it probes the densest part of the lens. High surface mass density corresponds also to the high surface mass density of microlenses, which combined with the large magnification factors makes microlensing especially relevant in this case. Microlensing effects are studied in more detail in section \ref{sec_micro}.

   \begin{figure} 
   \centering
   \includegraphics[width=9.0cm]{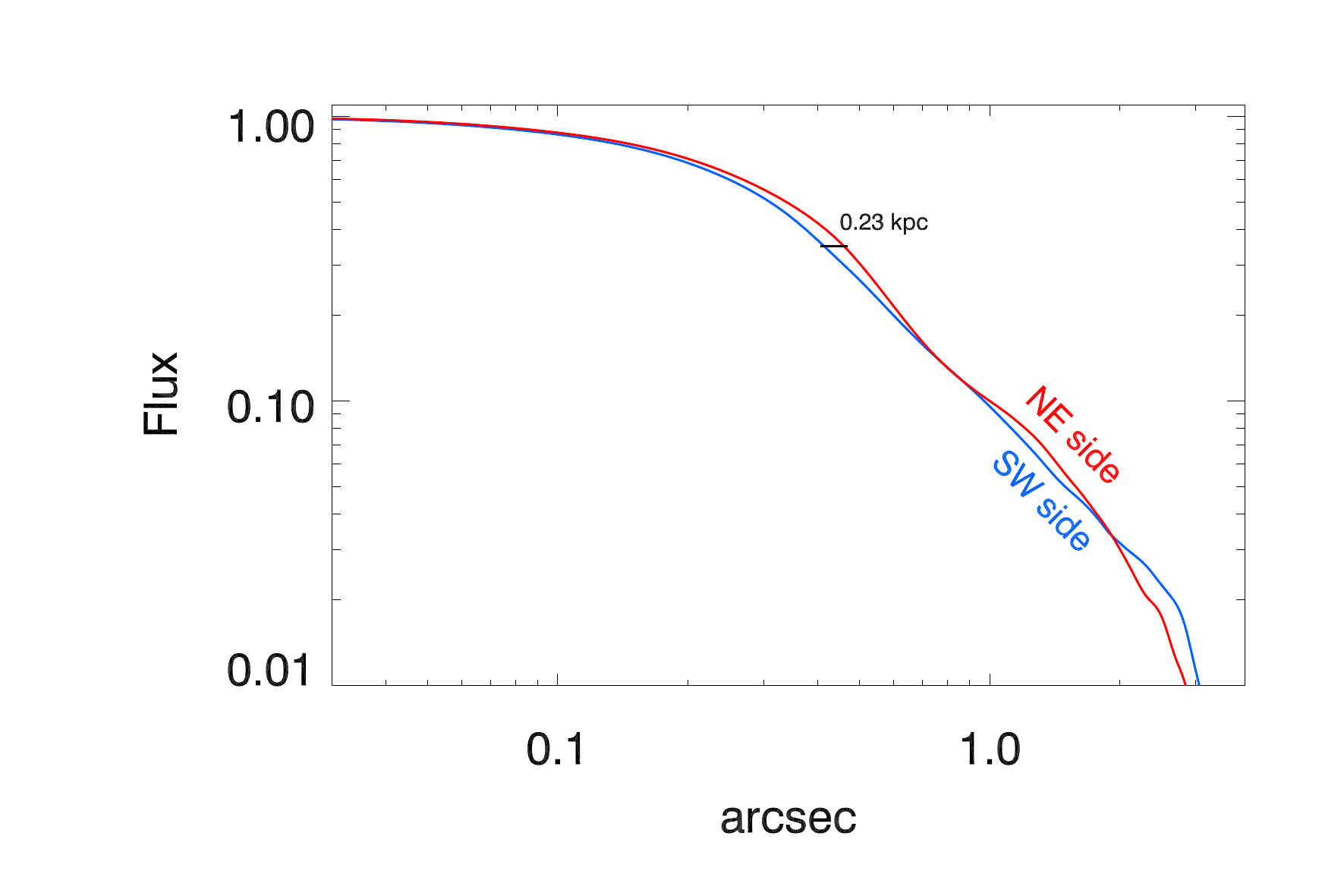}
      \caption{Profile of the galaxy after projecting along the major axis. The centre of the profile is taken at the maximum of the observed flux, in the centre of the galaxy. The map is divided in two sectors, NE and SW. 
               The profile of the SW sector is marked with a red curve and the profile for the SW sector is shown in blue. Note the excess in the NE sector with respect to the SW sector at $\approx 0.4"$  (or $\approx 1.6$ kpc). The small black solid line is equal to the size of the displacement (0.23 kpc) along the major axis in the NE direction applied to the dark matter halo in Figure~\ref{Fig_Galaxy}.   
              }
         \label{Fig_ProfileEastWest}
   \end{figure}

\section{Macrolens model}\label{sec_macro}
In this section we present the model used to describe the macrolens. The full model contains also microlenses, but these are discussed in section \ref{sec_micro}. As mentioned in the previous section, a robust lens model of this system must include the baryonic component as well as a dark matter halo. Including the baryonic component is trivial since the projected baryonic mass is expected to trace closely the observed light (as most of the baryonic mass is expected to be in the form of stars). 
In addition to the baryonic component, the lens model includes also a dark matter halo that is parameterized as an elliptical potential. Below, we discuss briefly these two  components.

\subsection{Baryonic component}

   \begin{figure*} 
   \centering
   \includegraphics[width=18.0cm]{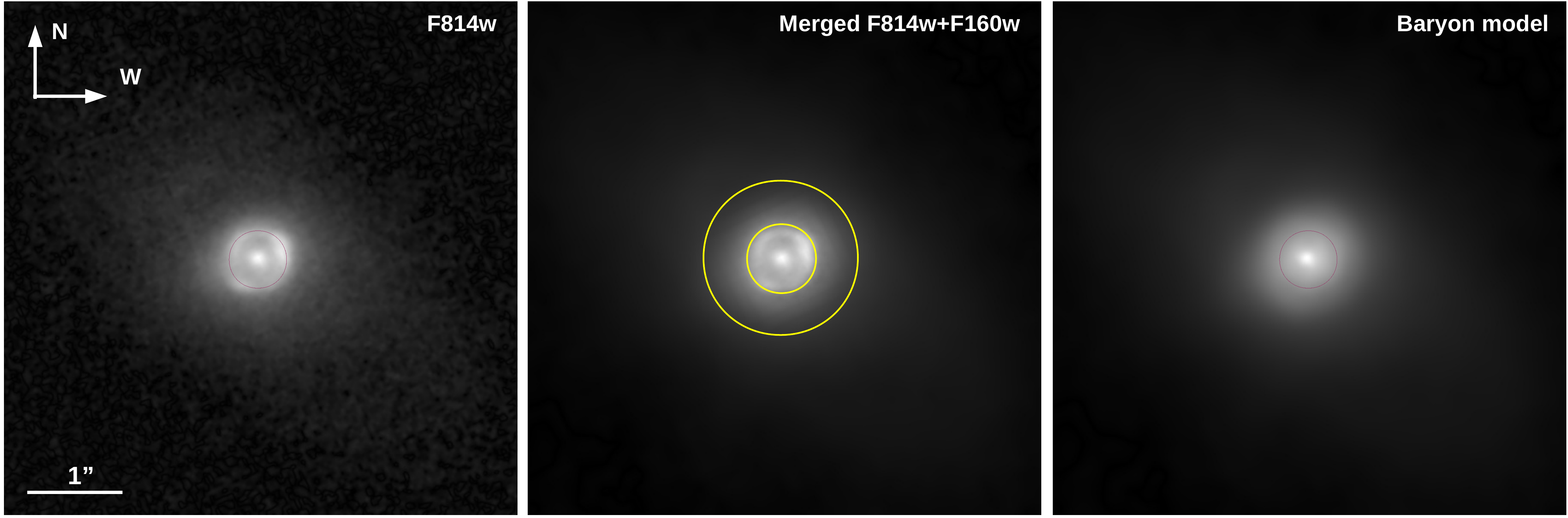}
      \caption{The left panel shows the 2018 HST data in the F814W band (slightly smoothed and to the power 1/2 to increase the detail at larger radii). The circle on top of the Einstein ring has 0.3" radius. The middle panel shows the merged F814W+F160W image. The yellow circles mark the region where the two images are weighted and combined. The right panel shows the baryon model based on a combination of images in the F160W and F814W bands and after subtracting the background galaxy. For reference, we overlay the 0.3" radius circle marking the position of the Einstein ring.   
              }
         \label{Fig_BaryonModel}
   \end{figure*}

The baryonic component is mostly composed of stars, especially in the central region, and is assumed to trace the light distribution. We make a model of the baryonic component from a combination of images taken in the F160W and F814W bands. By construction this model assumes that the mass-to-light ratio is the same in these two filters. The F160W image traces better the light distribution at larger radii (beyond 0.3" radius from the centre of the lens) while the F814W band has better resolution in the central region (at radii below 0.3"). The improved resolution of F814W enables a better removal of the lensed image. As shown in the left panel of Figure~\ref{Fig_BaryonModel}, at radii $\approx 0.3"$, the observed light is dominated by the lensed background galaxy. In order to get a model for the light distribution of the lens at this radius, the background galaxy needs to be subtracted. This is done in two steps; 
i) The first step is to merge the F814W image with the F160W image in order to produce a map of the light distribution where the outer region is given by the deeper F160W image, while the inner region is given by the shallower, but higher resolution F814W image. The two images are merged between radii 0.36" and 0.81" from the centre . Between these two radii, a composite image is formed as the combination $w(r)\times F814{\rm w} + (1-w(r))\times F160{\rm w}$, where $w(r)=1$ if $r<0.36"$ and $w(r)= exp[-(r-r_o)^2/b^2]$ if $r>0.36"$, with $r_o=0.36"$ and $b=0.12"$. The Gaussian weights guarantee a smooth transition between the F814W image and the F160W image. The merged image is shown in the middle panel of Figure~\ref{Fig_BaryonModel}. The two radii at 0.36" and 0.81" are shown as two yellow circles. At radii above  $\approx 0.6"$ the merged image is mostly given by the F160W image. At radii below 0.4" the merger image is mostly given by the F814w image. 
ii) The next step is to remove the background galaxy from the merged image. We apply a high-pass filter (an à trous wavelet) to the merged image to isolate the ring. The ring is later smoothed and subtracted under the constraint that the residual under the ring must be comparable to the surface brightens outside the masked region. These residuals still contain substructure from the ring. To reduce their contribution, the resulting image is smoothed again to produce a locally interpolated map in the masked region. The smoothing results in a shallower baryon model, but given the fact that there is no central image, our results are insensitive to the central slope. Formally, smoothing is not the same as an interpolation but produces similar results by substituting the value in a given pixel with a Gaussian weighted local average value, that takes into account the surface brightness of neighbouring pixels. The FWHM of the smoothing Gaussian is set to 0.12" in order to guarantee that areas outside the masked region have significant weight. The masked region is then substituted with this smooth version that has all small scale fluctuations removed. A final smoothing of FWHM=0.03" is applied in order to remove discontinuities at the edges of the masked region and also to reduce additional small scales fluctuations (noise) elsewhere in the baryon model. The final result is shown in the right panel of Figure~\ref{Fig_BaryonModel}. This map constitutes our baryon model. Note that Figure~\ref{Fig_BaryonModel} shows only the central 5.4" but we consider a larger field of view of 8.1 arcseconds as shown in Figure~\ref{Fig_Galaxy}. Also, as shown in Figure~\ref{Fig_Galaxy}, and in order to eliminate edge effects, the mass of the baryon model beyond the radius 4.05" is set to zero. 
We tested our results against an alternative baryon model that was built by substituting the region under the Einstein ring by a power law fit to both the F814W and F160W observed profiles outside the Einstein ring region. Using this alternative baryon model renders similar results. 
The process described above avoids fitting the galaxy with symmetric models, usual in galaxy fitting algorithms, and retains any possible asymmetries in the light distribution.  

\subsection{Dark matter component}
For the dark matter component, we assume an elliptical halo. In order to facilitate an easier comparison with earlier models, we follow \cite{Mortsell2020} and adopt a generalized isothermal ellipsoid, for which the convergence is:
\begin{equation}
\kappa(x,y)=\frac{\rho _o}{\left[ r_o^2 + (1-e)x^2 + (1+e)y^2  \right]^{1-\alpha/2} }
\label{Eq_kappa_macro}
\end{equation}
where $e$ is the ellipticity, $r_o$ a core radius, and $\alpha$ a slope parameter. For $\alpha=1$ and $e=0$ one recovers the standard cored isothermal profile. In \cite{Mortsell2020}, the authors establish an upper limit for $r_o$ based on the absence of a central image. For the different models they consider, they find that $r_o$ must be smaller than 0.08"=290 pc. 
In the same work, and after fitting the Hubble constant to $h=0.7$, they find evidence for a slope $\alpha>1$ from the observed time delays. Our model includes also a baryonic component based on the light distribution. For this model, and at the position of the Einstein ring, the circularly averaged light profile falls with radius as $r^{-1}$. At twice the Einstein radius, the circularly averaged light profile falls like $r^{-2}$, and at 4 times the Einstein radius it falls like $r^{-3}$.
We have tested the case where a central SMBH is added near the centre of the galaxy, but found that the improvement in the lens model was very small. Based on this small improvement and the Akaike information criterion, we do not include a central SMBH in our macrolens model. In addition, due to the lack of central image, the mass of the SMBH is degenerate with the radius $r_o$. In other words, the small radius $r_o$ plays a role similar to a possible central SMBH. The possibility that central SMBH contribute to the small magnification of the central image has been tested in earlier work with quasar central images \citep{Winn2004}. 

\subsection{Joint macromodel}\label{subsect_jointmacro}
Combining the baryon and DM models we form the macrolens model by simply adding the convergence of the baryon model with the convergence of the DM halo. During the lens optimization process, we explore a model with 6 parameters. For the baryon model, since its shape and orientation are fixed, there is only one parameter that corresponds to the mass of the galaxy, $M_{\rm gal}$. For the DM model we fit for its mass, $M_{\rm DM}$, halo ellipticity, $e$, orientation with respect to the North direction, $\psi$, and its relative position with respect to the centre of the galaxy, $dx$, and $dy$. We fix the core radius, $r_o$, of the  DM halo to a value smaller than the upper limit derived by \cite{Mortsell2020} and check that values smaller than this upper limit have no impact on the derived best fitting models. 
This is expected since the lack of a central image prevents us from constraining the inner slope and only an upper limit can be derived as discussed in \cite{Mortsell2020}. 
The slope is fixed to the isothermal value $\alpha=1$, and study the impact of varying $\alpha$ in Appendx B. 
The parameters $dx$ and $dy$ defining the relative position of the DM halo with respect to the galaxy play a double role. On one hand, they allow us to consider models where a real shift exists between the DM and baryonic distributions, as predicted by some models where DM is allowed to scatter (i.e., self-interacting models). More importantly, by introducing an offset between the galaxy and DM halo, we can account for an  asymmetric distribution of DM, since the lens model will contain more DM in the direction of the offset than in the opposite direction. 
We compute the light profile in the NE and SW sectors by masking the NE sector when computing the SW sector profile and vice versa. The profiles of the NE and SW sectors are shown in Figure~\ref{Fig_ProfileEastWest}. The two profiles look very similar, suggesting a high degree of symmetry in the galaxy between the NE and SW parts, but a small excess can be appreciated in the NE side with respect to the SW side at $\approx 1.6$ kpc from the centre. This excess can be also reinterpreted as a relative displacement of $\approx 0.23$ kpc along the major axis as shown in Figure~\ref{Fig_ProfileEastWest}. 
Since the light distribution presents a small asymmetry between the NE and SW sectors, one would naively assume that a similar asymmetry (or displacement of the DM halo with respect to the galaxy) may be also present in the DM halo. 

Once the convergence of the macromodel is simulated $\kappa=\kappa_{\rm gal} + \kappa_{\rm DM}$, we compute the deflection field in Fourier space as the convolution of the convergence with the two lensing kernels, one for the deflection in the direction-x and the other one for the deflection in the y-direction. To minimize edge effects introduced by the periodicity of the FFT, we consider a field of view twice as large as the targeted region and centered in the lens, with the mass in this buffer zone area set to zero.  

\section{Macrolens model optimization}\label{Sect_ModelSelection}
As mentioned earlier, iPTF16geu offers the unique opportunity of providing reliable magnification measurements at 4 positions in the image plane. In principle, this extra information can be used to further constrain the macrolens model. However, as pointed out in earlier work, but also from the anomalously large signed sum of the magnifications, it is likely that the magnification values are affected by microlensing effects. 
Selecting the best model when adding magnification constraints in the presence of microlensing requires a special treatment, since microlensing can have a significant effect on magnification. On the other hand, the position of the SN is not affected by microlensing (changes in the position of the SNe images due to microlensing are much smaller than the positional accuracy of the observations). 
Due to the influence of microlensing in the magnification of the SN images, we consider two scenarios to derive the macromodel. In the first one, we derive the macrolens model ignoring the inferred magnification in the four images of the SN, and using only their position as constraints. In the second scenario we use both the position and magnification at the four SN images to derive the macrolens model.   
Since microlensing is likely affecting the magnification, and images with negative parity (images 1 and 3) are in general more affected by microlensing than images with positive parity \citep[see for instance][]{Diego2018,Diego2019}, in the second scenario we use magnification constraints only from the two images with positive parity (images 2 and 4), since they are the most reliable in terms of magnification (i.e., where microlensing effects can be more safely ignored).  
These two scenarios are explored in the next subsections. 

\subsection{Positional constraints only}\label{Subsect_PosOnly}
Let's assume that  $\delta\beta_{x_i}$ and $\delta\beta_{y_i}$ are the relative distances in the source plane between the model predicted positions of the 4 SN images ($i=1,2,3,4$) and the real (unknown) source position, $\beta_x^t$, and $\beta_y^t$. The true position is unknown, but for a valid lens model the true position is approximated by the average of the 4 reconstructed positions. That is,
\begin{equation}
\beta_x^t=\frac{\sum_i^4 \beta_{x_i}}{4},
\label{Eq_Bx}
\end{equation}
and 
\begin{equation}
\beta_y^t=\frac{\sum_i^4 \beta_{y_i}}{4},
\label{Eq_By}
\end{equation}
and subtract it from the predicted positions, i.e., $\delta\beta_{x_i}=\beta_{x_i} - \beta_x^t$ where $\beta_{x_i}$ are the predicted positions by the model (and similarly for the predicted $\beta_y$ positions). 

Then, for an ideal lens model,  $\delta\beta_{x_i}=\delta\beta_{y_i}=0$ and the dispersion $\sum_i^4 (\delta\beta_{x_i})^2 + (\delta\beta_{y_i})^2$ must be also zero. Since the lens model can never be perfect (for instance, due to unmodeled substructures in the lens plane, or asymmetries in the lens), instead of a vanishing dispersion, we should expect a small value for the dispersion between the four predicted positions in the source plane.  
One can then define the likelihood of the lens model as follows,
\begin{equation}
-2 ln(\mathcal{L})=  \sum_i^4\frac{\delta\beta_{x_i}^2 +\delta\beta_{y_i}^2}{\sigma_i^2}
\label{Eq_Lkhd1}
\end{equation}
where $\sigma_i$ is the positional error for image $i$ in the source plane. 

The likelihood in Eq.~\ref{Eq_Lkhd1} optimizes the model in the source plane. This is a convenient choice since it is significantly faster than optimizing in the image plane. Also, there is no need to fit for the unknown source position (see discussion above), hence reducing the number of free parameters. Instead, the source position is derived a posteriori from equations \ref{Eq_Bx} and \ref{Eq_By}. 
In this work we do not explicitly fit for the lensed galaxy as a whole (except in Appendix C), but features like the position of the gaps, and bright features like the nucleus are naturally reproduced by the best lens models, and a simple toy model for the source. 

 For the error in the position of the SN, we adopt a conservative value of $\sigma_i=5$ mas. This value is smaller than the positional accuracy of HST ($\approx 12$ mas), but larger than the corresponding (demagnified) accuracy in the source plane. 
 In principle, $\sigma_i$ should be different for each image and should be of the order of the positional accuracy in the lens plane divided by the square root of the magnification from the macromodel (since a unit area in the source plane is a factor $\mu_{macro}$ times smaller in the source plane), that is $\sigma_i\approx$3--4 mas if one adopts $\mu_{macro}\approx 10$, as found in earlier work. However, since any modelling effort inevitably misses real features in the lens plane, insisting on a perfect reproduction of the observed positions with imperfect models results in systematic biases, with respect to the true underlying mass distribution in the model \citep[this was beautifully illustrated for a galaxy cluster scale lens in][]{Ponente2011}. These biases can be reduced if one relaxes the positional constraints by increasing the error in the fit.  Since the real magnification from the macromodel is also unknown (due to microlensing distortions), we hence assume a conservative value $\sigma_i =5$ mas for all SN images.

   \begin{figure} 
   \centering
   \includegraphics[width=9.0cm]{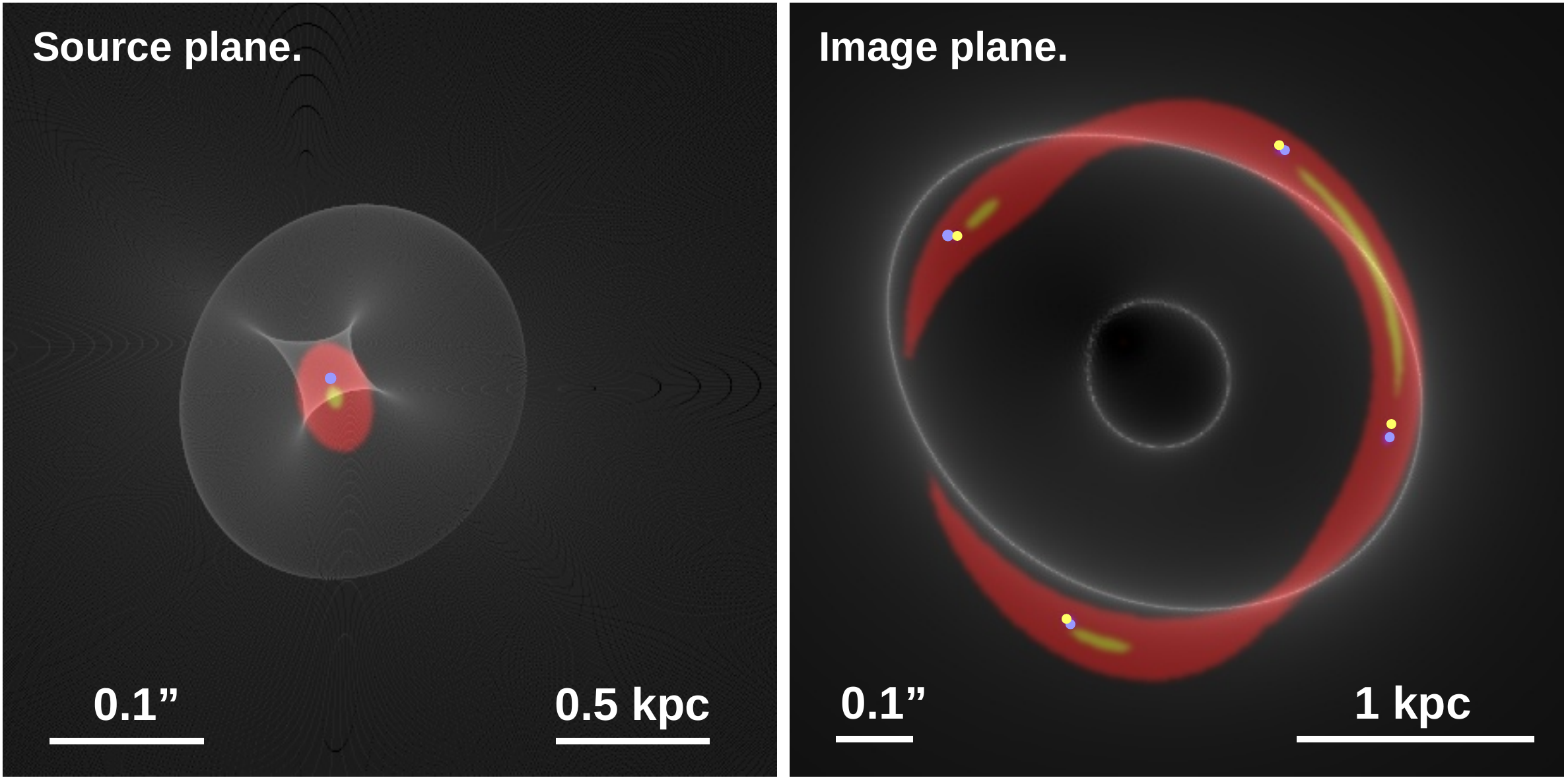}
      \caption{Caustics (left) and critical curves (right) for the model where only position is used. A simple source with just three components, nucleus (orange), halo (red) and SN (blue) is shown in the left panel (source plane). The predicted image is shown in the right panel (image plane). The yellow dots in the right panel mark the observed position of the four SN images. 
              }
         \label{Fig_Caus_CC1}
   \end{figure}

 With this choice of likelihood, we explore the space of parameters ($e,\psi,dx,dy,M_{\rm gal},M_{\rm DM}$). We fix the slope of the DM halo to $\alpha=1$ and refer to the best model derived with this slope as the fiducial model. In appendix B we show alternative best models derived with different values of $\alpha$. In particular, we show how the likelihood of the best model improves slightly for values of $\alpha$ slightly larger than $\alpha=1$. However, the improvement in the likelihood is modest ($ln(\mathcal{L})=0.2642$ for $\alpha=1$ vs $ln(\mathcal{L})=0.2072$ for $\alpha=1.2$ and $ln(\mathcal{L})=0.1871$ for $\alpha=1.4$). 
The remaining of this paper focuses on models with $\alpha=1$, but we consider other values in the discussion section. The reader is directed also to appendix B for the specific results derived with  $\alpha<1 $, and $\alpha>1$. 
 
\begin{figure} 
   \centering
   \includegraphics[width=9.0cm]{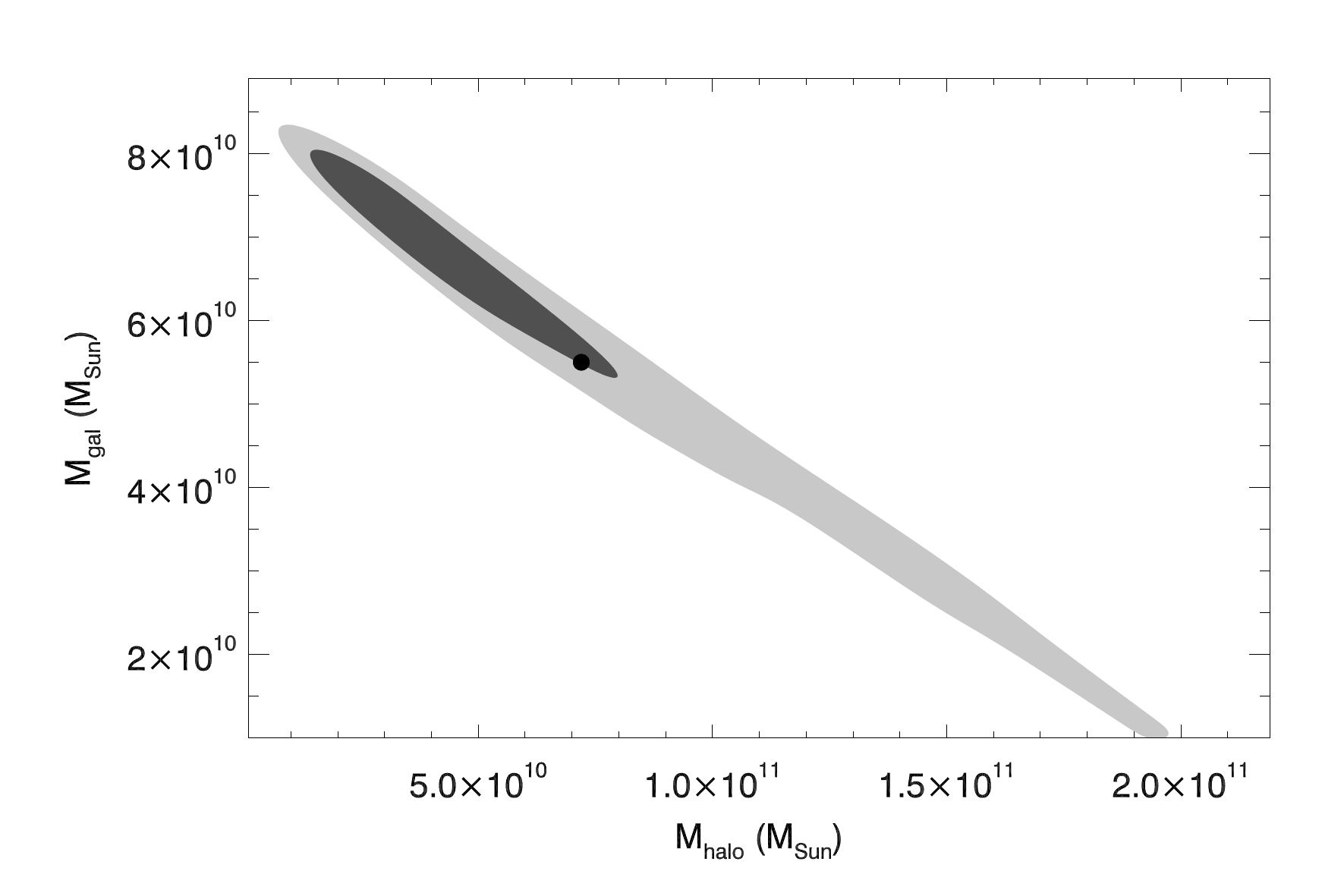}
      \caption{Marginalized probability in the ${\rm M}_{\rm halo}$--${\rm M}_{\rm gal}$ plane. The best model (black dot) is at ${\rm M}_{\rm halo}=7.2\times 10^{10}$ \Msun and ${\rm M}_{\rm gal}=5.5\times 10^{10}$ \Msun. It has a likelihood value of $ln(\mathcal{L}) = 0.2642$. Note how the marginalized probability favors models with a lower halo mass. The ellipticity of the best model is $e=0.4$, and the orientation is -56$^{\circ}$ anticlockwise with respect to North. (Shiftx=-19 and ShiftY=+20 pixels or 0.23 kpc NE from the centre of the galaxy at -9,0)  
              }
         \label{Fig_MhaloMgal1}
   \end{figure}

 Based on the fiducial model, in Figure~\ref{Fig_MhaloMgal1} we show the marginalized likelihood for the masses of the DM halo and galaxy (the constraints on $e,\psi, dx$, and $dy$ are weaker and not as well defined). The best model with the maximum likelihood is marked with a black dot.

The best model has comparable masses in the baryonic and DM components (note that the DM halo model is restricted to the same aperture as the baryonic model so the mass of the DM halo corresponds to the one enclosed in this aperture, while the real DM halo will likely extend further). The orientation of the best DM halo is similar (within a few degrees) to the observed orientation of the galaxy on large scales. However, the best DM halo is shifted with respect to the centre of the galaxy by $\approx 0.23$ kpc in the NE direction. The DM halo shape (and its offset with respect to the galaxy) is shown as a red ellipse (and a red dot) in Figure~\ref{Fig_Galaxy}. Interestingly, this displacement of 0.23 kpc is similar to the offset between the NE and SW sector profiles shown in  Figure~\ref{Fig_ProfileEastWest}, suggesting that the reason behind the apparent asymmetry between the NE and SW sectors discussed in section \ref{subsect_jointmacro} may be related to the 0.23 kpc offset between the DM halo and the galaxy. 

The critical curves, caustics, and predicted source (based on a simple toy model) of the best model are shown in Figure~\ref{Fig_Caus_CC1}. The critical curves (right panel) trace the shape and orientation of the galaxy and DM halo. The corresponding caustics (left panel) show the traditional diamond shape of elliptical potentials. The radial caustic (external elliptical curve) is relatively far away from the tangential caustic (diamond shape), as expected for a lens with a very dense central region (possibly hosting a  SMBH). We model the source using a simple toy consisting of a small nucleus, an elliptical halo and a point-like bright source, representing the SN. The source model is shown in the right panel with the nucleus crossing the caustic and the SN position (blue dot) north of it. This simple model is able to reproduce the main features in the observed image as shown in the left panel, where we show the predicted image of the source model. The lensed SN positions fall very close to the observed positions (yellow dots), the lensed nucleus produces a bright feature between images 1-4 and fainter images between the other images, similar to the features observed in the data. The gaps between the images are also well reproduced. 
Due to the small core radius of the halo model, and compact central configuration of the baryon model, the predicted fifth central image has very small magnification and is not observed, as in the observations.

   \begin{figure} 
   \centering
   \includegraphics[width=9.0cm]{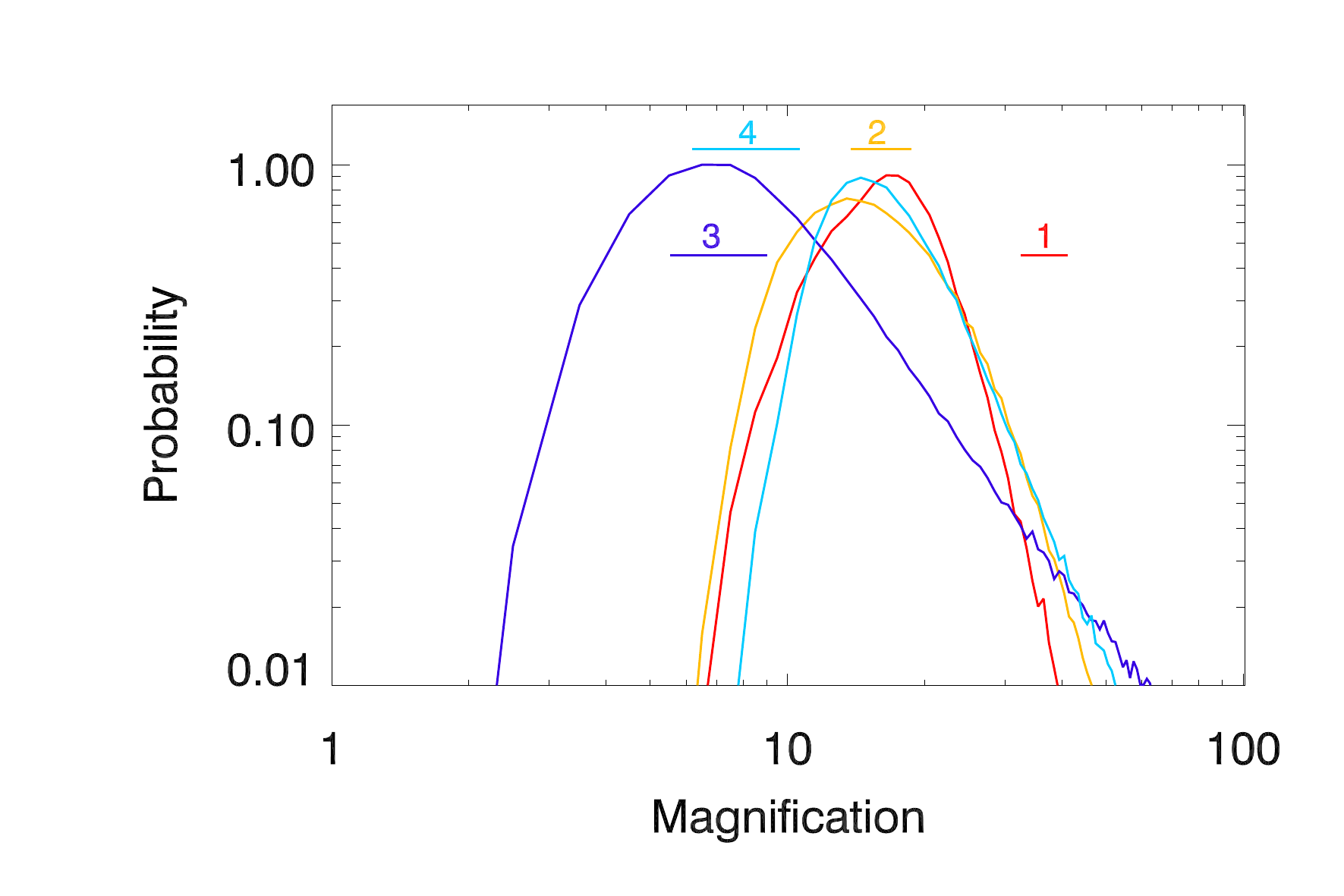}
      \caption{Marginalized probability distribution of the magnification obtained when only the 4 positions of the SNe images are used as constraints. The curves show the marginalized probability for each of the 4 SNe images. The horizontal colored lines show the inferred range of magnification from observations. 
              }
         \label{Fig_ProbMu1}
   \end{figure}

The marginalized probability for $M_{\rm gal}$ and $M_{\rm halo}$ is shown in Figure~\ref{Fig_MhaloMgal1}. The contours correspond to the 68\% and 95\% intervals and the black dot marks the position of the best model. Note how the best model is outside the 68\% region of the marginalized probability. The N-dimensional likelihood near the best model forms a shallow valley that extends towards the 68\% confidence region. The best models clearly prefer a relatively narrow region in the  $M_{\rm gal}$--$M_{\rm DM}$ space. An interesting result from this figure, is that a baryon-only model (i.e., $M_{\rm DM}=0$) does not perform much worse than a model with a DM halo, although the data prefers a model with a DM halo. The inferred mass of the galaxy (baryons) in the best model ($5.4\times10^{10}$ \Msun) is marginally consistent with the estimated stellar mass derived from the velocity dispersion -- stellar mass correlation \citep{Hyde2009,Zahid2016,Cannarozzo2019}. Assuming the velocity dispersion in \cite{Mortsell2020}, and the models in  \citep{Zahid2016,Cannarozzo2019}, the predicted stellar mass is $\approx 2.5\times10^{10}$ \Msun. This is about a factor two less than the stellar mass inferred from the lens model. However,  a factor 2 uncertainty is typical in mass estimations from the velocity dispersion (see the references above). Another possibility is that the gas mass is similar to that of the stars, increasing the baryonic mass by a factor 2 compared with the stellar mass. This is in principle feasible based on the gas to stellar mass ratios, which is close to 1 for early type galaxies (like the lens considered in this work) and stellar masses $\sim 10^{10}$ \Msun \citep{Calette2018}. However, in the central part of the galaxy the stellar component is still expected to dominate, especially if a bulge is present, as suggested by the data. An alternative way of estimating the contribution from the gas is by comparing with galaxies of similar morphological type. According to \cite{Casasola2020}, and considering a morphological type for the lens galaxy between Sa and Sb, the gas fraction for this type of galaxy is $\approx 20\%$.   
In \cite{Schruba2011}, gas and molecular surface densities as high as O(100) \Msun pc$^{-2}$ can be found in areas with large star formation rates. In the most extreme cases of star formation rates, gas surface densities of $\approx 1000$ \Msun pc$^{-2}$ can be found. However, even at these extreme star formation rates, the contribution from the gas to the baryonic mass is still subdominant with respect to the stellar contribution from our fiducial model. 

\begin{table}
 \centering
  \begin{minipage}{90mm}
    \caption{Magnification, total convergence, shear (for the lens and source redshifts), and surface mass densities of the galaxy and DM halo at the 4 SN positions. The critical surface mass density for the redshifts of the lens and source is 5025.6 \Msun pc$^{-2}$.}
 \label{tab_1}
 \begin{tabular}{|c|ccccc|}   
 \hline
   Image   & $\mu$  & $\kappa$   &     $\gamma$   & $\Sigma _{\rm gal} \left ( \frac{\rm M_{\odot}}{{\rm pc}^2} \right )$ & $\Sigma _{\rm halo} \left (\frac{\rm M_{\odot}}{{\rm pc}^2} \right )$ \\
 \hline
 1    &  14.95  &   0.681  &  0.411   &   2467.6  &   965.6     \\    
 2    &  11.64  &   0.551  &  0.340   &   2077.4  &   689.2     \\    
 3    &   6.42  &   0.683  &  0.506   &   1962.1  &  1462.1     \\    
 4    &  13.27  &   0.550  &  0.357   &   1999.9  &   765.3     \\    
 \hline
\end{tabular}
\end{minipage}
\end{table}

For each model we compute the magnification in each of the four SN positions. The models are then marginalized as a function of their magnification, after being weighted by their likelihood. Figure\ref{Fig_ProbMu1} shows the marginalized probability (solid lines) as a function of magnification, at each of the 4 SN positions, with a  different color being used for each SN image. The horizontal bars labelled with numbers 1--4 show the range of observed magnification for each SN image as estimated in \cite{Dhawan2020}. The peak of probability from the lens models agrees well with the observed magnifications for images 2 and 3 but not so well but images 1 and 4, with image 1 having the largest deviation between the model prediction and the observation.  
The predicted magnifications from the best model at the four SN positions are listed in \ref{tab_1}, together with the values of the convergence, shear and surface mass densities of the galaxy and DM halo at the same positions. 

\begin{figure} 
   \centering
   \includegraphics[width=9.0cm]{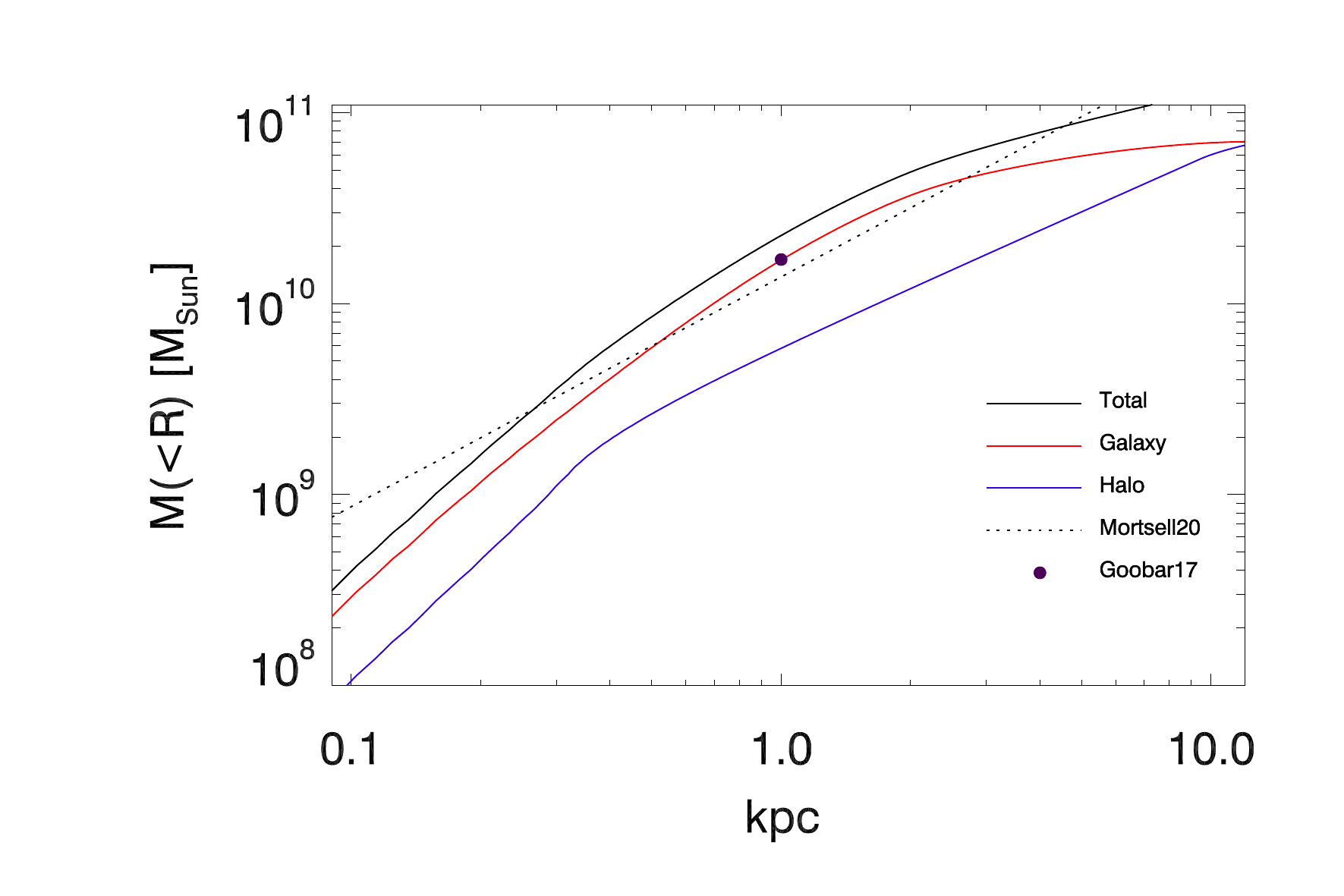}
      \caption{Circularly averaged integrated mass profiles of the best model for the case where only positional constraints are used. The profile of the galaxy is shown in red and the profile of the halo is shown in blue. The profile of the total mass is shown as a black curve. The black dot at 1 kpc is the estimated mass in the original lens model of \cite{Goobar2017}. The dotted line is the new estimate by \cite{Mortsell2020} using a circular aperture. An isothermal profile scales its integrated mass as the radius. 
              }
         \label{Fig_Profiles}
   \end{figure}

It is interesting to compare the best model with earlier estimates. In Figure \ref{Fig_Profiles} we show the integrated mass as a function of distance from the centre of the galaxy. We show the profiles for both components, the baryon (red curve) and dark matter (blue curve). The total mass is shown as a black line. The original model by  \cite{Goobar2017} estimated a mass of $\approx 1.7\times10^{10}$\Msun within the critical curve. This estimate is shown as a black circle in Figure \ref{Fig_Profiles}. 
The same team produced a new lens model in \cite{Mortsell2020}. Using the parameters of their singular isothermal model, where $\alpha=1.2$, we compute the integrated mass as a function of circular aperture. The result is shown as a dotted line in the same figure. 
An additional estimate, compatible with the ones above, is presented in appendix \ref{sect_James}.
In table \ref{tab_1}, the magnifications of image 1 (the most discrepant with earlier estimates) is predicted to be 14.95. This is comparable to the value predicted by the model in  \cite{Mortsell2020} with slope $\alpha=1.2$ (see their Table 4) that predicts $\mu_1=14.2$\footnote{Note that \cite{Mortsell2020} uses $\eta$, a different definition for the slope, but the two slopes are related by $\alpha=3-\eta$}.

In \cite{More2017}, several models are produced using two different algorithms. The magnification of image 1 from these models is in the range $\mu_1=5.2$--8.2. 
A larger magnification for image 1 of $\mu=18.41$ is predicted by the recent M3 model in \cite{Williams2020} (model M3 is selected by the authors as the best in terms of predicting the magnification of the first image). In this work, the authors use a model that resemble ours since the lens model is composed of two components. However, in contrast with our lens model, none of the components is required to trace the light of the observed galaxy.

\begin{figure} 
   \centering
   \includegraphics[width=9.0cm]{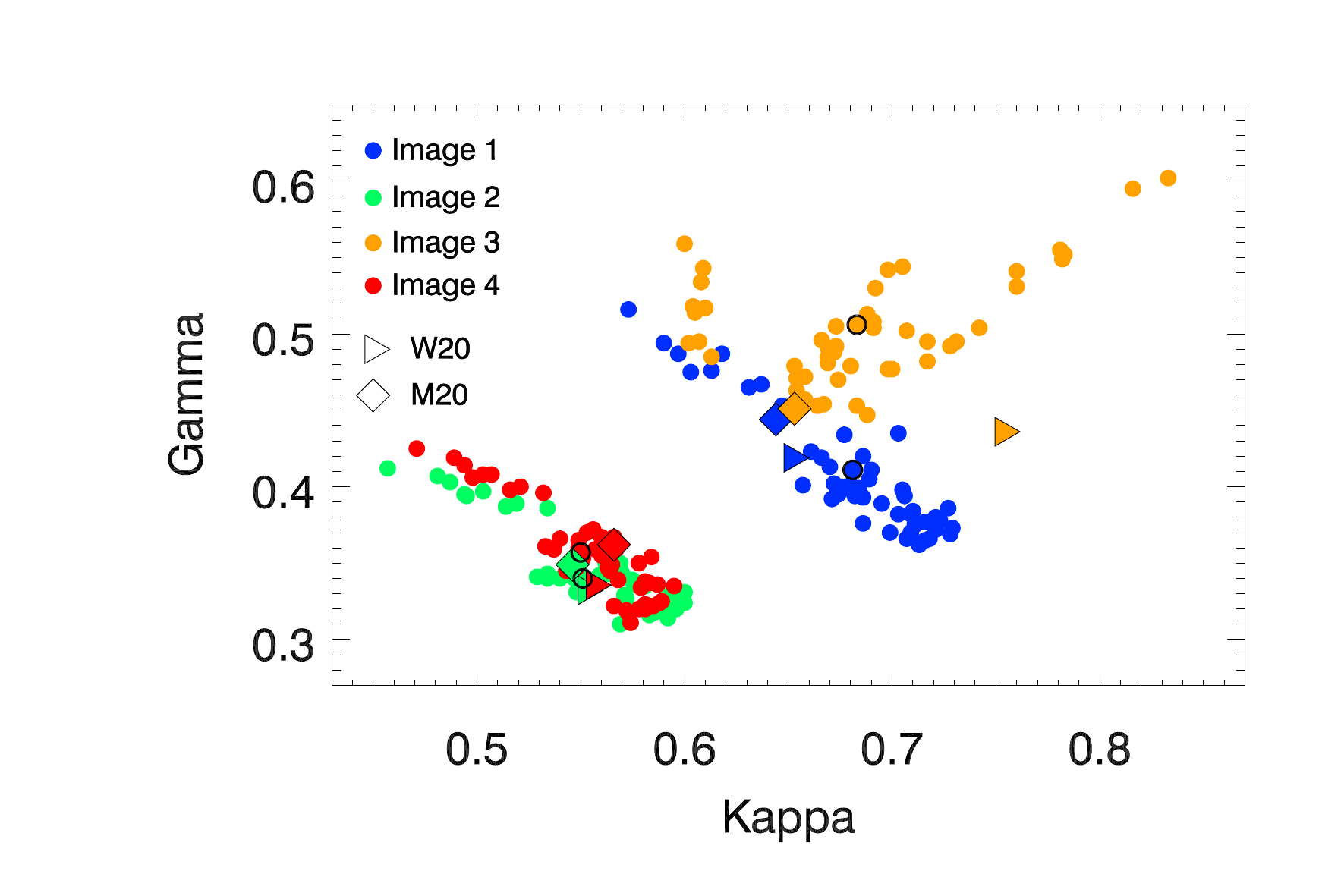}
      \caption{The colored disks show the values of $\kappa$ and $\gamma$ for 50 models which are randomly selected among the best likelihoods according to Equation \ref{Eq_Lkhd1}. Different colors are used for each image position. The disks surrounded by the thick black circles correspond to the best model in Table \ref{tab_1}. The triangles are the model M3 in \cite{Williams2020}, while the diamonds are the model in \cite{Mortsell2020}.  
              }
         \label{Fig_Comparison_WilliamsMortsell}
   \end{figure}

In order to compare our best models with recent independent estimates, we select 50 random models among the ones having the best likelihoods from Equation \ref{Eq_Lkhd1}, with likelihood values at least 0.25 times the maximum likelihood. The values of $\kappa$ and $\gamma$ for these 50 models is shown in Figure \ref{Fig_Comparison_WilliamsMortsell}, for the positions of the 4 images. The best model in Table \ref{tab_1} is highlighted with black circles surrounding the colored dots. In the same plot, we show the corresponding values for the best model in \cite{Mortsell2020}, and model M3 in \cite{Williams2020}. Remarkably, our best models are comparable, in terms of $\kappa$ and $\gamma$, with these earlier models, despite being derived under different assumptions. The largest disagreement appears in image 3 with the model in \cite{Williams2020}, which predicts a larger value for the convergence, probably due to a displacement of one of the two halos in \cite{Williams2020} towards the position of image 3. In contrast, our model predicts a larger value for the shear at the position of image 3. 
We also find a similar direction and offset of the mass asymmetry in the mass distribution as in \cite{Williams2020}. In that work, the second mass component is displaced from the center of light by about 0.2-0.25 kpc in the NE direction, which is consistent with our findings (0.23 kpc in the NE direction) using an alternative approach. This coincidence in results supports the idea that the centers of baryonic and dark matter are offset. 
Finally, in both lens models, the total density (DM+baryons) flattens near the center (r$<0.5$ kpc), and steepens towards the isothermal at larger radii (see \ref{Fig_Profiles}.

The larger macromodel magnification at the position of image 1 predicted by our model, helps to alleviate the tension with the observed value ($\mu_1>30$), but is not enough to explain it. As pointed out by other authors, we expect microlensing (or millilensing) to be responsible for this discrepancy. Before studying the impact of microlenses, it is interesting to explore the second scenario where we use the observed magnification of images 2 and 4 as additional constraints, and explore the range of models that could reproduce the observed magnifications without the need of microlensing.

\subsection{Positional and magnification constraints}
Because of the relatively small angular size of the SN, the lensed images can be  affected by microlensing effects. Due to the relatively low redshift of the SN, the critical curves form relatively close to the centre of the lens, where microlensing effects are expected to be more considerable. The saddle points (images 1 and 3) are more likely to be affected by microlensing than the minima points \citep{Paczynski1986,Wambsganss1990}. At the  saddle points, demagnification by large factors ($>3)$ are more likely than at the minima points where only more modest demagnification factors are possible. In this section we consider the magnification of images 2 and 4 as additional constraints, since they are expected to be the least affected by microlensing. Based on the 4 observed positions (contributing with two constraints each) and the two constraints on the magnification of images 2 and 4,  we redefine the likelihood as follows,
\begin{equation}
-2 ln(\mathcal{L})= \frac{(\mu_2-15.7)^2 + (\mu_4-9.1)^2}{\sigma_{\mu}^2} + 
\sum_i^4\frac{\delta\beta_{x_i}^2 +\delta\beta_{y_i}^2}{\sigma_i^2}
\label{Eq_Lkhd2}
\end{equation}
where 15.7 and 9.1 are the observed magnifications at positions 2 and 4 respectively. The second term in Eq.~\ref{Eq_Lkhd2} is the same as in Eq.~\ref{Eq_Lkhd1}.
For the error in the magnification, $\sigma_{\mu}$, we adopt the uncertainty in the observed magnification found by \cite{Dhawan2020} which quotes $\sigma_{\mu}=1.1$. 
We emphasize again that Eq. \ref{Eq_Lkhd2} assumes no microlensing effects in images 2 and 4, which is likely an unrealistic scenario, but the results obtained in this section are illustrative of the existing problems when the magnification information from the lensed SN images is used to constrain the macromodel.

   \begin{figure} 
   \centering
   \includegraphics[width=9.0cm]{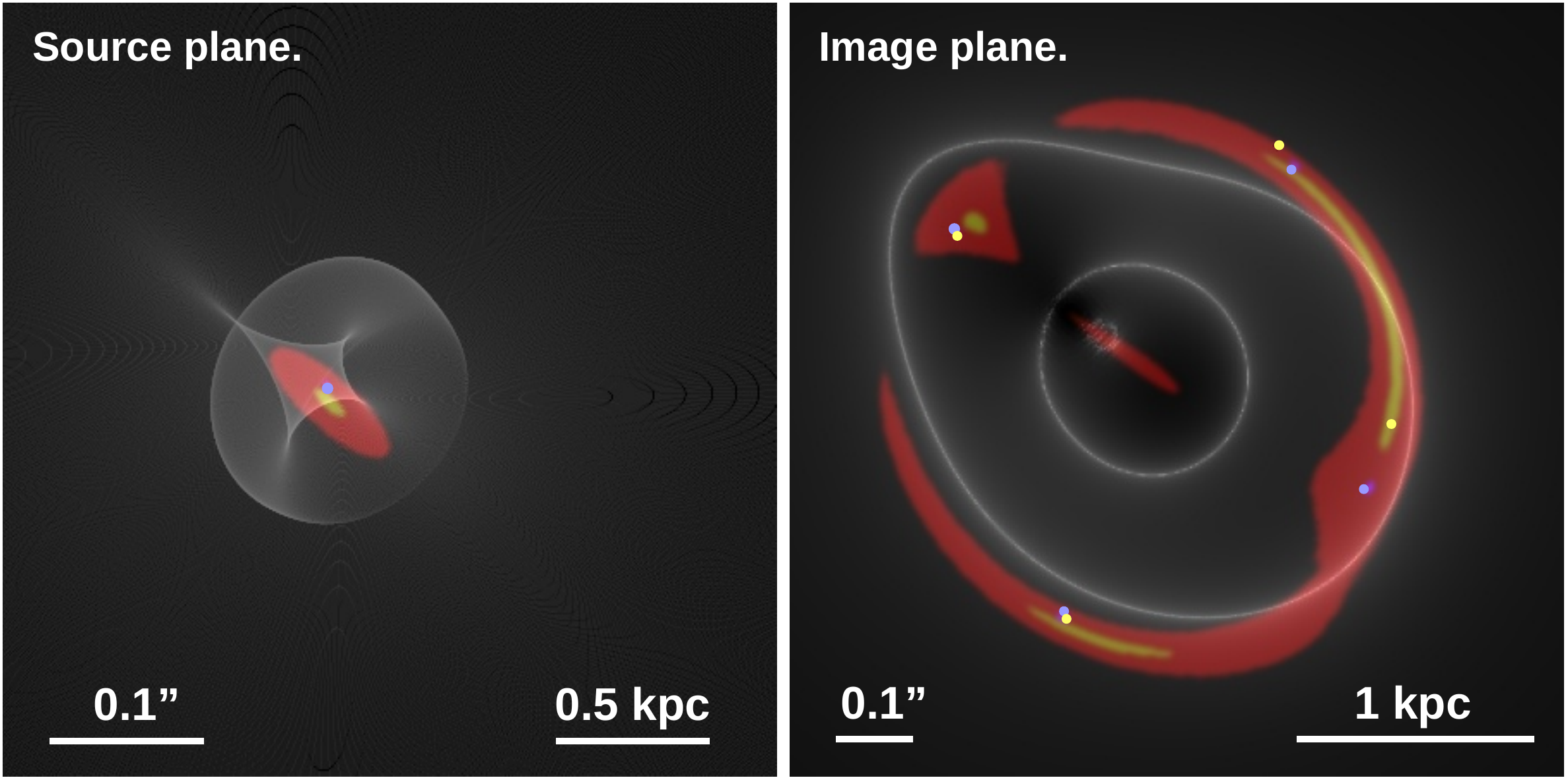}
      \caption{Caustics (left) and critical curves (right) for the model where both position of the 4 SNe and magnification of images 2 and 4 are used as constraints. A simple source with just three components, nucleus (orange), halo (red) and SN (blue) is shown in the left panel (source plane). The predicted image is shown in the right panel (image plane). The yellow dots in the right panel mark the observed position of the four SN images. Note how this model fails at reproducing the position of the SNe images 1 and 3. 
              }
         \label{Fig_Caus_CC2}
   \end{figure}

We use Eq.~\ref{Eq_Lkhd2} to find the best model, and the marginalized probabilities, for the case where the magnification in images 2 and 4 are used as extra constraints. The best model in this case is slightly different. Small change in the masses, ellipticity, and orientation of the best model (shown in Figure~\ref{Fig_Caus_CC2}) result in small, but non-negligible changes, in the predicted magnification of images 2 and 4. In terms of critical curves, and caustics, the changes are almost imperceptible as shown by comparing Figures~\ref{Fig_Caus_CC1} and \ref{Fig_Caus_CC2}. The tangential critical curve is now closer to image 2, bringing its magnification closer to the observed value. As a result, image 1 is also closer to the critical curve. The magnifications predicted by the best model are 33.1, 15.7, 7.6, and 10.0 from images 1,2,3 and 4 respectively. The source model is similar to the one presented in the previous subsection (with small adjustments in position). Overall, the main observed features are reproduced well with the exception of the SN positions which are significantly offset, especially for images 1 and 4. 

\begin{figure} 
   \centering
   \includegraphics[width=9.0cm]{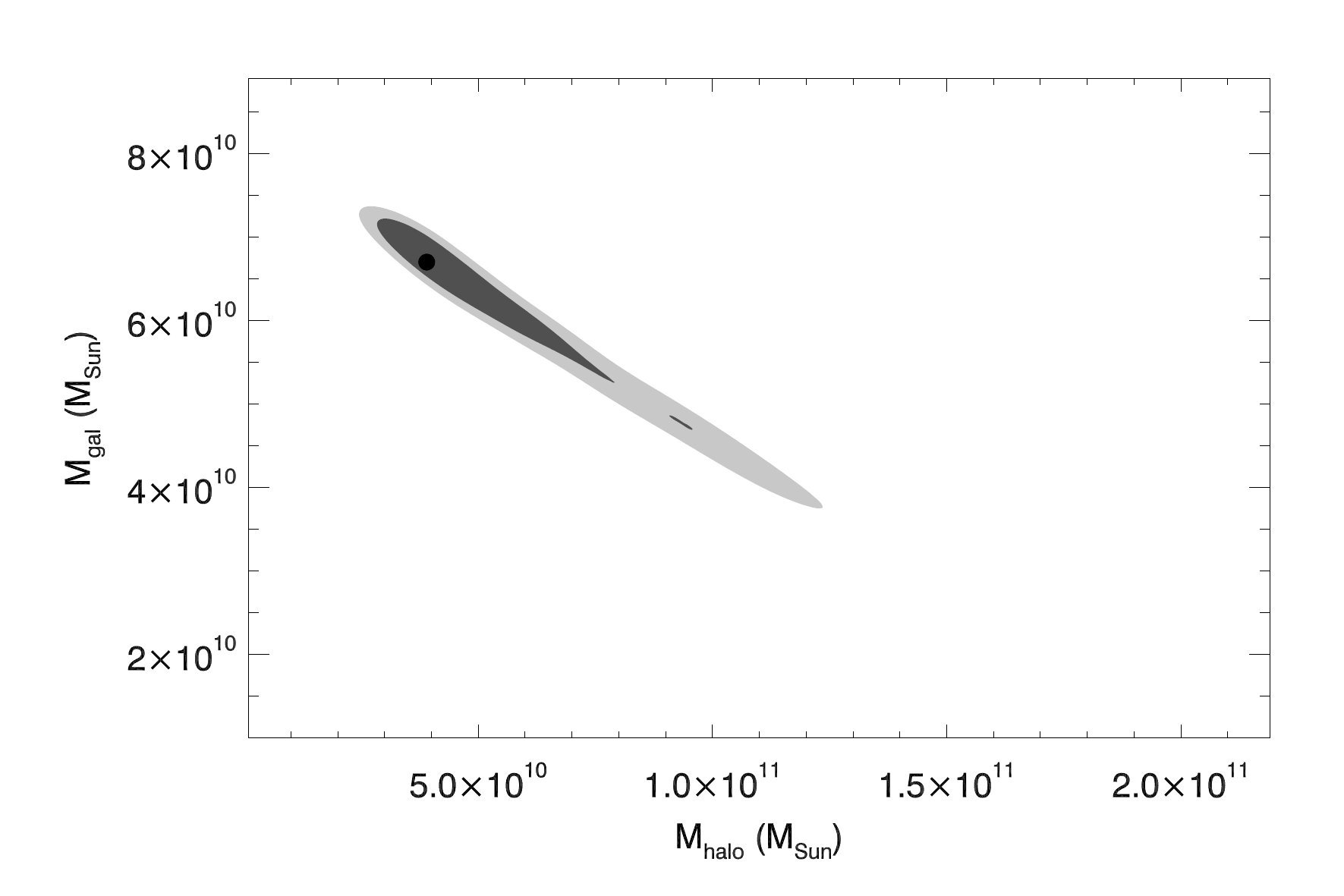}
      \caption{Marginalized probability in the ${\rm M}_{\rm halo}$--${\rm M}_{\rm gal}$ plane. 
              }
         \label{Fig_MhaloMgal2}
   \end{figure}

\begin{figure} 
   \centering
   \includegraphics[width=9.0cm]{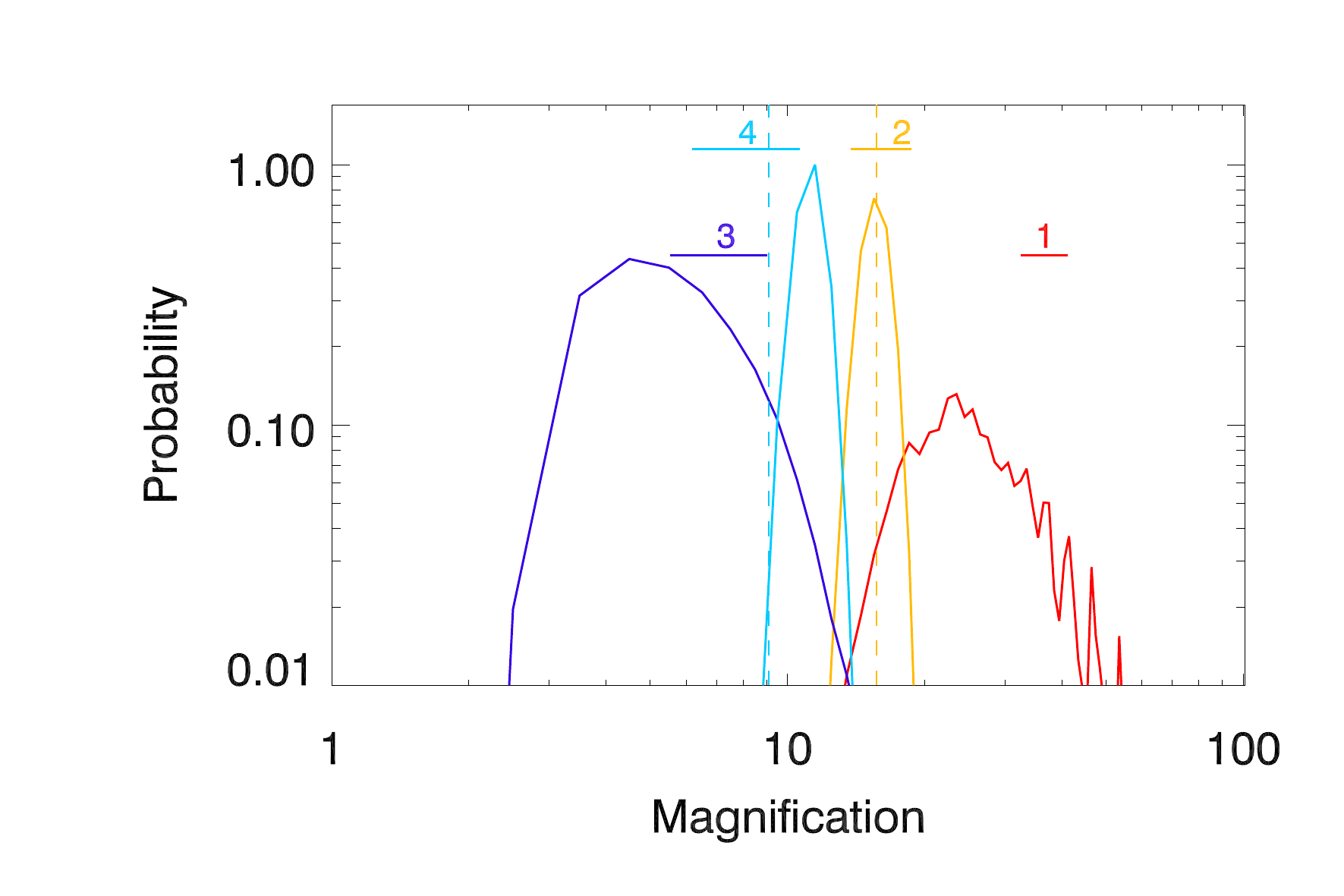}
      \caption{Like in Figure~\ref{Fig_ProbMu1} but when both position plus magnification (of images 2 and 4) are used as constraints. The curves show the marginalized probability for the magnification at the position of the 4 SNe images. The horizontal colored lines show the inferred range of magnification from observations. The two dashed lines indicate the magnification values used to constrain the lens model in images 2 and 4.  
              }
         \label{Fig_ProbMu2}
   \end{figure}

In terms of marginalized probabilities, DM halos having masses higher than the ones found in the previous subsection than are preferred (see Figure~\ref{Fig_MhaloMgal2}). The marginalized probabilities for the magnification show a better agreement with the observed values, as shown in Figure~\ref{Fig_ProbMu2}. For images 2 and 4 the probabilities concentrate around the values used as constraints (dashed lines).  
Even though the magnification of images 1 and 3 are not used as constraints, this model reproduces the observed magnification of these images relatively well. These models would in principle not require the existence of microlensing to explain the observed magnifications, but as shown in Figure~\ref{Fig_Caus_CC2}, the success in reproducing the magnification comes at the expense of degrading the overall performance of the model, in particular at reproducing the positions of the 4 SN images. Since the SN positions should be  unaffected by microlensing, one should trust more the model derived with the positional constraints only, that is able to reproduce well the observed SN positions. A similar result is found in \cite{Mortsell2020}, which concluded that independently of the assumed lens model (in particular its slope), the positions and flux ratios could not be reproduced simultaneously.  Hence substructures that are not included in the model are probably responsible for the flux discrepancy.  In the next section, we study if the discrepancy in magnification of the best model derived with positional constrains only (previous subsection) can be explained with microlensing, which must be ubiquitous given the proximity of the critical curve to the centre of the galaxy.

\section{Microlensing}\label{sec_micro}
In this section we explore the role that microlenses play on the magnification of the 4 lensed SN images. For smooth elliptical macromodels and quadruple images, one expects the signed sum of the magnifications (i.e., the sum of the magnifications $\mu _i$ weighted by their parity $p_i=\pm 1$), to be relatively small \citep[typically between 1 and 3][]{Witt2000}. Adopting the observed values in iPTF16geu \citep[from the fiducial model in][]{Dhawan2020,Mortsell2020} one obtains $\left | \sum _i p_i \mu _i \right | \approx 18$, far from the expected small values. This discrepancy suggests that some level of substructure, or departure from symmetry is expected in the lens plane. For comparison, the best model for the case where only positional constraints are used, results in a signed sum of 3.54, close to the expected value for smooth lenses. Also for comparison, using the values of the model in \cite{Mortsell2020} (see their table 4), one obtains a signed sum of 3.1. 

A simple explanation for the discrepancy with the observed magnification, is that stars in the lens are acting as microlenses. As discussed in the introduction, this hypothesis has been explored in earlier work for this particular lens. Since SNe are relatively small, their photospheres can be completely contained within the caustic regions of microlenses with stellar masses. For instance, a microlens with a  mass of 1 \Msun would form a caustic of size $\approx 10^{-2}$ pc, or about an order of magnitude larger than the typical photosphere size of a SN, one month after explosion.
In this section, we revisit the role played by microlenses, but subject to the constrain on their abundance imposed by the baryonic component in our macrolens model. Our aim is to test if a standard model for the stellar population, and that is consistent with the lens model, is able to reproduce the observed magnifications and light curves. We follow \cite{Diego2018,Diego2019} to compute the combined deflection field of the microlenses plus smooth component (macrolens). A brief description of the microlensing simulations is given in Appendix A.

We estimate the probability of magnification in the presence of microlenses by adding a distribution of stellar microlenses (stars and remnants) in the lens plane and around the line of sight of the 4 SN images. For the macromodel, we adopt the best model obtained with the positional constraints only. The values for the convergence, shear, and stellar surface mass densities used for the simulations are listed in table \ref{tab_1}. The combined deflection field is used to compute the magnification in the source plane by inverse ray tracing (details are provided in the appendix). 
Figure~\ref{Fig_Caustics_Pos1} shows the caustics in the source plane in one of our simulated fields. In particular, this case corresponds to the caustics produced by the stars at the position of image 1. The stars in the line of sight of the other 3 images produce additional caustics, which would overlap in the same source plane, but are not included in this figure for clarity purposes. 
The average magnification in the area shown in Figure~\ref{Fig_Caustics_Pos1} is 15.05, very close to the magnification predicted by the macromodel ($\mu_{macro}=14.95$). The small mismatch between the mean magnification and $\mu_{macro}$ is due to the limited area of the simulated region. The average magnification computed in a larger simulated area would be closer to the macromodel value.  For illustration purposes, we mark some positions in the image with numbers (in blue) indicating the magnification in that particular region. Note the large fluctuation in magnification values, ranging from a few to almost a hundred. The yellow dot near the centre is placed in an area with magnification $\approx$ 30--40, similar to the observed magnification of image 1. The size of the yellow dot is comparable to the expected size of the photosphere of a typical SN one month (in the rest frame) after the explosion, that is, $R\approx 10^{-3}$ pc, assuming typical expansion velocities $v \approx 10000$ km s$^{-1}$ \citep{Pan2020}.
   \begin{figure} 
   \centering
   \includegraphics[width=9.0cm]{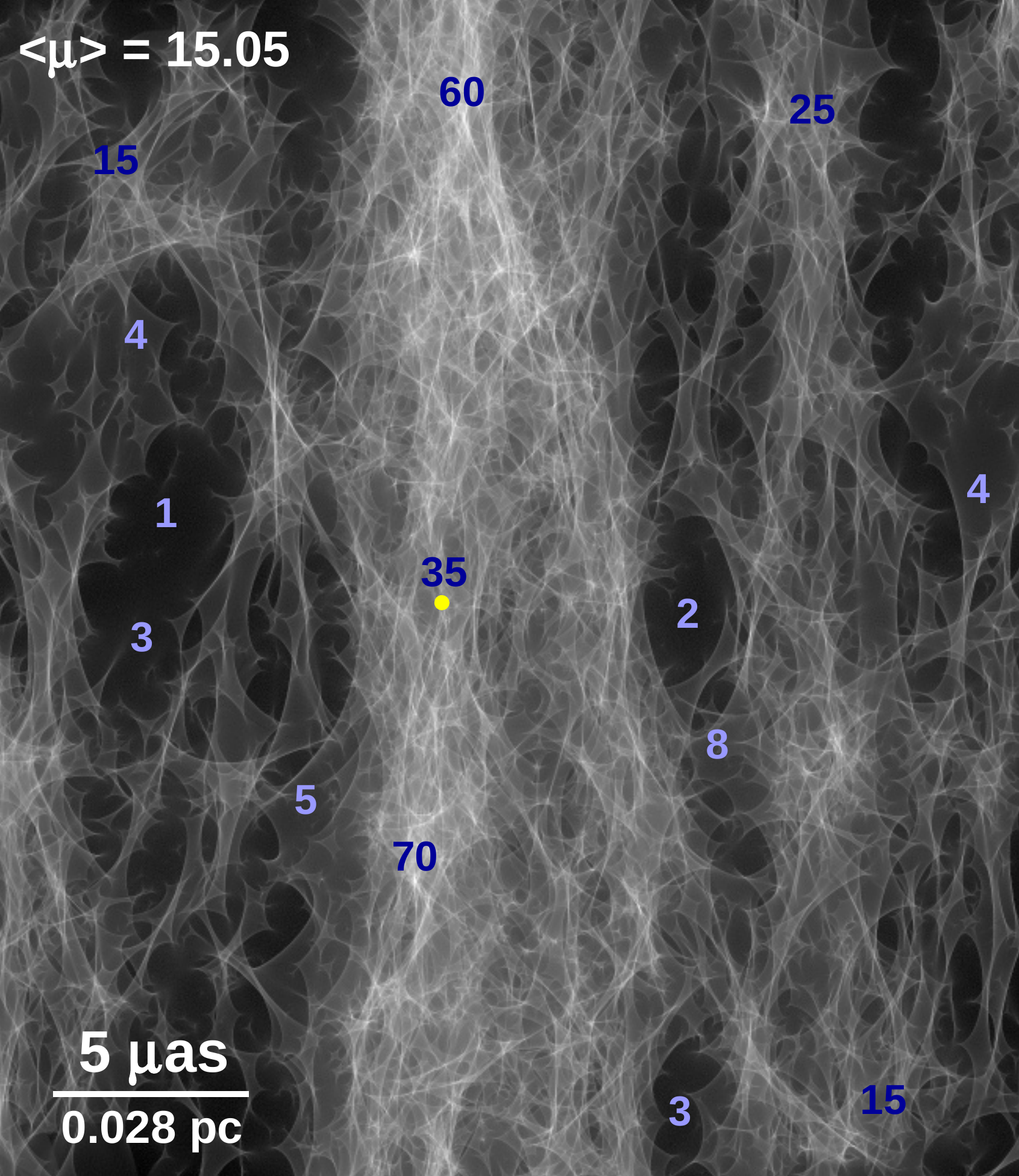}
      \caption{Caustics corresponding to position 1. The yellow circle near the centre has a radius of $\approx 10^{-3}$ pc at the redshift of the SN. The blue colored numbers indicate typical magnifications at those positions. Although this image represents the microcaustics around position 1, the SN is sensitive to the superposition of the 4 caustic planes shown in Appendix A (figure \ref{Fig_Caustics_4panels}).
              }
         \label{Fig_Caustics_Pos1}
   \end{figure}

\begin{figure} 
   \centering
   \includegraphics[width=9.0cm]{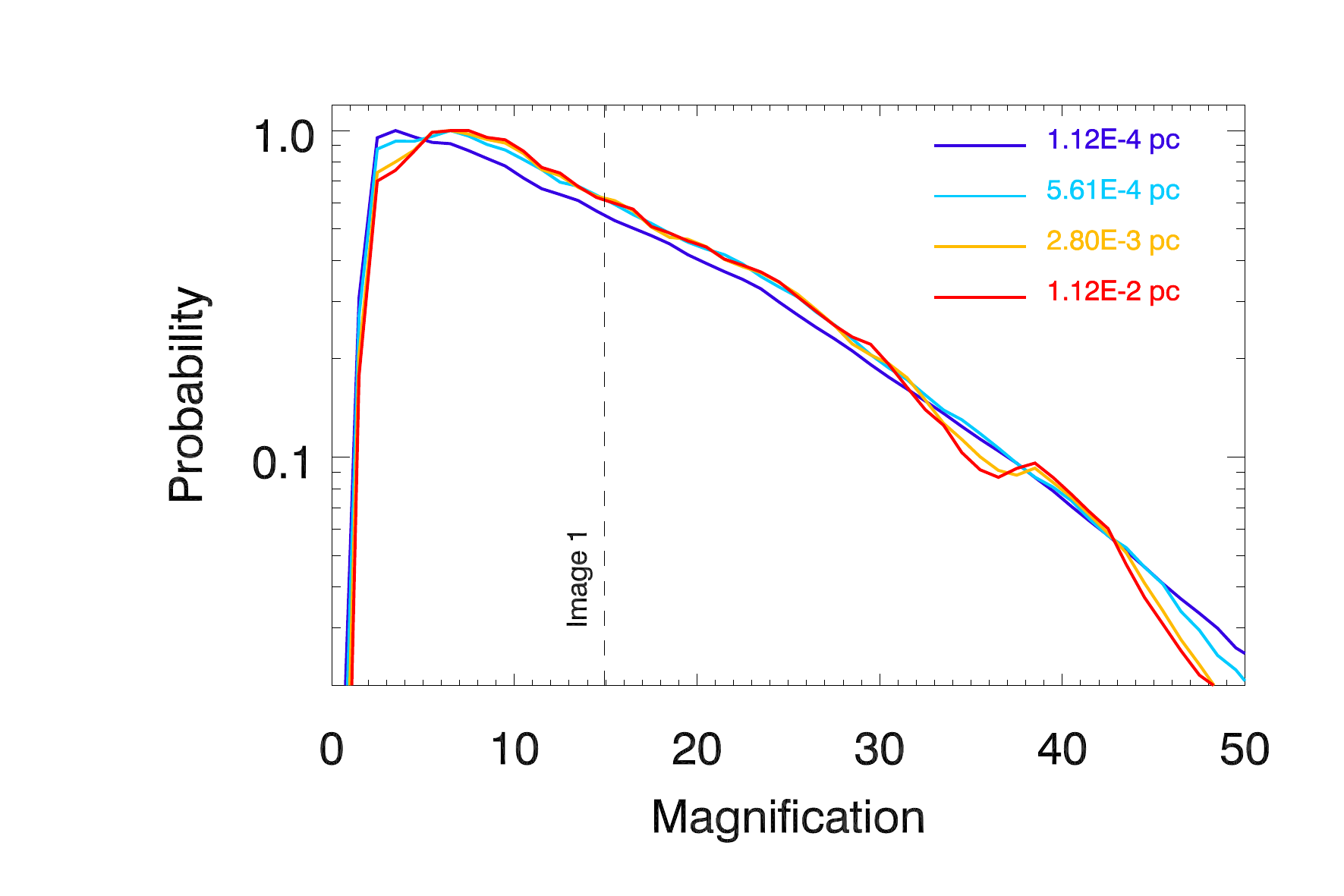}
      \caption{Probability of magnification (dN/d$\mu$) at the position 1 and with microlensing from stars. The surface mass density of stars is the one corresponding to the baryon model in position 1 listed in table~\ref{tab_1}. The IMF is a Salpeter model with a cutoff in 0.1 \Msun and includes remnants. The dashed vertical line marks the predicted magnification from the macromodel (without microlensing) at position 1. Each curve represents the magnification experienced by a source modelled as a Gaussian with different FWHM (expressed in pc). For typical expansion velocities of the photosphere, the smallest FWHM represents the size of an expanding photosphere after a few days the initial explosion. After two months, the photosphere is expected to have expanded up to a radius $R \approx 2\times 10^{-3}$ pc (orange curve).      
              }
         \label{Fig_ProbMu_MicroPos1}
   \end{figure}

In order to account for the size of the expanding photosphere, we convolve the magnification in the source plane with Gaussians of varying width. As discussed in, for instance, \cite{Pierel2019}, a Gaussian distribution is a simple approximation to the flux distribution of the SN photosphere.  The resulting magnification distribution for image 1 is shown in Figure~\ref{Fig_ProbMu_MicroPos1}, for different values of the FWHM of the Gaussian (expressed in parsecs in the legend of the figure). In the absence of microlensing, and assuming the macrolens model is correct, one would  expect to observe image 1 with the magnification indicated as a dashed line, that is, $\mu=14.95$. Microlenses perturb the magnification pattern in a manner shown in Figure~\ref{Fig_Caustics_Pos1}. The average magnification in the simulated region is very close to the magnification from the macromodel. Hence, a source that is several parsecs across in size, and that is homogeneous in flux, would be virtually insensitive to the fluctuations in the magnification at small scales.  However, sources that are sufficiently small are capable of probing only small regions in the source plane, where fluctuations in the magnification can be significant. As the small source moves relative to the lens and the observer, it crosses multiple microcaustics. An observer will measure fluctuations in the flux that depend on the strength of the microlens and the size of the background source. Equivalently, if the source expands over time, such as a SN, the growing photosphere intersects microcaustics and regions of different magnifications. Also, as the photosphere grows, the observed flux is the convolution of the surface brightness of the source with the magnification map. For large photosphere radii the convolved signal converges to the value expected for the average magnification in (i.e., the macromodel magnification). The effect of the expanding photosphere is shown in Figure~\ref{Fig_ProbMu_MicroPos1} as different curves, one curve for a different size (shown in parsecs). For the estimated velocity of iPTF16geu presented in \cite{Johansson2021}, we estimate that one month after the explosion of the SN, the photosphere reaches a radius of $R=1.2\times10^{-3}$ pc, increasing to $R=2.1\times10^{-3}$ pc two months after the explosion. 
In Figure~\ref{Fig_ProbMu_MicroPos1} we see that if the SN is initially located in a region in the source plane with magnification larger than $\mu=50$, after the first month, the typical magnification stars to decline. This is the average expected behaviour, but for particular positions in the source plane near microcaustics, the magnification could first grow and later decline. Also from the same plot, we can appreciate that the change in magnification is not substantial (typically less than a factor 2) during the first month after the explosion. However, changes in magnification as the photosphere expand do occur, and they can be used to constrain the microlens model as we discuss below. 

One can use results similar to Figure~\ref{Fig_ProbMu_MicroPos1} (at the different SN positions, and for different macro- and micro- models)  to estimate the probability that the observed magnification can be reproduced by a combination of a macromodel model like the ones presented in section \ref{Subsect_PosOnly}, and a microlens model like the one presented in this section (and the appendix A). Considering image 1, and based on the probability of magnification shown in Figure~\ref{Fig_ProbMu_MicroPos1}, we can compute the relative probability of a background source to have magnification between 30 and 40 (i.e., the observed magnification of image 1). Despite the rapid decline of probability at higher magnifications, this probability is found to be only a factor 2.5 times smaller than the probability of having magnification between 12 and 17, which is the range predicted by the macromodel without microlensing. If we compare the probabilities above and below 15 (i.e., the value predicted by the macromodel), We find that $P(\mu<15)/P(\mu>15)=1.6$, indicating that microlensing is more likely to demagnify image 1 (with respect to the case without microlensing) but the probability of relative magnification is not significantly smaller. 

\begin{figure} %
   \centering
   \includegraphics[width=9.0cm]{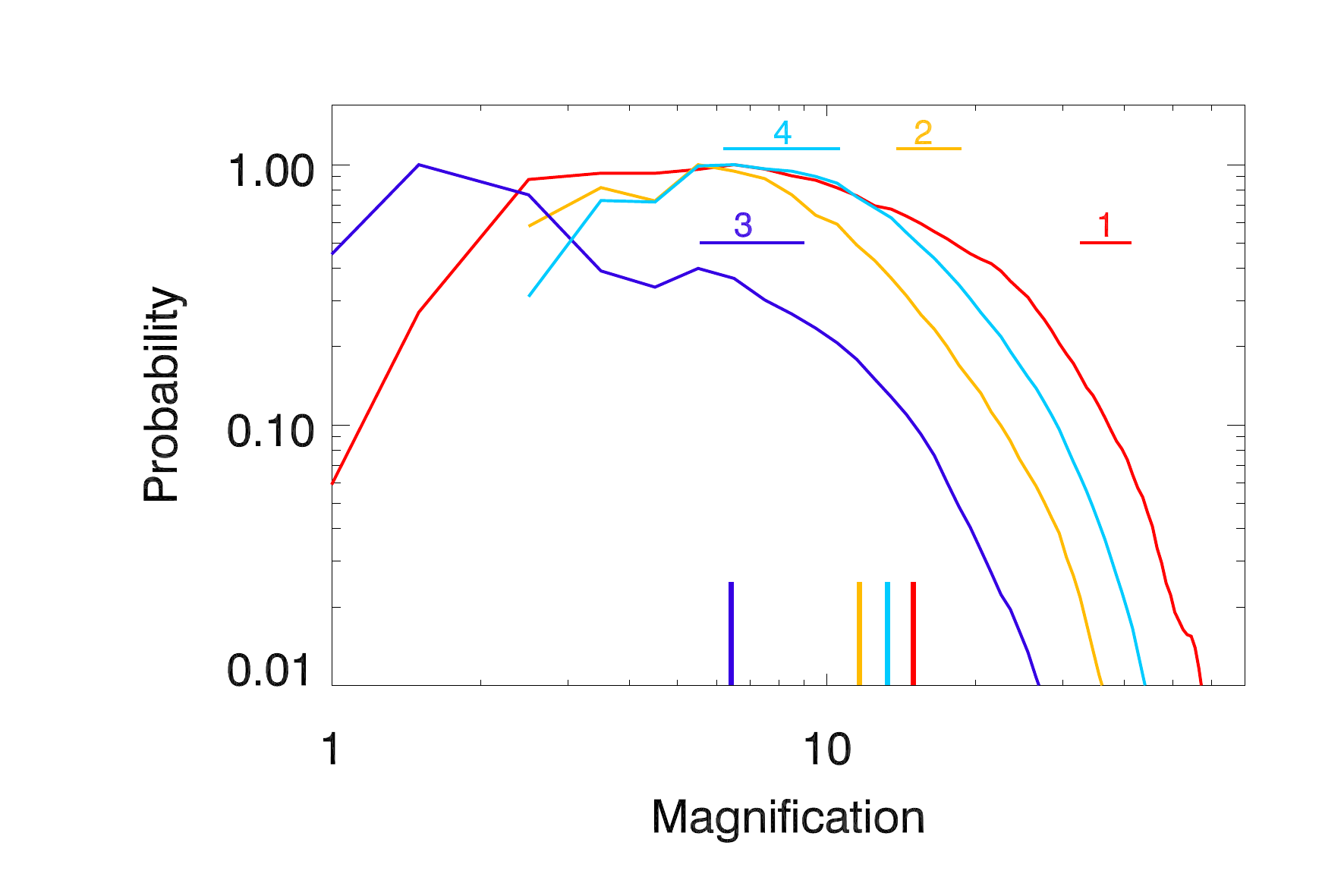} 
      \caption{Probability of magnification at the 4 image positions when microlenses are added. Each color corresponds to one of the positions. The horizontal bars show the observed magnification with its uncertainty. The four vertical lines in the bottom of the plot mark the predicted magnification from the macromodel (i.e., without microlenses). The curves show the probability of the magnification when microlenses are included. The curves are computed assuming the source is a Gaussian with FWHM=$5.6\times10^{-4}$ pc. This size is similar to the extent of the photosphere of a typical SN during its peak emission.  
              }
         \label{Fig_ProbMu_Micro4Positions}
   \end{figure}

Similar estimations can be done for the remaining 3 positions. Figure  \ref{Fig_ProbMu_Micro4Positions} shows the probability of magnification for the best macromodel in section \ref{Subsect_PosOnly} when microlenses are added. From this plot, the observed magnification of image 4 is at the maximum of the expected probability, while for the other images the probabilities are smaller, with image 1 showing the largest deviation. However, when compared with the case where no microlenses are present (see Figure \ref{Fig_ProbMu1}), the probability of being magnified by a value similar to the observed value increases by approximately an order of magnitude.   

In the next section we combine the available observables on iPTF16geu to constrain the combination of macro+micro lens model. 

\section{Joint macro+microlens model optimization}
In the previous sections we have shown how a macrolens model can explain the observed positions of the four SN images but not the magnifications. When the macromodel is required to explain also the observed magnification in two of the images (the ones where microlensing effects are expected to be smaller), the error in the predicted positions degrades significantly, especially for images 1 and 4. The likelihood of the best model is also appreciably worse ($\Delta[ln(\mathcal{L})] \approx  6.6$). When microlenses with the surface mass density of the fiducial model are included, the positional accuracy remains unchanged but there is a notable improvement in the probability of reproducing the observed magnifications. On the other hand, when the microlenses  are added, one expects changes not only in the magnification of the four images, but also in the light curve for a fraction of the simulated light curves. As the photosphere increases in size, parts of its surface intersect microcaustics, introducing small, but measurable changes in the magnification (or observed flux). Using the light curves as a way to constrain the microlensing component of the macro+micro model has not been explored in earlier work. 
In this section we combine the macrolens and microlens models. We explore a range of macrolens models, (all reproducing the observed configuration of the lensed galaxy and SN images), and add microlensing models with varying amounts of stellar surface mass density. The different amounts of microlensing  encode our relative ignorance on the contribution of the stellar mass to the total mass. 
We combine all the available information from the 4 SN images. Namely the four SN positions, magnifications, and lack of relative (to the model) fluctuations larger than 0.2 magnitudes in the light curves. 
By comparing the predictions with then observations we aim at extracting information about the amount of microlensing, and the stellar component. 

\begin{figure} 
   \centering
   \includegraphics[width=9.0cm]{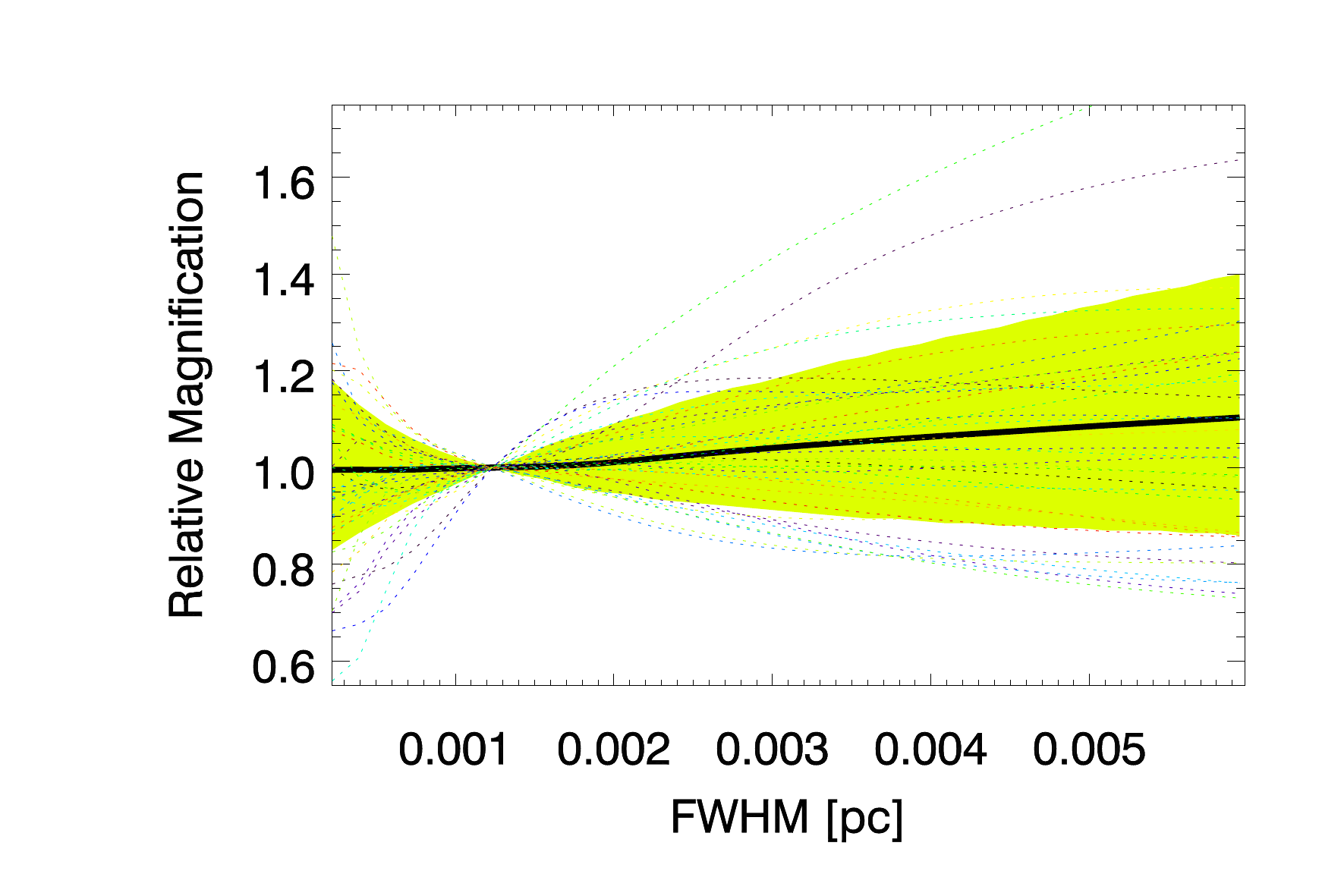}
      \caption{Photosphere weighted magnification distortion acting over the light curves, computed at the position of image 1. The distortion is plotted as a function of the FWHM of the expanding photosphere, and each realization has been re-scaled to the moment where the SN reaches maximum luminosity (FWHM$\approx 1.2\times10^{-3}$ pc). The green region marks the 68\% confidence interval from over 7 million simulated distortion curves. The colored dotted curves show 35 individual random realizations out of the 7 million. The thick black curve corresponds to the average of the 7 million curves. 
              }
         \label{Fig_LightCurvesPos1}
   \end{figure}

\begin{figure} 
   \centering
   \includegraphics[width=9.0cm]{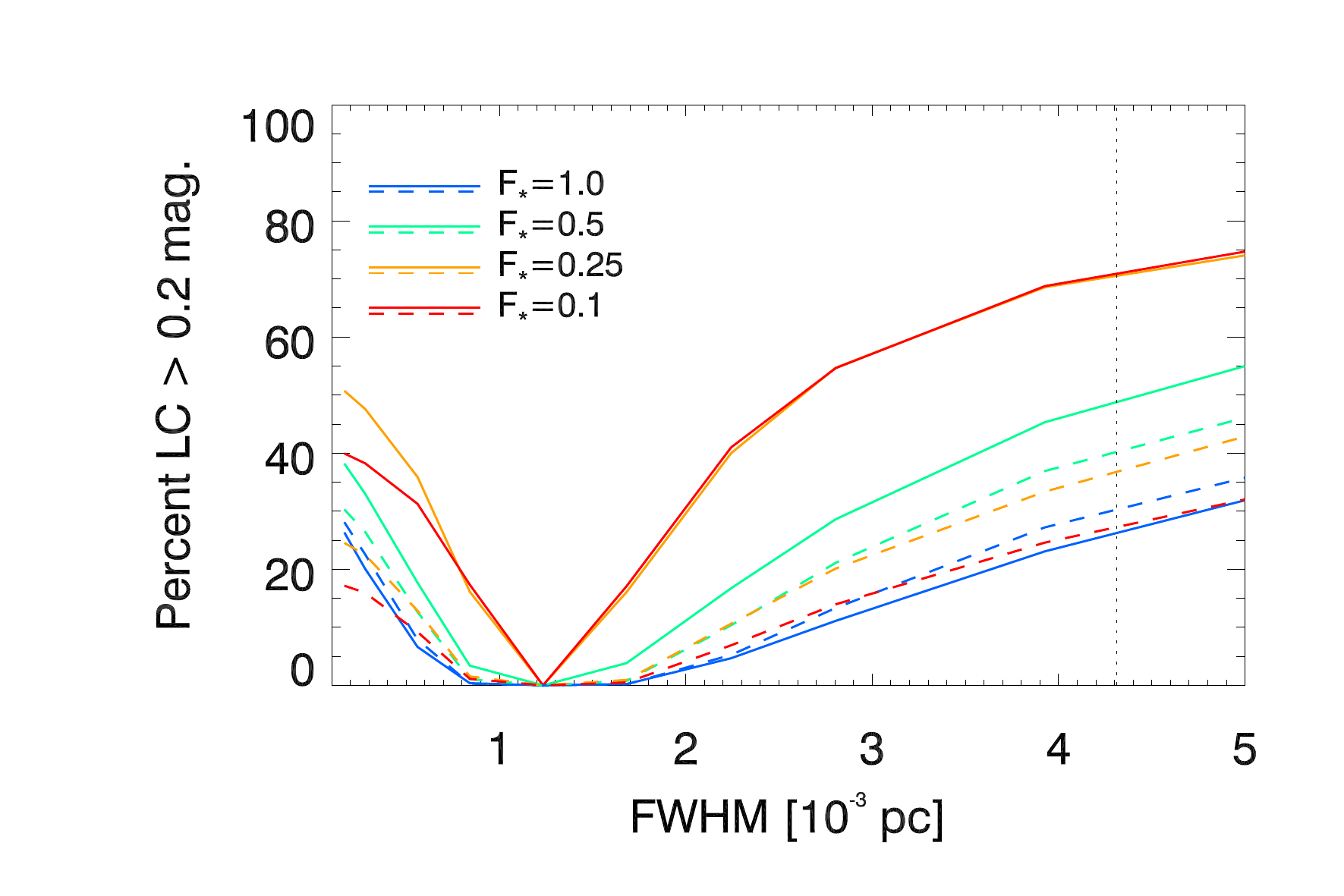}
      \caption{Percentage of lightcurves with deviations above 0.2 magnitudes as a function of photosphere size. Each color represents a different fraction of the stellar component with ${\rm F}_{*}=1$ corresponding to the fiducial model and ${\rm F}_{*}=0.1$ with a model having ten times less microlens surface mass density. The solid lines are for the image 1 and the dashed lines are for image 2. Note how, especially for image 1, reducing the mass in the microlens model can result in more frequent distortions in the light curve. The vertical dotted line marks the FWHM 65 days after the explosion (FWHM$=2*R=4.31\times10^{-3}$ pc, in the SN rest frame, see section~\ref{Sect_Photosphere}). 
              }
         \label{Fig_LightCurvesFractions1}
   \end{figure}

\begin{figure} 
   \centering
   \includegraphics[width=9.0cm]{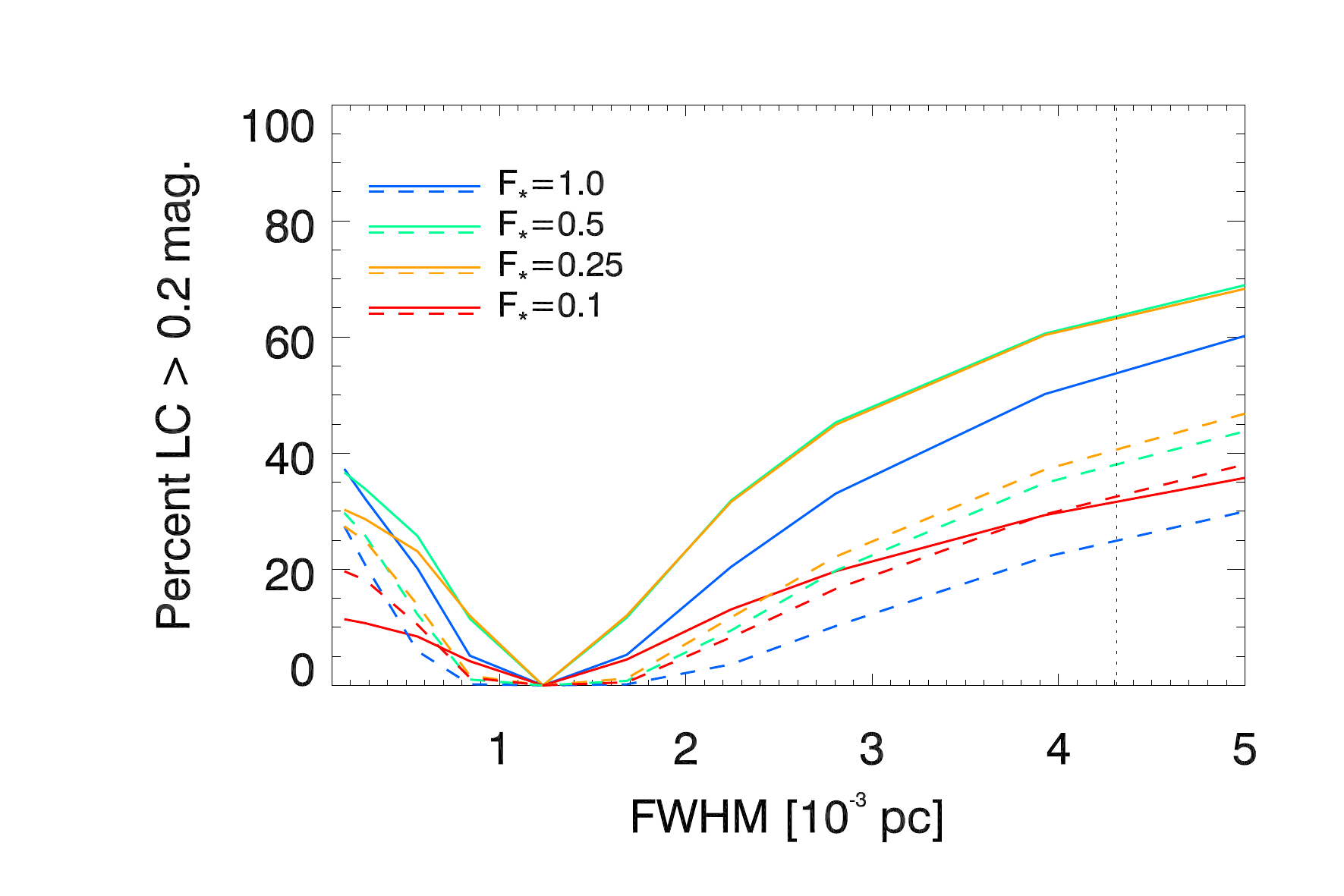}
      \caption{Like figure~\ref{Fig_LightCurvesFractions1} but for images 3 (solid lines) and 4 (dashed lines).
              }
         \label{Fig_LightCurvesFractions2}
   \end{figure}

First, we use macro+micro lensing simulations to study the relative change in time of the light curves. 
In Figure \ref{Fig_LightCurvesPos1} we show examples of distortions in the light curves induced by microlensing for image 1, as a function of photosphere size. The curves are normalized to the size 0.0012 pc, or approximately the time when the SN reaches maximum luminosity. Relative to this moment, the average flux of all realizations, marked as thick solid line, increases by $\approx 10\,\%$ due to microlensing between the moment of maximum luminosity and $\approx 2$ months after explosion (or FWHM $\approx 0.004$ pc). This trend is expected since image 1 has negative parity and the SN is more likely to begin expanding in a region of relative demagnification with respect to the macromodel. As the photosphere expands, the average magnification converges to the larger macro-model magnification. The green colored interval marks the region containing 68\,\% of all simulated expanding photospheres. By setting a limit of 0.2 magnitudes on the amount of relative variability due to microlensing (allowed in the observed light curves as shown in figure~\ref{Fig_Photometry}), we can quantify the fraction of simulated light curves that would exceed this limit.  Figures \ref{Fig_LightCurvesFractions1} and \ref{Fig_LightCurvesFractions2} show the percentage of simulated light curves with deviations of 0.2 magnitudes due to microlensing for images 1,2 and 3,4 respectively, and for different stellar mass fractions.

We find that the percentage of light curves that get distorted by microlensing is largest for stellar surface densities that make the effective optical depth $\tau_{\rm eff}\gtrapprox 1$ (see Eq.\ref{Eq_TauEff} for a definition of $\tau_{\rm eff}$). In order to compare different models with varying stellar mass fractions, we define $F_{*}$ as the relative stellar mass fraction, with $F_{*}=1$ corresponding to models with the same stellar mass fraction of the fiducial model (see table \ref{tab_1}). For the values of the total and radial magnification at position 1 in our fiducial model, $\tau_{\rm eff}  \approx 10.5$. From Fig.~\ref{Fig_LightCurvesFractions1} the fraction of curves with distortions larger than 0.2 magnitudes is maximum for $F_{*}=0.1$ and $F_{*}=0.25$ which would have  $\tau_{\rm eff} \approx$1 and 2.5 respectively. Interestingly, models with larger stellar fractions predict smaller distortions, and hence are favoured by the lack of distortion in the observed LCs. This is a counter-intuitive result but can be explained by the increased level of caustic overlapping that is created in the source plane, once the saturation regime ($\tau_{\rm eff} \approx 1$), is surpassed. Larger values of  $\tau_{\rm eff}$ result in larger overlapping of relatively low magnification regions with large magnification caustics.
At position 2,  see Fig.~\ref{Fig_LightCurvesFractions2}, both the macromodel magnification and $\Sigma$ are smaller, reducing the value of $\tau_{\rm eff}$. We find $\tau_{\rm eff} \approx 7$. In this case, the maximum percentage of LC with distortion $>0.2$ magnitudes is found for $F_{*}=0.25$, with $\tau_{\rm eff} \approx 1.7$,  while for smaller values, like $F_{*}=0.1$, the fraction is now clearly smaller.  
 
In the remaining of this section we select the best combination of macro+micro lens models by combining all available observables at the position of the 4 SN images; i) the positions of the four SN images (constraining the macrolens model), ii) the magnification of the 4 images (constraining both the macrolens and microlens model), and iii) the apparent lack of features in the light curves of the four images (which constrains also the combination macro+micro model). In order to satisfy the positional constraint, the values of the convergence, $\kappa$, and shear, $\gamma$ are taken from the models studied in section \ref{Subsect_PosOnly}. The fraction of stellar mass is varied in order to explore the role played by microlenses, which are parameterized by the stellar surface mass density.

We start by defining the magnification probability of a certain macro+micro model $\mathcal{M}$ as: 
\begin{equation}
P(\mathcal{M}|\mu_{obs}) = \left[ \Pi _i P(\mu_{obs}|\mathcal{M})_i \right] \times P(\mathcal{M}),
\label{eq_Prob1}
\end{equation}
where $P(\mathcal{M})$ is a prior for model $\mathcal{M}$, and $P(\mu_{obs}|\mathcal{M})_i$ is the probability of observing $\mu_{obs}$ for the model $\mathcal{M}$ at position $i$. The prior is given by Equation~\ref{Eq_Lkhd1}, that is determined by the constraints on the four positions of the SN. In other words, we assume a flat prior for the microlensing part of the model and the prior is based solely on how well the macrolens model is able to reproduce the four SN image positions. 
Equation \ref{eq_Prob1} can be more easily understood if we consider the case of the fiducial model shown in  Figure~\ref{Fig_ProbMu_Micro4Positions}. For this model the probability $P(\mathcal{M}|\mu_{obs})$ corresponds to the product of the values of the four colored curves at the observed values of the magnification (middle points in the 4 horizontal bars), multiplied by $P(\mathcal{M})$, as given by Equation~\ref{Eq_Lkhd1}. 

In addition to the magnification probability, we define a light curve probability. For the definition of this probability, we construct a function that penalizes models that predict light curves with significant changes  in the flux during the first two months after the explosion, as the photosphere expands. To first order, one can model this penalty function as a Heaviside function, with models showing variability in a significant percentage of the simulated light curves having zero probability (as this is not observed), and models with little or no variability having probability 1. A smooth version of the Heaviside function is more adequate since it is not clear what fraction of the light curves should show no variability. We set this fraction to 25\% based on the fact that none of the four images shows significant variability (i.e., more than 0.2 magnitudes). 

For the penalty function we use a form inspired on the logistic function, 
\begin{equation}
C_{\mathcal{M}}(f) = 1 - \frac{1}{1+e^{-A(f-b)}}.
\label{eq_Prob2}
\end{equation}
The parameter $b$ is such that $C(f=b)=0.5$ (with $0<f<1$). The parameter $A$ controls the rate at which the transition takes place between $C=0$ and $C=1$. For $b=0.4$ and $A=20$, and at the desired fraction of 25\% (f=0.25), we have $C_{\mathcal{M}}(0.25)=0.95$. That is, the models $\mathcal{M}$ that predict a fraction $f<0.25$, are penalized very little. For values $f>0.25$, $C_{\mathcal{M}}(f)$ decreases rapidly towards zero. 
The final probability for a given model is given by the product of equations~\ref{eq_Prob1} and ~\ref{eq_Prob2}. 

After combining these two equations, we can compute the total probability for every possible combination of macro+micro models. Since estimating the probability for every model considered in the previous sections is computationally prohibitive, we restrict the calculation to 50 macromodels selected among the best ones in terms of the likelihood given in Eq. \ref{Eq_Lkhd1}. All selected 50 models are within a factor 3.9 times the maximum likelihood according to equation \ref{Eq_Lkhd1} (These 50 models are also shown in  Figure \ref{Fig_Comparison_WilliamsMortsell}). The first model within this sample is the fiducial model described in section \ref{Subsect_PosOnly}, which is also the model with the maximum likelihood in Eq. \ref{Eq_Lkhd1}. 
For the microlensing model we adopt the fiducial values of the stellar surface mass density listed in table\ref{tab_1}, and also models with a smaller fixed fraction of stellar masses relative to the fiducial model. These are parameterized by the constant $F_{*}$. The fiducial model presented in table \ref{tab_1} corresponds to a stellar fraction with $F_{*}=1$, and models with  smaller stellar masses have $F_{*}<1$.  

\begin{figure} 
   \centering
   \includegraphics[width=9.0cm]{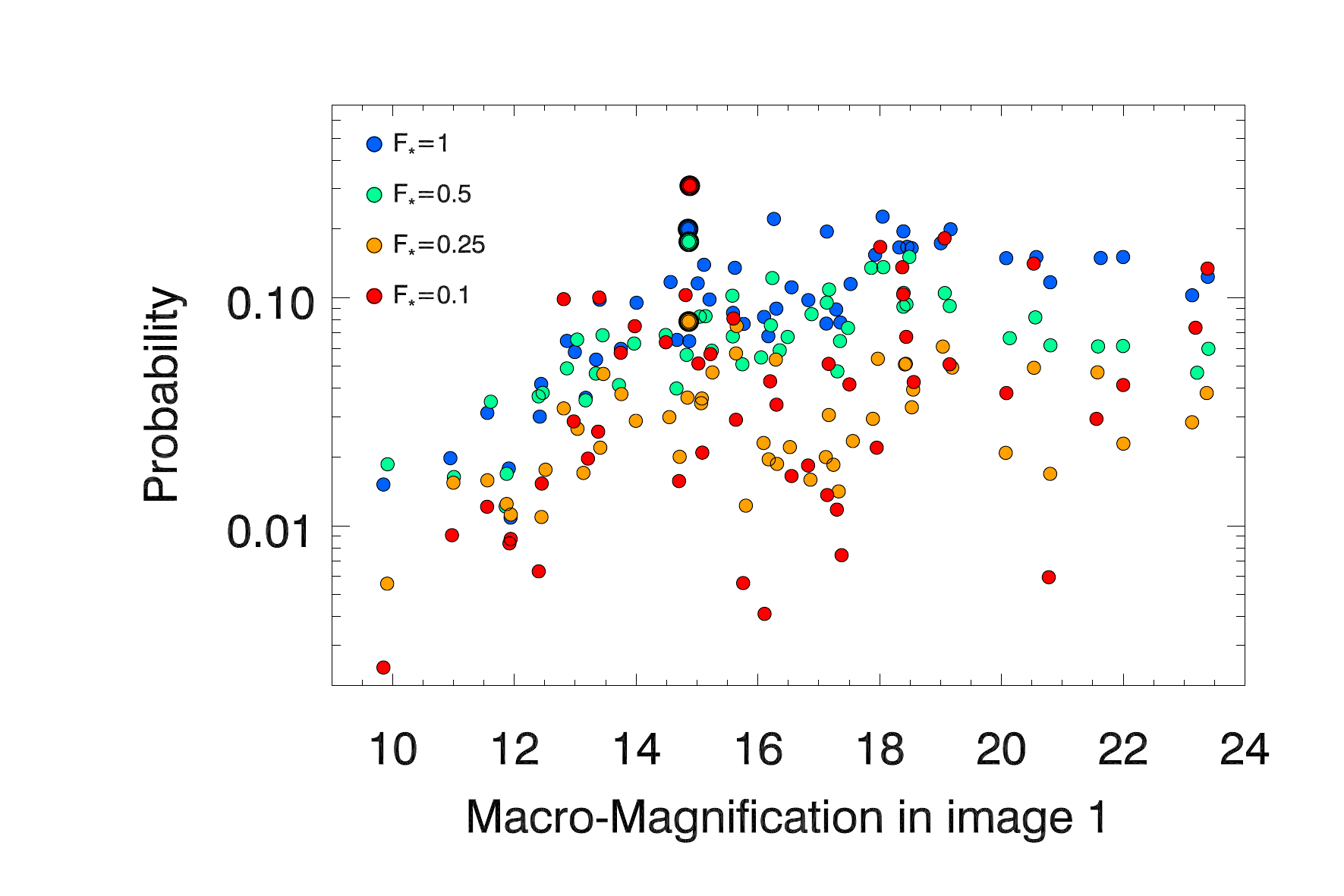}
      \caption{Probability of the 50 randomly selected macro+micro models as a function of the macromodel magnification in image 1. The probability is the product of equations \ref{eq_Prob1} and \ref{eq_Prob2}. The probabilities are color coded based on the amount of microlensing. The fiducial macro model is highlighted by larger symbols for the four different values of F$_*$. 
              }
         \label{Fig_FinalProbability}
   \end{figure}

\begin{figure} 
   \centering
   \includegraphics[width=9.0cm]{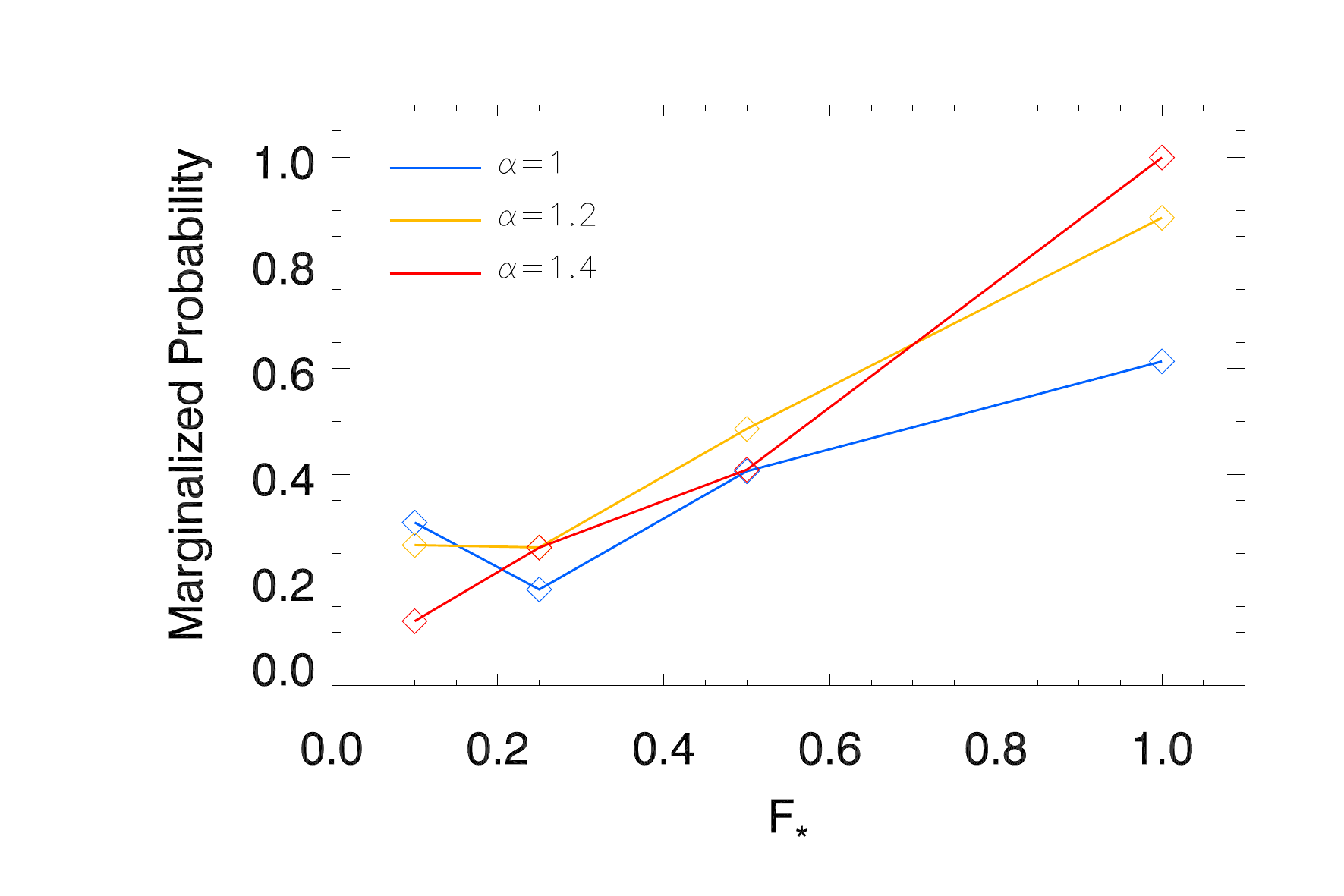}
      \caption{Marginalized probability as a function of F$_*$. We show the marginalized probability for three macromodel slopes, the fiducial $\alpha=1.0$ is shown in blue, while the shallower profiles with $\alpha=1.2$, and $\alpha=1.4$ are shown in yellow and red respectively. In all cases, larger values of F$_*$ are favoured.  
              }
         \label{Fig_MarginalizedProbability}
   \end{figure}

The resulting probability of the 50 selected models is shown in figure \ref{Fig_FinalProbability}, where we represent the final probability, as a function of the macrolens model magnification at the position of image 1. 
Each dot corresponds to one of the 50 models. Different colors are used for different stellar fractions, or $F_{*}$. The fiducial model, for the four considered values of  $F_{*}$, is marked with a thick black circle. 
As expected, macrolens models which predict a smaller magnification for image 1 have a smaller likelihood. Interestingly, macrolens models that predict larger magnifications for image 1 have also smaller probability. This is due to the prior, which penalizes more these models since they can not reproduce the four positions as well. Another interesting result is that the models that have the largest probabilities have either  $F_{*}=1$ or  $F_{*}=0.1$ (red and blue points). The dependency with  $F_{*}$ is better appreciated in figure \ref{Fig_MarginalizedProbability}), where we marginalize over all 50 macro models. 
We find that models with small values of $F_{*}$ are least favoured by the data. These models result in values of $\tau_{\rm eff}\approx 1$, where microlensing effects are expected to be largest. 
In figure \ref{Fig_MarginalizedProbability}) we show also the resulting marginalized probabilities for two shallower DM halo profiles, with exponent $\alpha=1.2$ and $\alpha=1.4$. The marginalized probability is normalized to the maximum overall probability, that in this case corresponds to the model with $\alpha=1.4$ and $F_{*}=1$. As shown in Appendix B, these shallower DM models are able to reproduce the four SN positions slightly better, mostly images 2 and 3. The models with  $\alpha=1.2$ and $\alpha=1.4$ predict also larger macromodel magnifications in all four images, a feature that is characteristic of shallower profiles. Interestingly, this increase in magnification from the macromodel results in an increase in the marginalized probability in figure \ref{Fig_MarginalizedProbability}), except for values of $F_{*}=0.1$, for which the effective optical depth is now larger, and closer to the saturation regime. But most importantly, our main conclusion remains the same whether $\alpha=1.0$, $\alpha=1.2$ or $\alpha=1.4$; models with larger $F_{*}$, and consistent with the values of the fiducial model are preferred over models with smaller stellar masses.

\section{Discussion}\label{sec_discus}

The result from the last section suggests that the baryonic component in our fiducial lens model helps explain the observed magnifications in the four images of iPTF16geu. The baryons in the galaxy accomplish this in two ways. First, by introducing an additional degree of freedom in the distribution of mass, especially inside the Einstein ring, where the baryonic component dominates over the dark matter. The small misalignment in the lens model between the baryonic component and dark matter halo helps boost the magnification from the macromodel in image 1. Second, we find evidence that the abundance of microlenses from the baryonic model at the position of the four SN images help explain both the magnification of the 4 images and the apparent lack of variability in the light curves. Our results show how a less numerous population of microlenses (i.e., a value of  $F_{*}<1$) would introduce larger fluctuations in the magnifications and light curves, which are not observed. This counter-intuitive result can be understood after realizing that the amount of microlensing predicted by the fiducial lens model results in an effective optical depth of mcicrolensing larger than 1, that is, we are in the optically thick regime. In this regime, multiple microcaustics are usually overlapping in any given position in the source plane. As a moving (or expanding) background source moves in the source plane, it can cross one of these microcaustics from one of the microlenses. In the optically thick regime, the microcaustic being crossed is surrounded by other microcaustics that are also contributing to the magnification. The sudden increase in flux due to the crossing of one microcaustic is smaller in relative terms (typically a 10\% -- 30\% increase in total flux, depending on the number of overlapping microcaustics) than in a situation where the microcaustics do not overlap. In this case, in the optically thin regime, one caustic crossing can result in relative flux changes of order 100\%. The situation in the optically thick regime is similar to the conceptually easier to understand case of microlensing of a background globular cluster. 
This case was recently studied in detail in \cite{Dai2020}, where a single microcaustic may simultaneously act over a number of background stars from a globular cluster. At any given time there is some probability that one of the stars is crossing the microcaustic increasing the flux of that star by a factor of order 10 in a period of days, but the observed flux from the entire cluster changes only slightly at the percent level since the rest of the stars are not varying their flux significantly over the same period. In our case, we have a single source and multiple microcaustics but the explanation of why there is a relatively small change in flux is analogous to the globular cluster case studied in \cite{Dai2020}. 

We find that the smallest distortions in the light curve correspond to $F_{*}=1$ (beyond the saturation regime) and $F_{*}=0.1$ (below the saturation regime). However, for $F_{*}=0.1$, the total stellar mass would be 10 times smaller than for the fiducial model and in tension with the stellar mass derived from the velocity dispersion. On the other hand, the fiducial model is able to reproduce the observed magnification, lack of noticeable fluctuations in the light curve, and is consistent with (although larger by a factor $\approx 2$ than) the stellar mass inferred from the velocity dispersion. In contrast, the model with $F_{*}=0.5$ which would correspond to stellar masses similar to those derived from the velocity dispersion is not favoured by our analysis. Our results favour models with a larger stellar mass fraction (see table \ref{tab_1}). Our result can be reconciled with the velocity dispersion analysis if one adopts a bottom-heavy initial mass function for the stellar population in the lens, which would increase the number of small mass microlenses, without affecting the photometric observations.

Our results demonstrate how lensed SNe can be used to constraint the amount of microlensing. This opens the possibility of using strongly lensed SNe to study a candidate of dark matter, primordial black holes. In particular, primordial black holes in the mass regime probed by LIGO would act as powerful microlenses, adding to the perturbation in magnification and light curves studied in the previous sections. It is estimated that a fraction of dark matter as small as 1\% is sufficient to explain the LIGO observations with primordial black holes. However, in the case of iPTF16geu (where 1\% of dark matter represents less than 1\% the stellar mass) no useful limits to the fraction of dark matter in the form of primordial black holes can be obtained, given the large optical depth from stellar microlenses, which overwhelms the signal and impedes to constrain the fraction of primordial black holes to better than the percent level (the current upper limit on their abundance). In order to exploit lensed SNe as dark matter probes, one needs to observe high redshift SNe, for which the multiple images form further away from the negative influence of stellar microlenses and where the surface mass density of dark matter dominates clearly over the baryonic surface mass density.

\section{Conclusion}\label{sec_concl}
iPTF16geu is the first observed multiply lensed type Ia supernova, and for which magnification, time delays, and light curves have been measured. In this work we attempt to explain the observed position, flux and light curves with a realistic lens model that includes a baryonic component that traces the observed light distribution, and a dark matter halo. We test the hypothesis that stars in the baryonic component are responsible for microlensing effects, needed to explain the anomalous flux ratios between the counterimages.

Since the observed magnification values in the four SN positions are expected to be distorted by microlensing, we constrain the macrolens model using the position of the four SN images only, which are insensitive to microlensing. 
We consider a two-component model for the macrolens model. The spatial distribution of one of the components (baryons) is fixed by the observed distribution of light and we fit only for its amplitude (or total mass). The second component, a dark matter halo, is allowed to vary in mass, orientation, ellipticity and centroid position with respect to the baryonic component. We find that the best model shifts the centroid of the dark matter halo by $\approx 0.23 kpc$ with respect to the peak of the light distribution.  The ellipticity and orientation of the dark matter halo is similar to that of the baryonic component (on the largest scales). We find that lens models that incorporate the baryonic component are able to reproduce well the positions of the 4 observed SNe images. Within the relatively small Einstein radius of the lens, the mass is dominated by the baryonic component. As in previous work, the best macrolens models can not reproduce the observed magnification at the four SN positions. 

In order to reproduce the observed fluxes,  we consider a model that includes also microlensing. The number density of microlenses is given by the stellar surface mass density inferred from the baryonic component in the macrolens model, and after assuming that  most of the baryonic mass in the central region of the lens is in stellar form. In addition to magnification and position constraints, we use also information form the light curves. In particular, we require that models that include microlensing must not produce distortions in the light curve of the SN which are greater than 0.2 magnitudes. Such distortions are  not present in the observed light curves. 

We find that models that include a level of microlensing that is consistent with the stellar mass fraction from the macrolens model baryonic component, explains the observed positions, magnifications, and the apparent lack of fluctuations in the light curves. Stellar mass fractions which are 4 times smaller are not favoured by our results, since they would produce distortions in the light curves, which are not observed. Our results favour bottom-heavy initial mass functions for the stellar component, that can simultaneously reproduce the unusual flux ratios while keeping the light curves relatively undisturbed. 

Future observations of type Ia SNe at higher redshift, can be used to constrain exotic models of dark matter. In particular the abundance of primordial black holes can be constrained with strongly lensed type Ia SN. These constraints must be obtained at larger distances from the central region of the lens, that those considered in this work, and where the negative impact of stellar microlenses can be diminished.   

\begin{acknowledgements}
 The authors thank Liliya Williams for very useful comments and feedback.  J.M.D. acknowledges the support of project PGC2018-101814-B-100 (MCIU/AEI/MINECO/FEDER, UE) Ministerio de Ciencia, Investigaci\'on y Universidades.  This project was funded by the Agencia Estatal de Investigaci\'on, Unidad de Excelencia Mar\'ia de Maeztu, ref. MDM-2017-0765.  
 A.~G. acknowledges support from the Swedish National Space Agency grant 110/18 and the Swedish Research Council grant 2020-03444.
 E.M. acknowledges support from the Swedish Research Council under Dnr VR 2020-03384.
P.K. is supported by NSF grant AST-1908823, and HST GO-15936, GO-16278, and AR-15791. 
 JWN is supported by the UK Space Agency, through grant ST/N001494/1 and by InnovateUK through grant TS/V002856/1.
  J.M.D. acknowledges the hospitality of the Physics Department at the University of Pennsylvania for hosting him during the preparation of this work, despite the fact that most of this work was done under strict social distancing rules. 
\end{acknowledgements}

\bibliographystyle{aa} 
\bibliography{MyBiblio} 

\begin{thebibliography}{64}
\expandafter\ifx\csname natexlab\endcsname\relax\def\natexlab#1{#1}\fi

\bibitem[{{Anguita} {et~al.}(2008){Anguita}, {Schmidt}, {Turner}, {Wambsganss},
  {Webster}, {Loomis}, {Long}, \& {McMillan}}]{Anguita2008}
{Anguita}, T., {Schmidt}, R.~W., {Turner}, E.~L., {et~al.} 2008, \aap, 480, 327

\bibitem[{{Astropy Collaboration} {et~al.}(2013){Astropy Collaboration},
  {Robitaille}, {Tollerud}, {Greenfield}, {Droettboom}, {Bray}, {Aldcroft},
  {Davis}, {Ginsburg}, {Price-Whelan}, {Kerzendorf}, {Conley}, {Crighton},
  {Barbary}, {Muna}, {Ferguson}, {Grollier}, {Parikh}, {Nair}, {Unther},
  {Deil}, {Woillez}, {Conseil}, {Kramer}, {Turner}, {Singer}, {Fox}, {Weaver},
  {Zabalza}, {Edwards}, {Azalee Bostroem}, {Burke}, {Casey}, {Crawford},
  {Dencheva}, {Ely}, {Jenness}, {Labrie}, {Lim}, {Pierfederici}, {Pontzen},
  {Ptak}, {Refsdal}, {Servillat}, \& {Streicher}}]{astropy1}
{Astropy Collaboration}, {Robitaille}, T.~P., {Tollerud}, E.~J., {et~al.} 2013,
  \aap, 558, A33

\bibitem[{Bolton {et~al.}(2008)Bolton, Burles, Koopmans, Treu, Gavazzi,
  Moustakas, Wayth, \& Schlegel}]{Bolton2008}
Bolton, A.~S., Burles, S., Koopmans, L. V.~E., {et~al.} 2008, The Astrophysical
  Journal, 682, 964

\bibitem[{{Calette} {et~al.}(2018){Calette}, {Avila-Reese},
  {Rodr{\'\i}guez-Puebla}, {Hern{\'a}ndez-Toledo}, \&
  {Papastergis}}]{Calette2018}
{Calette}, A.~R., {Avila-Reese}, V., {Rodr{\'\i}guez-Puebla}, A.,
  {Hern{\'a}ndez-Toledo}, H., \& {Papastergis}, E. 2018, \rmxaa, 54, 443

\bibitem[{{Cannarozzo} {et~al.}(2019){Cannarozzo}, {Sonnenfeld}, \&
  {Nipoti}}]{Cannarozzo2019}
{Cannarozzo}, C., {Sonnenfeld}, A., \& {Nipoti}, C. 2019, arXiv e-prints,
  arXiv:1910.06987

\bibitem[{{Carr} {et~al.}(2017){Carr}, {Raidal}, {Tenkanen}, {Vaskonen}, \&
  {Veerm{\"a}e}}]{Carr2017}
{Carr}, B., {Raidal}, M., {Tenkanen}, T., {Vaskonen}, V., \& {Veerm{\"a}e}, H.
  2017, \prd, 96, 023514

\bibitem[{{Casasola} {et~al.}(2020){Casasola}, {Bianchi}, {De Vis}, {Magrini},
  {Corbelli}, {Clark}, {Fritz}, {Nersesian}, {Viaene}, {Baes}, {Cassar{\`a}},
  {Davies}, {De Looze}, {Dobbels}, {Galametz}, {Galliano}, {Jones}, {Madden},
  {Mosenkov}, {Tr{\v{c}}ka}, \& {Xilouris}}]{Casasola2020}
{Casasola}, V., {Bianchi}, S., {De Vis}, P., {et~al.} 2020, \aap, 633, A100

\bibitem[{{Chang} \& {Refsdal}(1979)}]{Chang1979}
{Chang}, K. \& {Refsdal}, S. 1979, \nat, 282, 561

\bibitem[{{Chang} \& {Refsdal}(1984)}]{Chang1984}
{Chang}, K. \& {Refsdal}, S. 1984, \aap, 132, 168

\bibitem[{{Chen} \& {Huang}(2018)}]{Chen2018}
{Chen}, Z.-C. \& {Huang}, Q.-G. 2018, \apj, 864, 61

\bibitem[{{Dai}(2020)}]{Dai2020}
{Dai}, L. 2020, arXiv e-prints, arXiv:2007.01301

\bibitem[{{Dhawan} {et~al.}(2020){Dhawan}, {Johansson}, {Goobar}, {Amanullah},
  {M{\"o}rtsell}, {Cenko}, {Cooray}, {Fox}, {Goldstein}, {Kalender},
  {Kasliwal}, {Kulkarni}, {Lee}, {Nayyeri}, {Nugent}, {Ofek}, \&
  {Quimby}}]{Dhawan2020}
{Dhawan}, S., {Johansson}, J., {Goobar}, A., {et~al.} 2020, \mnras, 491, 2639

\bibitem[{{Diego}(2019)}]{Diego2019}
{Diego}, J.~M. 2019, \aap, 625, A84

\bibitem[{{Diego} {et~al.}(2018){Diego}, {Kaiser}, {Broadhurst}, {Kelly},
  {Rodney}, {Morishita}, {Oguri}, {Ross}, {Zitrin}, {Jauzac}, {Richard},
  {Williams}, {Vega-Ferrero}, {Frye}, \& {Filippenko}}]{Diego2018}
{Diego}, J.~M., {Kaiser}, N., {Broadhurst}, T., {et~al.} 2018, \apj, 857, 25

\bibitem[{Diemer(2018)}]{colossus}
Diemer, B. 2018, The Astrophysical Journal Supplement Series, 239, 35

\bibitem[{Foreman-Mackey(2016)}]{corner}
Foreman-Mackey, D. 2016, The Journal of Open Source Software, 1, 24

\bibitem[{{Goobar} {et~al.}(2017){Goobar}, {Amanullah}, {Kulkarni}, {Nugent},
  {Johansson}, {Steidel}, {Law}, {M{\"o}rtsell}, {Quimby}, {Blagorodnova},
  {Brand eker}, {Cao}, {Cooray}, {Ferretti}, {Fremling}, {Hangard}, {Kasliwal},
  {Kupfer}, {Lunnan}, {Masci}, {Miller}, {Nayyeri}, {Neill}, {Ofek},
  {Papadogiannakis}, {Petrushevska}, {Ravi}, {Sollerman}, {Sullivan}, {Taddia},
  {Walters}, {Wilson}, {Yan}, \& {Yaron}}]{Goobar2017}
{Goobar}, A., {Amanullah}, R., {Kulkarni}, S.~R., {et~al.} 2017, Science, 356,
  291

\bibitem[{Hunter(2007)}]{matplotlib}
Hunter, J.~D. 2007, Computing in Science \& Engineering, 9, 90

\bibitem[{{Hyde} \& {Bernardi}(2009)}]{Hyde2009}
{Hyde}, J.~B. \& {Bernardi}, M. 2009, \mnras, 394, 1978

\bibitem[{{Johansson} {et~al.}(2021){Johansson}, {Goobar}, {Price}, {Sagu{\'e}s
  Carracedo}, {Della Bruna}, {Nugent}, {Dhawan}, {M{\"o}rtsell},
  {Papadogiannakis}, {Amanullah}, {Goldstein}, {Cenko}, {De}, {Dugas},
  {Kasliwal}, {Kulkarni}, \& {Lunnan}}]{Johansson2021}
{Johansson}, J., {Goobar}, A., {Price}, S.~H., {et~al.} 2021, \mnras, 502, 510

\bibitem[{{Kayser} {et~al.}(1986){Kayser}, {Refsdal}, \&
  {Stabell}}]{Kayser1986}
{Kayser}, R., {Refsdal}, S., \& {Stabell}, R. 1986, \aap, 166, 36

\bibitem[{Kelly(2020)}]{pyquad}
Kelly, A.~J. 2020, pyquad

\bibitem[{{Kochanek}(2004)}]{Kochanek2004}
{Kochanek}, C.~S. 2004, \apj, 605, 58

\bibitem[{Lam {et~al.}(2015)Lam, Pitrou, \& Seibert}]{numba}
Lam, S.~K., Pitrou, A., \& Seibert, S. 2015, Proceedings of the Second Workshop
  on the LLVM Compiler Infrastructure in HPC - LLVM '15, 1

\bibitem[{{Liu} {et~al.}(2018){Liu}, {Guo}, \& {Cai}}]{Liu2018}
{Liu}, L., {Guo}, Z.-K., \& {Cai}, R.-G. 2018, arXiv e-prints
  [\eprint[arXiv]{1812.05376}]

\bibitem[{{Liu} {et~al.}(2019){Liu}, {Guo}, \& {Cai}}]{Liu2019}
{Liu}, L., {Guo}, Z.-K., \& {Cai}, R.-G. 2019, arXiv e-prints
  [\eprint[arXiv]{1901.07672}]

\bibitem[{Ludlow {et~al.}(2016)Ludlow, Bose, Angulo, Wang, Hellwing, Navarro,
  Cole, \& Frenk}]{Ludlow2016}
Ludlow, A.~D., Bose, S., Angulo, R.~E., {et~al.} 2016, Monthly Notices of the
  Royal Astronomical Society, 460, 1214

\bibitem[{{More} {et~al.}(2017){More}, {Suyu}, {Oguri}, {More}, \&
  {Lee}}]{More2017}
{More}, A., {Suyu}, S.~H., {Oguri}, M., {More}, S., \& {Lee}, C.-H. 2017,
  \apjl, 835, L25

\bibitem[{{M{\"o}rtsell} {et~al.}(2020){M{\"o}rtsell}, {Johansson}, {Dhawan},
  {Goobar}, {Amanullah}, \& {Goldstein}}]{Mortsell2020}
{M{\"o}rtsell}, E., {Johansson}, J., {Dhawan}, S., {et~al.} 2020, \mnras, 496,
  3270

\bibitem[{Navarro {et~al.}(1997)Navarro, Frenk, \& White}]{Navarro1997}
Navarro, J.~F., Frenk, C.~S., \& White, S. D.~M. 1997, The Astrophysical
  Journal, 490, 493

\bibitem[{Nightingale {et~al.}(2021{\natexlab{a}})Nightingale, Hayes, Kelly,
  Amvrosiadis, Etherington, He, Li, Cao, Frawley, Cole, Enia, Frenk, Harvey,
  Li, Massey, Negrello, \& Robertson}]{Nightingale2021}
Nightingale, J., Hayes, R., Kelly, A., {et~al.} 2021{\natexlab{a}}, Journal of
  Open Source Software, 6, 2825

\bibitem[{Nightingale \& Dye(2015)}]{Nightingale2015}
Nightingale, J.~W. \& Dye, S. 2015, Monthly Notices of the Royal Astronomical
  Society, 452, 2940

\bibitem[{Nightingale {et~al.}(2018)Nightingale, Dye, \&
  Massey}]{Nightingale2018}
Nightingale, J.~W., Dye, S., \& Massey, R.~J. 2018, Monthly Notices of the
  Royal Astronomical Society, 478, 4738

\bibitem[{Nightingale {et~al.}(2021{\natexlab{b}})Nightingale, Hayes, \&
  Griffiths}]{pyautofit}
Nightingale, J.~W., Hayes, R.~G., \& Griffiths, M. 2021{\natexlab{b}}, Journal
  of Open Source Software, 6, 2550

\bibitem[{Nightingale {et~al.}(2021{\natexlab{c}})Nightingale, Hayes, Kelly,
  Amvrosiadis, Etherington, He, Li, Cao, Frawley, Cole, Enia, Frenk, Harvey,
  Li, Massey, Negrello, \& Robertson}]{pyautolens}
Nightingale, J.~W., Hayes, R.~G., Kelly, A., {et~al.} 2021{\natexlab{c}},
  Journal of Open Source Software, 6, 2825

\bibitem[{Nightingale {et~al.}(2019)Nightingale, Massey, Harvey, Cooper,
  Etherington, Tam, \& Hayes}]{Nightingale2019}
Nightingale, J.~W., Massey, R.~J., Harvey, D.~R., {et~al.} 2019, Monthly
  Notices of the Royal Astronomical Society, 489, 2049

\bibitem[{{Oguri} {et~al.}(2018){Oguri}, {Diego}, {Kaiser}, {Kelly}, \&
  {Broadhurst}}]{Oguri2018}
{Oguri}, M., {Diego}, J.~M., {Kaiser}, N., {Kelly}, P.~L., \& {Broadhurst}, T.
  2018, \prd, 97, 023518

\bibitem[{{Paczynski}(1986)}]{Paczynski1986}
{Paczynski}, B. 1986, \apj, 301, 503

\bibitem[{{Pan}(2020)}]{Pan2020}
{Pan}, Y.-C. 2020, arXiv e-prints, arXiv:2004.14544

\bibitem[{Pedregosa {et~al.}(2011)Pedregosa, Varoquaux, Gramfort, Michel,
  Thirion, Grisel, Blondel, Prettenhofer, Weiss, Dubourg, Vanderplas, Passos,
  Cournapeau, Brucher, Perrot, \& Duchesnay}]{scikit-learn}
Pedregosa, F., Varoquaux, G., Gramfort, A., {et~al.} 2011, Journal of Machine
  Learning Research, 12, 2825

\bibitem[{{Pierel} \& {Rodney}(2019)}]{Pierel2019}
{Pierel}, J.~D.~R. \& {Rodney}, S. 2019, \apj, 876, 107

\bibitem[{{Piro} \& {Nakar}(2014)}]{Piro2014}
{Piro}, A.~L. \& {Nakar}, E. 2014, \apj, 784, 85

\bibitem[{{Ponente} \& {Diego}(2011)}]{Ponente2011}
{Ponente}, P.~P. \& {Diego}, J.~M. 2011, \aap, 535, A119

\bibitem[{{Price-Whelan} {et~al.}(2018){Price-Whelan}, {Sip{\H{o}}cz},
  {G{\"u}nther}, {Lim}, {Crawford}, {Conseil}, {Shupe}, {Craig}, {Dencheva},
  {Ginsburg}, {VanderPlas}, {Bradley}, {P{\'e}rez-Su{\'a}rez}, {de Val-Borro},
  {Paper Contributors}, {Aldcroft}, {Cruz}, {Robitaille}, {Tollerud},
  {Coordination Committee}, {Ardelean}, {Babej}, {Bach}, {Bachetti}, {Bakanov},
  {Bamford}, {Barentsen}, {Barmby}, {Baumbach}, {Berry}, {Biscani}, {Boquien},
  {Bostroem}, {Bouma}, {Brammer}, {Bray}, {Breytenbach}, {Buddelmeijer},
  {Burke}, {Calderone}, {Cano Rodr{\'\i}guez}, {Cara}, {Cardoso}, {Cheedella},
  {Copin}, {Corrales}, {Crichton}, {D{\textquoteright}Avella}, {Deil},
  {Depagne}, {Dietrich}, {Donath}, {Droettboom}, {Earl}, {Erben}, {Fabbro},
  {Ferreira}, {Finethy}, {Fox}, {Garrison}, {Gibbons}, {Goldstein}, {Gommers},
  {Greco}, {Greenfield}, {Groener}, {Grollier}, {Hagen}, {Hirst}, {Homeier},
  {Horton}, {Hosseinzadeh}, {Hu}, {Hunkeler}, {Ivezi{\'c}}, {Jain}, {Jenness},
  {Kanarek}, {Kendrew}, {Kern}, {Kerzendorf}, {Khvalko}, {King}, {Kirkby},
  {Kulkarni}, {Kumar}, {Lee}, {Lenz}, {Littlefair}, {Ma}, {Macleod},
  {Mastropietro}, {McCully}, {Montagnac}, {Morris}, {Mueller}, {Mumford},
  {Muna}, {Murphy}, {Nelson}, {Nguyen}, {Ninan}, {N{\"o}the}, {Ogaz}, {Oh},
  {Parejko}, {Parley}, {Pascual}, {Patil}, {Patil}, {Plunkett}, {Prochaska},
  {Rastogi}, {Reddy Janga}, {Sabater}, {Sakurikar}, {Seifert}, {Sherbert},
  {Sherwood-Taylor}, {Shih}, {Sick}, {Silbiger}, {Singanamalla}, {Singer},
  {Sladen}, {Sooley}, {Sornarajah}, {Streicher}, {Teuben}, {Thomas},
  {Tremblay}, {Turner}, {Terr{\'o}n}, {van Kerkwijk}, {de la Vega}, {Watkins},
  {Weaver}, {Whitmore}, {Woillez}, {Zabalza}, \& {Contributors}}]{astropy2}
{Price-Whelan}, A.~M., {Sip{\H{o}}cz}, B.~M., {G{\"u}nther}, H.~M., {et~al.}
  2018, \aj, 156, 123

\bibitem[{{Raidal} {et~al.}(2019){Raidal}, {Spethmann}, {Vaskonen}, \&
  {Veerm{\"a}e}}]{Raidal2019}
{Raidal}, M., {Spethmann}, C., {Vaskonen}, V., \& {Veerm{\"a}e}, H. 2019,
  \jcap, 2, 018

\bibitem[{Ravasi \& Vasconcelos(2019)}]{pylops}
Ravasi, M. \& Vasconcelos, I. 2019 [\eprint[arXiv]{1907.12349}]

\bibitem[{{Schruba} {et~al.}(2011){Schruba}, {Leroy}, {Walter}, {Bigiel},
  {Brinks}, {de Blok}, {Dumas}, {Kramer}, {Rosolowsky}, {Sandstrom},
  {Schuster}, {Usero}, {Weiss}, \& {Wiesemeyer}}]{Schruba2011}
{Schruba}, A., {Leroy}, A.~K., {Walter}, F., {et~al.} 2011, \aj, 142, 37

\bibitem[{Shajib(2019)}]{Anowar2019}
Shajib, A.~J. 2019, Monthly Notices of the Royal Astronomical Society, 488,
  1387

\bibitem[{Speagle(2020)}]{dynesty}
Speagle, J.~S. 2020, Monthly Notices of the Royal Astronomical Society, 493,
  3132

\bibitem[{{Suyu} {et~al.}(2020){Suyu}, {Huber}, {Ca{\~n}ameras}, {Kromer},
  {Schuldt}, {Taubenberger}, {Y{\i}ld{\i}r{\i}m}, {Bonvin}, {Chan}, {Courbin},
  {N{\"o}bauer}, {Sim}, \& {Sluse}}]{Suyu2020}
{Suyu}, S.~H., {Huber}, S., {Ca{\~n}ameras}, R., {et~al.} 2020, \aap, 644, A162

\bibitem[{Tessore {et~al.}(2016)Tessore, Bellagamba, \& Metcalf}]{Tessore2016}
Tessore, N., Bellagamba, F., \& Metcalf, R.~B. 2016, Monthly Notices of the
  Royal Astronomical Society, 463, 3115

\bibitem[{{van der Walt} {et~al.}(2011){van der Walt}, {Colbert}, \&
  {Varoquaux}}]{numpy}
{van der Walt}, S., {Colbert}, S.~C., \& {Varoquaux}, G. 2011, Computing in
  Science Engineering, 13, 22

\bibitem[{Van~der Walt {et~al.}(2014)Van~der Walt, Sch{\"o}nberger,
  Nunez-Iglesias, Boulogne, Warner, Yager, Gouillart, \& Yu}]{scikit-image}
Van~der Walt, S., Sch{\"o}nberger, J.~L., Nunez-Iglesias, J., {et~al.} 2014,
  PeerJ, 2, e453

\bibitem[{Van~Rossum \& Drake(2009)}]{python}
Van~Rossum, G. \& Drake, F.~L. 2009, Python 3 Reference Manual (Scotts Valley,
  CA: CreateSpace)

\bibitem[{{Virtanen} {et~al.}(2020){Virtanen}, {Gommers}, {Oliphant},
  {Haberland}, {Reddy}, {Cournapeau}, {Burovski}, {Peterson}, {Weckesser},
  {Bright}, {van der Walt}, {Brett}, {Wilson}, {Jarrod Millman}, {Mayorov},
  {Nelson}, {Jones}, {Kern}, {Larson}, {Carey}, {Polat}, {Feng}, {Moore}, {Vand
  erPlas}, {Laxalde}, {Perktold}, {Cimrman}, {Henriksen}, {Quintero}, {Harris},
  {Archibald}, {Ribeiro}, {Pedregosa}, {van Mulbregt}, \&
  {Contributors}}]{scipy}
{Virtanen}, P., {Gommers}, R., {Oliphant}, T.~E., {et~al.} 2020, Nature
  Methods, 17, 261

\bibitem[{{Wambsganss} {et~al.}(1990){Wambsganss}, {Paczynski}, \&
  {Schneider}}]{Wambsganss1990}
{Wambsganss}, J., {Paczynski}, B., \& {Schneider}, P. 1990, \apjl, 358, L33

\bibitem[{{Weisenbach} {et~al.}(2021){Weisenbach}, {Schechter}, \&
  {Pontula}}]{Weisenbach2021}
{Weisenbach}, L., {Schechter}, P., \& {Pontula}, S. 2021, arXiv e-prints,
  arXiv:2105.08690

\bibitem[{{Williams} \& {Zegeye}(2020)}]{Williams2020}
{Williams}, L. L.~R. \& {Zegeye}, D. 2020, arXiv e-prints, arXiv:2006.09391

\bibitem[{{Winn} {et~al.}(2004){Winn}, {Rusin}, \& {Kochanek}}]{Winn2004}
{Winn}, J.~N., {Rusin}, D., \& {Kochanek}, C.~S. 2004, \nat, 427, 613

\bibitem[{{Witt} \& {Mao}(2000)}]{Witt2000}
{Witt}, H.~J. \& {Mao}, S. 2000, \mnras, 311, 689

\bibitem[{{Wyithe} \& {Turner}(2001)}]{Wyithe2001}
{Wyithe}, J.~S.~B. \& {Turner}, E.~L. 2001, \mnras, 320, 21

\bibitem[{{Wyithe} {et~al.}(2000){Wyithe}, {Webster}, \& {Turner}}]{Wyithe2000}
{Wyithe}, J.~S.~B., {Webster}, R.~L., \& {Turner}, E.~L. 2000, \mnras, 318,
  1120

\bibitem[{{Yahalomi} {et~al.}(2017){Yahalomi}, {Schechter}, \&
  {Wambsganss}}]{Yahalomi2017}
{Yahalomi}, D.~A., {Schechter}, P.~L., \& {Wambsganss}, J. 2017, arXiv
  e-prints, arXiv:1711.07919

\bibitem[{{Zahid} {et~al.}(2016){Zahid}, {Geller}, {Fabricant}, \&
  {Hwang}}]{Zahid2016}
{Zahid}, H.~J., {Geller}, M.~J., {Fabricant}, D.~G., \& {Hwang}, H.~S. 2016,
  \apj, 832, 203

\end{thebibliography}

\appendix
\section{Microlensing simulations}

   \begin{figure*} 
   \centering
   \includegraphics[width=18.0cm]{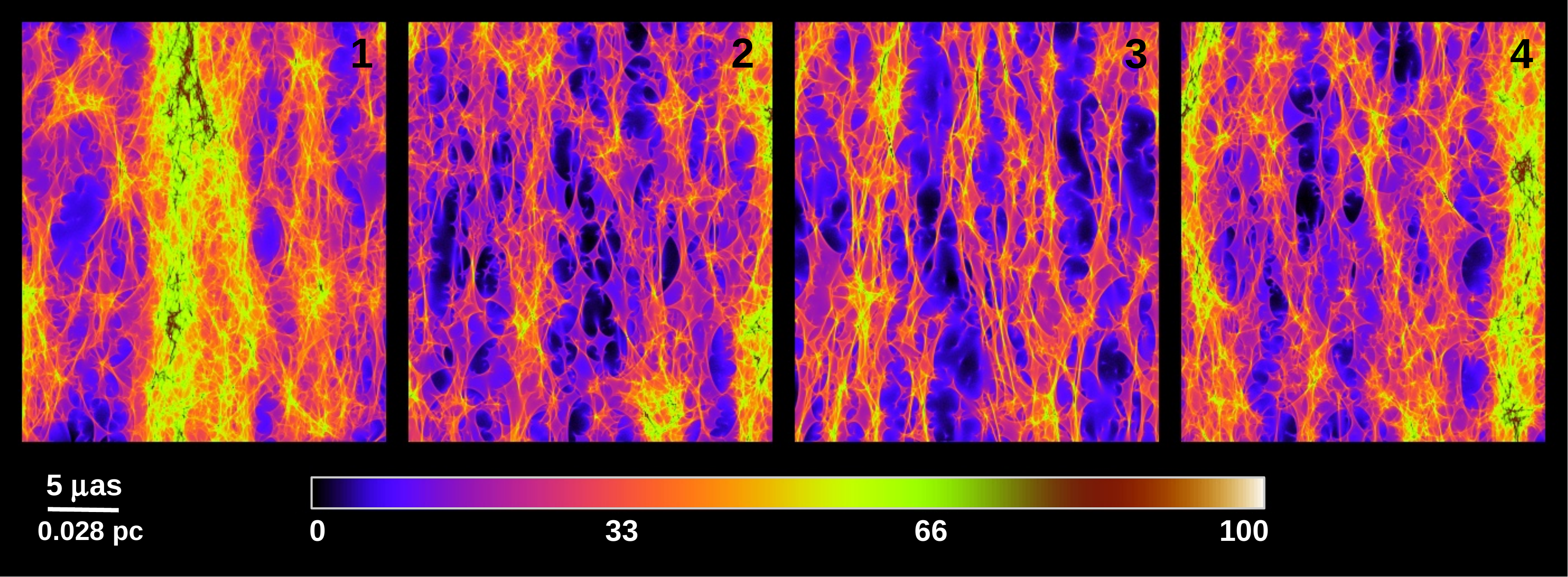}
      \caption{A portion of the simulated region for each one of the 4 SN positions (labelled in the top-right corner of each panel). The magnification maps have been smoothed with a Gaussian of $FWHM=2.8\times10^{-4}$ pc (or roughly, the extension of a typical SN photosphere around the time of peak emission). The area and color scale are the same for all panels. The region shown in the left panel is the same as the one in Figure~\ref{Fig_Caustics_Pos1}.
              }
         \label{Fig_Caustics_4panels}
   \end{figure*}

In this section we describe the microlensing simulations, and review briefly the lensing formalism for microlenses embedded in the macromodel potential. This topic has been widely covered in the literature \citep{Chang1979,Chang1984,Kayser1986,Paczynski1986}. 
For simplicity, we consider only a very small area around the positions where the lensed SN images form. In this case, 
the macromodel can be described with just two parameters, the macromodel magnification factors in the radial and tangential direction, or $\mu_r$ and $\mu_t$ respectively. Equivalently, one can describe the macromodel with the convergence $\kappa$ and shear $\gamma$, since they are related through $\mu_r^{-1}= 1-\kappa+\gamma$, and  $\mu_t^{-1}= 1-\kappa-\gamma$. 
Without loss of generality, we assume that the main direction of the shear, $\gamma$, is oriented in the horizontal direction, that is $\gamma_2=0$ and $\gamma=\sqrt{\gamma_1^2 + \gamma_2^2}=\gamma_1$. 
For a given choice of $\kappa$, and $\gamma$, the lens equation ($\vec{\beta} = \vec{\theta} - \vec{\alpha}(\vec{\theta})$) of the macromodel can be expressed as 
 \begin{equation}
 \vec{\beta}=  \vec{\theta} - \vec{\alpha}(\vec{\theta}) =
\begin{pmatrix} 1-\kappa-\gamma_1 & -\gamma_2 \\ -\gamma_2 & 1-\kappa+\gamma_1 \end{pmatrix} \vec{\theta},
\end{equation}
where the positions in the source plane are given by the coordinates $\beta=(\beta_x,\beta_y)$ and the positions in the image plane are given by the coordinates $\theta=(\theta_x,\theta_y)$. 

The lensing potential of the macromodel, $\phi$, is given by 
\begin{equation}
\phi(\theta_x,\theta_y)=\frac{\kappa}{2}(\theta_x^2 + \theta_y^2) + \frac{\gamma_1}{2}(\theta_x^2 - \theta_y^2) - \gamma_2\theta_x\theta_y = \frac{\kappa}{2}(\theta_x^2 + \theta_y^2) + \frac{\gamma}{2}(\theta_x^2 - \theta_y^2)
\label{Eq_Phi}
\end{equation}
where we remind the reader that $\gamma_2=0$ in our reference system, $\theta_x$ and $\theta_y$ are given in radian, and we ignore a constant additive term (i.e., the potential is identically zero at the origin of coordinates of $\theta$). The deflection field from the macromodel is obtained from the derivatives of the lensing potential
\begin{eqnarray}
\alpha_x(\theta_x,\theta_y) & = & S^x_{\kappa,\gamma}\theta_x\\ 
\alpha_y(\theta_x,\theta_y) & = & S^y_{\kappa,\gamma}\theta_y
\end{eqnarray}
Where $S^x_{\kappa,\gamma}$ and $S^y_{\kappa,\gamma}$ are the slopes of the deflection field in the x and y direction respectively. For sufficiently small regions, one can approximate these slopes to their first order expansion,
\begin{eqnarray}
S^x_{\kappa,\gamma} & = & (\kappa_o+\gamma_o) + \frac{\partial (\kappa+\gamma)}{\partial x}\big |_o \Delta x  =  (\kappa_o+\gamma_o) + s^x_o \Delta x \label{Eq_Sx} \\
S^y_{\kappa,\gamma} & = & (\kappa_o-\gamma_o) + \frac{\partial (\kappa-\gamma)}{\partial y}\big |_o \Delta y  =  (\kappa_o-\gamma_o) + s^y_o \Delta y \label{Eq_Sy}
\end{eqnarray}
where $\kappa_o$ and $\gamma_o$ are the values of the convergence and shear in a reference point (for instance the central pixel in the simulated region), the derivatives of $(\kappa+\gamma)$ and $(\kappa-\gamma)$ are computed at that point, and $\Delta x$ and $\Delta y$ are the relative distances to that point in the $x$ and $y$ directions respectively. Since near critical curves, it is satisfied that $(\kappa+\gamma)\approx 1$, and only small changes in $(\kappa+\gamma)$ result in significant changes in the magnification, one can simply ignore the derivative of $(\kappa-\gamma)$ in Eq. \ref{Eq_Sy}, and focus on $s^x_o$, the derivative of $(\kappa+\gamma)$, where for typical macromodels, and near the critical curves, $s^x_o$ is (in absolute value) in the range 0.01 to 0.1 (when expressed in arcseconds$^{-1})$\footnote{For the spherical isothermal model, this derivative evaluated at the critical curve is $\theta_E^{-1}$, where $\theta_E$ is the Einstein ring in arcseconds.}. When the reference point is at the critical curve, one can relate $ s^x_o$  with the magnification in the image and source plane;
\begin{equation}
\mu(\theta) = \frac{1}{(1-\kappa+\gamma)}\frac{1}{s^x_o\theta}
\label{eq_mu}
\end{equation}
with $\theta$ being the distance in the image plane to the critical curve, and 
\begin{equation}
\mu(\beta) = \frac{1}{(1-\kappa+\gamma)}\frac{1}{\sqrt{s^x_o\beta}}
\end{equation}
with $\beta$ being the distance to the caustic in the source plane.

Since both the deflection field and lensing potential are linear with the addition of new masses, if a population of $N$ point masses are present, the deflection, $\vec{\alpha}_{PS}(\vec{\theta})$,  and potential, $\phi_{PS}(\vec{\theta})$, from the distribution of point masses  can be  added to the above equations with;  
\begin{equation}
\vec{\alpha}_{PS}(\vec{\theta})=\sum_i^N\frac{4GM_iD(z_l,z_s)}{c^2}\frac{\delta\vec{\theta}_i}{|\delta\vec{\theta}_i|^2},
\end{equation}
and,
\begin{equation}
\phi_{PS}(\vec{\theta})=\sum_i^N\frac{4GM_iD(z_l,z_s)}{c^2}ln(|\delta\vec{\theta_i}|),
\label{Eq_Phi_ps}
\end{equation}
where $\delta\vec{\theta}_i=\vec{\theta}-\vec{\theta}_i$ is the distance to the point mass $i$ at $\vec{\theta}_i$ and with mass $M_i$, $D(z_l,z_s)$ is the geometric factor $D(z_l,z_s)=D_{ls}(z_l,z_s)/(D_l(z_l)D_s(z_s))$ with $D_{ls}(z_l,z_s)$, $D_l(z_l)$ and $D_s(z_s)$ the angular diameter distances between the lens and the source, between the observer and the lens, and between the observer and the source respectively. 

A quantity of interest, is the effective optical depth, $\tau_{\rm eff}$ introduced by 
\begin{equation}
\tau_{\rm eff}=(4.2\times 10^{-4})\Sigma\frac{\mu}{\mu_r}
\label{Eq_TauEff}
\end{equation}
where the total magnification ($\mu$) is the product of the tangential and radial magnifications (i.e., $\mu=\mu_t\times\mu_r$), and $\Sigma$ (expressed in units of $M_{\odot}/pc^2$ in the expression above) is the microlens surface mass density. When $\tau_{\rm eff}\approx 1$, the saturation regime is reached. In this regime, caustics constantly overlap in the source plane, and any source moving across a field with $\tau_{\rm eff} > 1$ will always be experiencing microlensing \citep{Diego2018,Diego2019}. 

For each one of the 4 observed lensed positions, we simulate an area $0.3\times0.3$ mas$^2$ in the lens plane. The pixel scale is set to 10 nanoarcsec (nas). Inverse ray tracing is used to compute the caustic region. In general, the usable caustic region is reduced by a factor $\mu$, where $\mu$ is the macromodel magnification at that particular position. We use the values in table~\ref{tab_1} for the total convergence, $\kappa_T$, and shear of the macromodel. The convergence of the microlenses, $\kappa_*$, is obtained from the baryon surface mass density in the same table, and after dividing by the critical surface mass density. The convergence from the smooth component is then obtained as $\kappa_s=\kappa_t-\kappa_*$. 
To simulate the microlenses, we adopt a late Salpeter initial mass function (IMF), that includes remnants. We set a cutoff in the IMF at 0.1 \Msun.

\section{Macromodels with varying $\alpha$}
In this appendix we show the marginalized probabilities for two alternative models where we vary the slope of the macromodel, $\alpha$, in Eq.~\ref{Eq_kappa_macro}. In addition to the fiducial case with $\alpha=1$, in this appendix we explore the cases with $\alpha=0.8$, $\alpha=1.2$,  $\alpha=1.4$. In all cases, we use only the four positional constraints of the four SN images in order to compute the marginalized likelihoods. 
The slope plays an important role in determining the magnification at a given separation from the critical curve. In general, shallower profiles (larger $\alpha$) predict larger magnifications, since the magnification scales with the distance to the critical curve, $\delta \theta_{cc}$, as $\mu(\delta \theta_{cc})=A_o/\delta \theta_{cc}$, and $A_o$ is inversely proportional to the derivative of the potential (see Eq. \ref{eq_mu}). 
Having larger magnifications relaxes the need for microlenses especially for image 1. On the other hand, the constrain on the observed magnification of images 3 and 4 limits the degree of shallowness of the potential. In addition, large magnification factors from the macromodel increase the effective optical depth of microlensing, so smaller stellar surface mass densities can mimic the effect of larger stellar surface mass densities. 


   \begin{figure} 
   \centering
   \includegraphics[width=9.0cm]{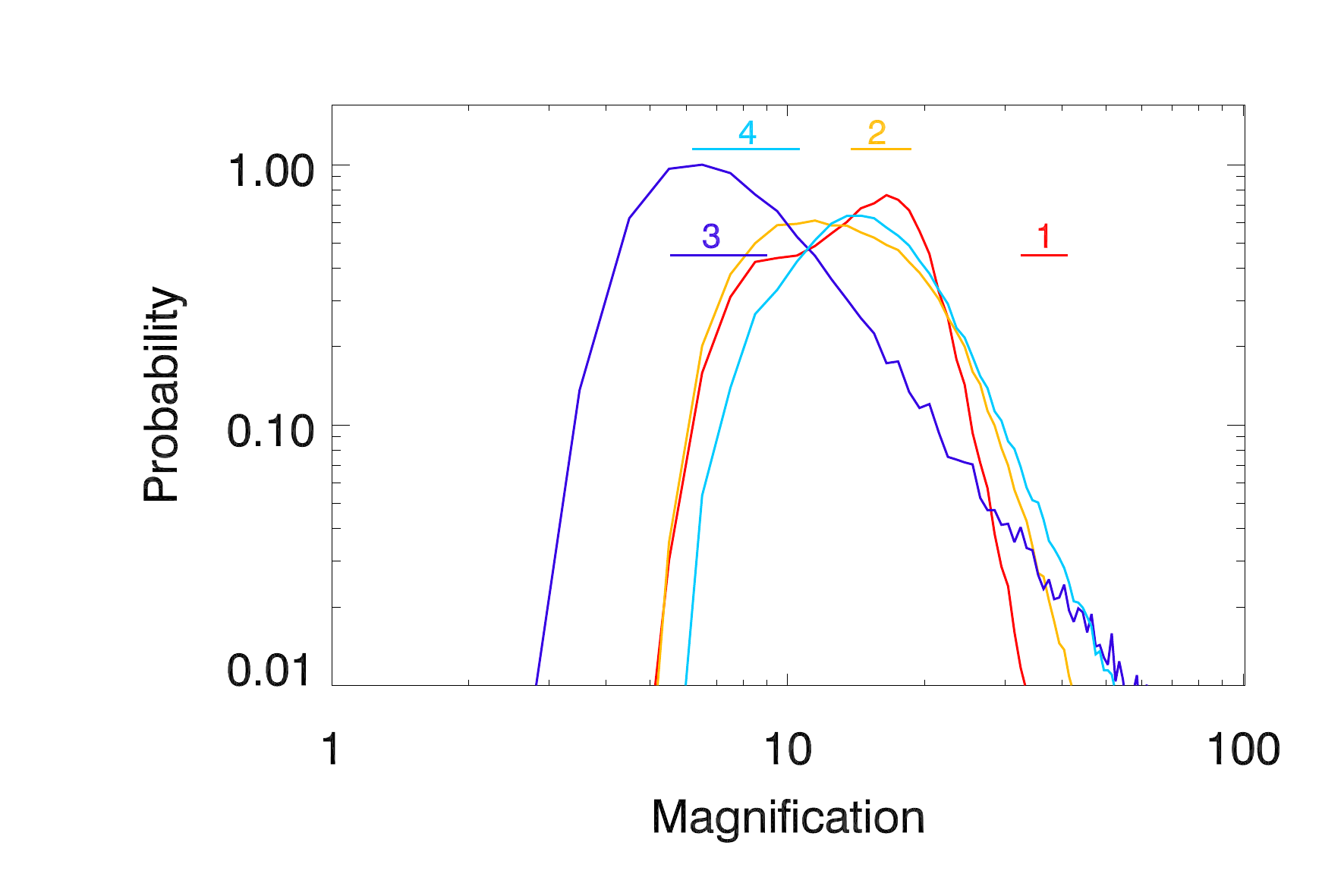}
      \caption{Marginalized probability distribution of the magnification for the case $\alpha=0.8$.  Only the 4 positions of the SNe images are used as constraints. The curves show the marginalized probability for each of the 4 SNe images. The horizontal colored lines show the inferred range of magnification from observations. 
              }
         \label{Fig_ProbMu_a0p8}
   \end{figure}

   \begin{figure} 
   \centering
   \includegraphics[width=9.0cm]{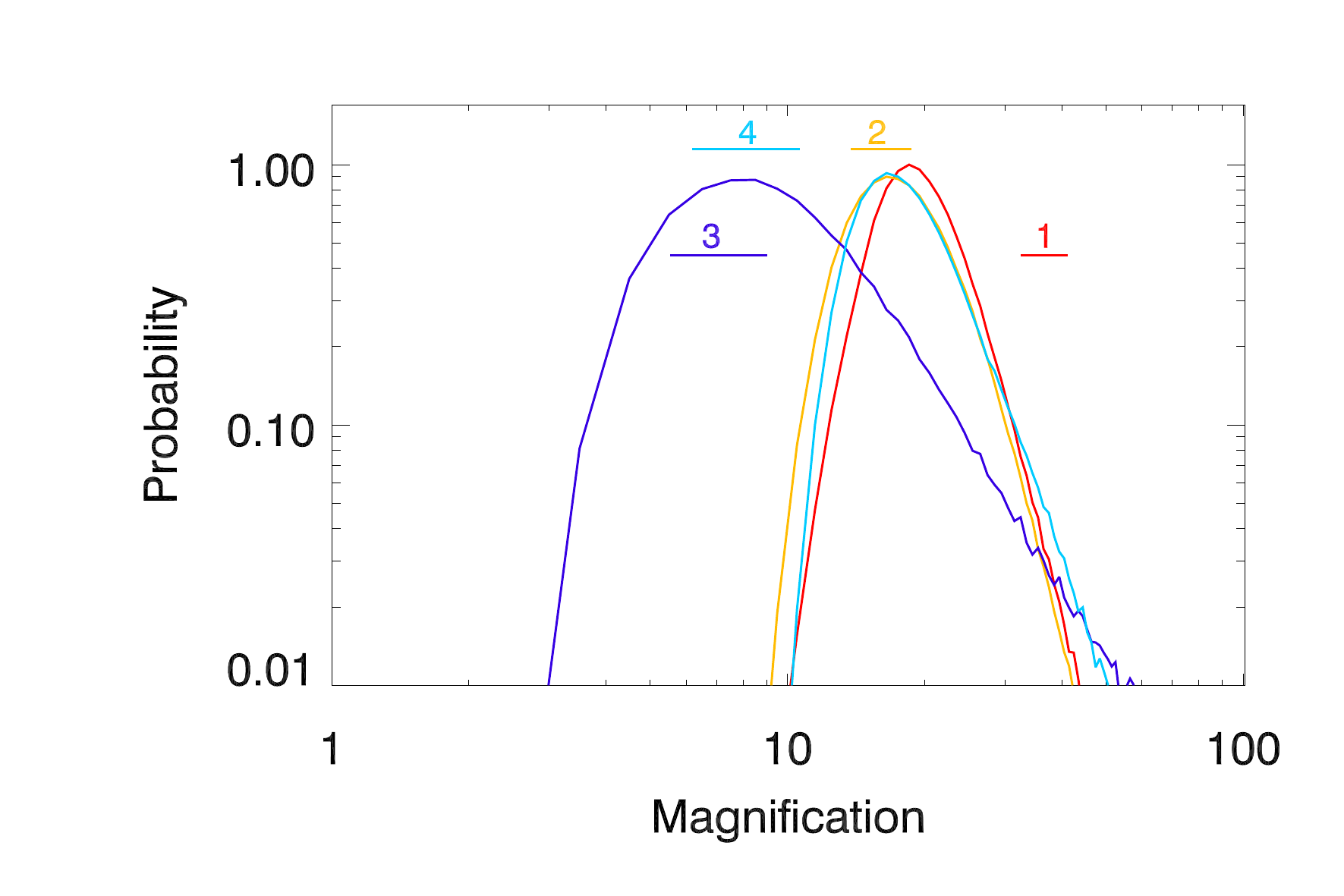}
      \caption{Like kin Figure \ref{Fig_ProbMu_a0p8} but for the case with $\alpha=1.2$. 
              }
         \label{Fig_ProbMu_a1p2}
   \end{figure}

   \begin{figure} 
   \centering
   \includegraphics[width=9.0cm]{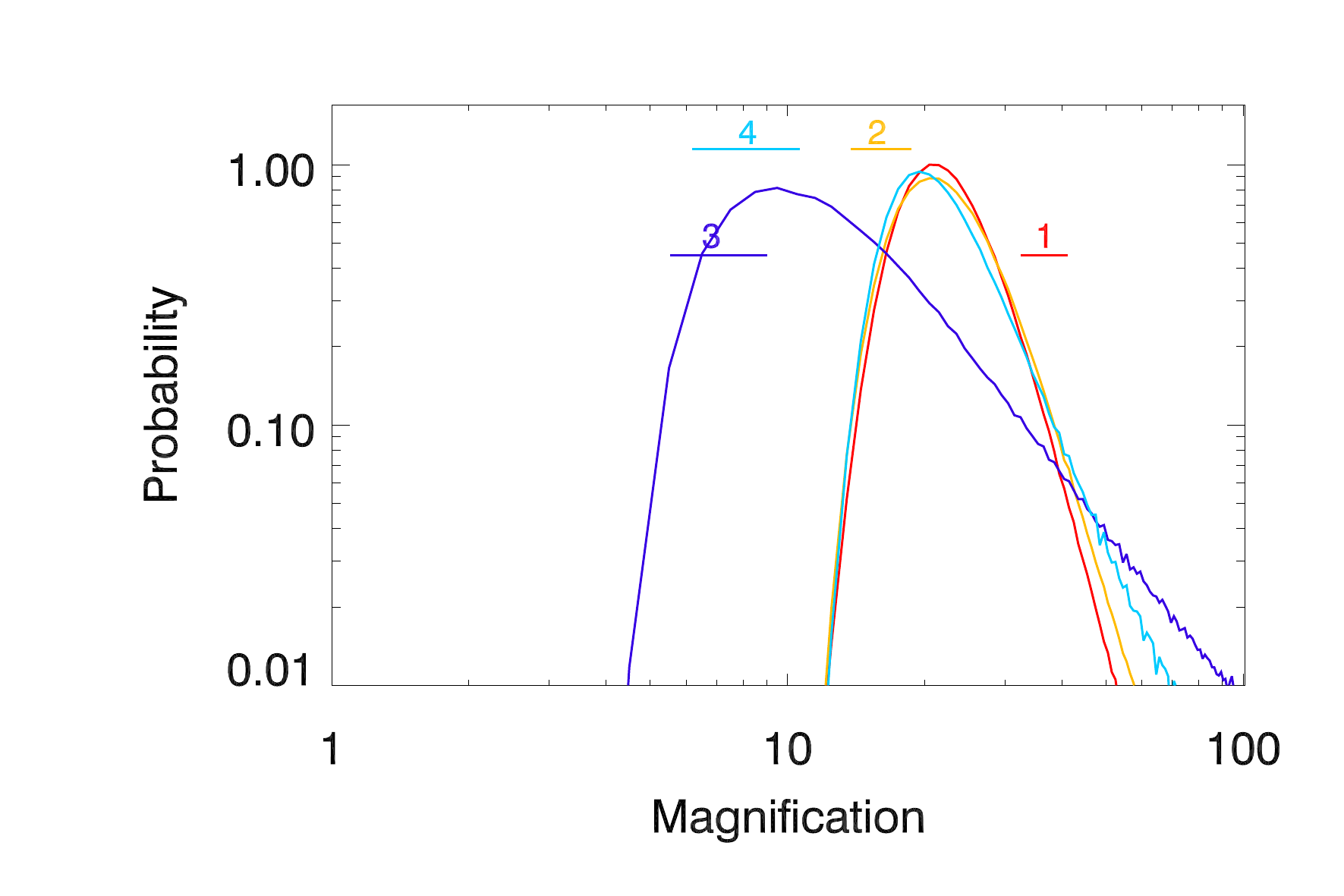}
      \caption{Like kin Figure \ref{Fig_ProbMu_a0p8} but for the case with $\alpha=1.4$.
              }
         \label{Fig_ProbMu_a1p4}
   \end{figure}

\begin{figure} 
   \centering
   \includegraphics[width=9.0cm]{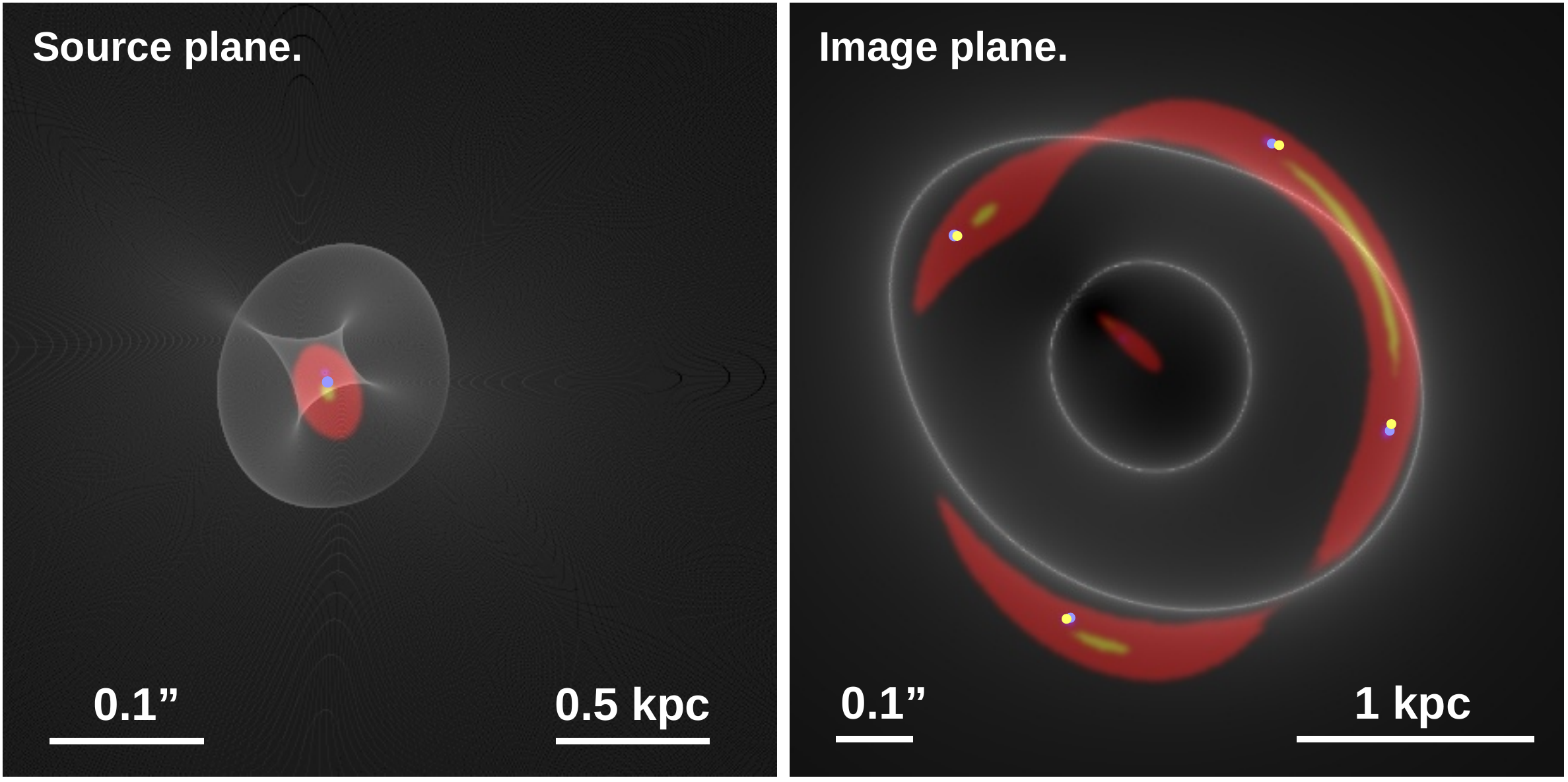}
      \caption{Caustics (left) and critical curves (right) for the best model with $\alpha=1.2$, where only the four SN positions are used as constraints. A simple source with just three components, nucleus (orange), halo (red) and SN (blue) is shown in the left panel (source plane). The predicted image is shown in the right panel (image plane). The yellow dots in the right panel mark the observed position of the four SN images. The predicted position for images 2 and 3 overlap with the yellow dots. Note the additional fifth image behind the centre of the galaxy.   
              }
         \label{Fig_Caus_CC3}
   \end{figure}

We observer that as the slope increases, so it does the typical magnification from the macromodel at the position of the four SN. The case with $\alpha=0.8$ has a very small probability of reproducing the magnification of image 1 (see Figure \ref{Fig_ProbMu_a0p8}). On the contrary, the model with $\alpha=1.2$ increases the probability for image 1 but at the expense of reducing the probability of image 4, that in this case is predicted to have a magnification $\approx$ 2 times larger than the observed value (see Figure \ref{Fig_ProbMu_a1p2}). For a model with $\alpha=1.4$ (see Figure \ref{Fig_ProbMu_a1p4}), the magnification increases further, with the peak of the probability for image 1 above $\mu=20$. However, in this case image 4 is more in tension with the observations, with the most likely magnification for image 4 being a factor more than two times larger than the observed value. In terms of the positional constraints only, the best model in the case where  $\alpha=1.2$ has a likelihood value of $-2ln(\mathcal{L}) = 0.2072$ (see Eq.~\ref{Eq_Lkhd1}), that should be compared with the best likelihood for our fiducial model ($\alpha=1$),  $-2ln(\mathcal{L}) = 0.2642$. However, this difference is small, and the effect on the macromodel is smaller than in  \cite{Mortsell2020} because the baryonic component has a fixed slope. The likelihood improves a bit more if $\alpha$ is increased. For $\alpha=1.4$, we find a best model with a likelihood $-2ln(\mathcal{L}) = 0.1871$. Like in the case with $\alpha=1.2$, the improvement in likelihood is mostly due to a better match with positions 2 and 3 as shown by comparing Figure \ref{Fig_Caus_CC1} and Figure \ref{Fig_Caus_CC3}. Since these positions were already the ones that where best reproduced by the fiducial model (Figure \ref{Fig_Caus_CC1}), the gain in the model is not as relevant as if the improvement would have taken place in images 1 and 3, which are the ones with the larger deviation between model and observed positions. 

\begin{figure} 
   \centering
   \includegraphics[width=9.0cm]{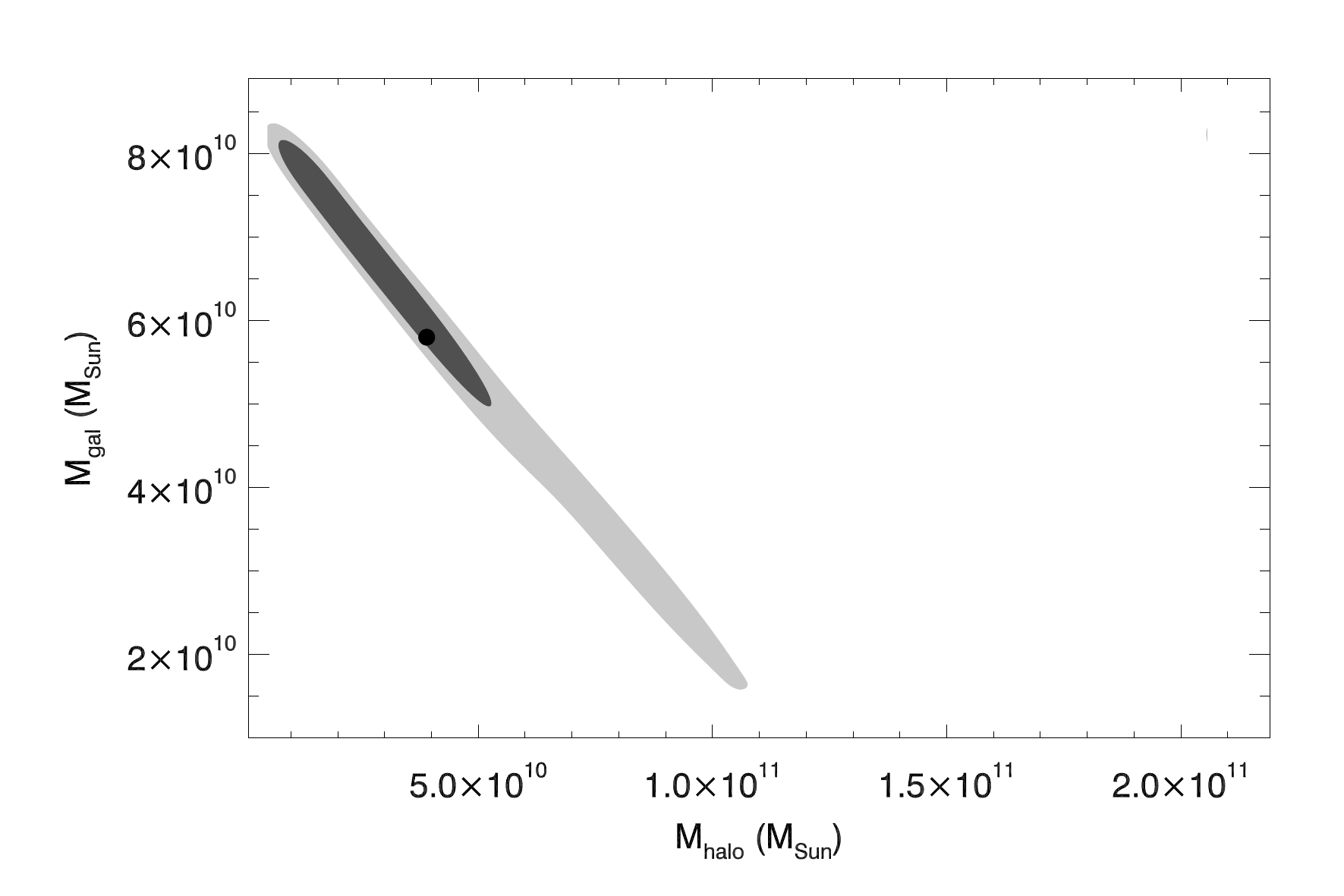}
      \caption{Marginalized probability in the ${\rm M}_{\rm halo}$--${\rm M}_{\rm gal}$ plane for the case $\alpha=0.8$. The best model is marked with a black dot and has a likelihood value of $ln(\mathcal{L}) = 1.2076$
              }
         \label{Fig_MhaloMgal_a0p8}
   \end{figure}

\begin{figure} 
   \centering
   \includegraphics[width=9.0cm]{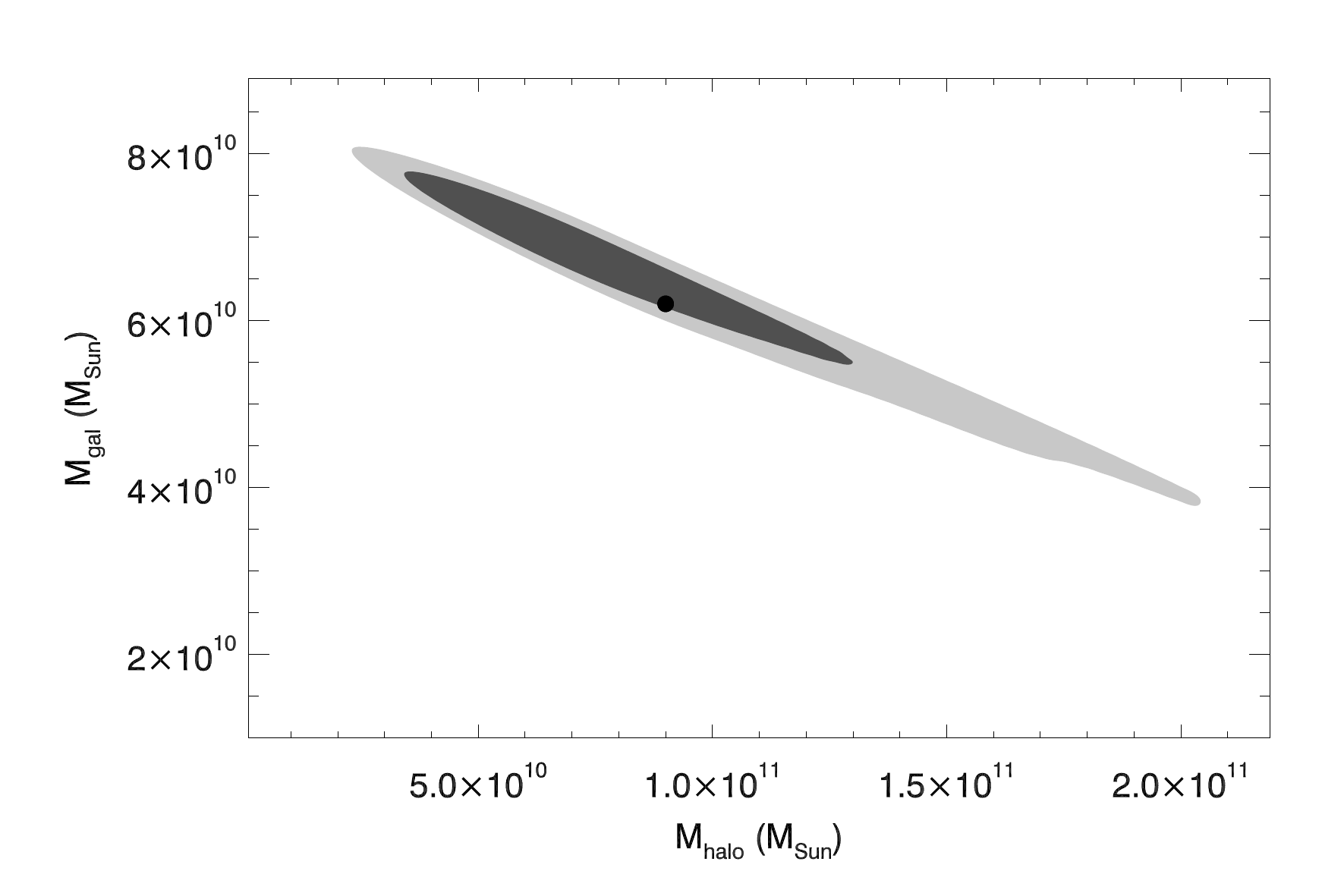}
      \caption{Marginalized probability in the ${\rm M}_{\rm halo}$--${\rm M}_{\rm gal}$ plane for the case $\alpha=1.2$. The best model is marked with a black dot and has a likelihood value of $ln(\mathcal{L}) = 0.2072$.
              }
         \label{Fig_MhaloMgal_a1p2}
   \end{figure}

In terms of the halo and baryonic masses, we observe that the constraint on the baryonic mass is almost independent of the value of $\alpha$ as shown in Figures \ref{Fig_MhaloMgal_a0p8}, \ref{Fig_MhaloMgal_a1p2} and \ref{Fig_ProbMu_a1p4}. On the contrary, the mass of the DM halo shows a strong dependence with $\alpha$, with the shallower halos (greater $\alpha$) having the larger masses.

   \begin{figure} 
   \centering
   \includegraphics[width=9.0cm]{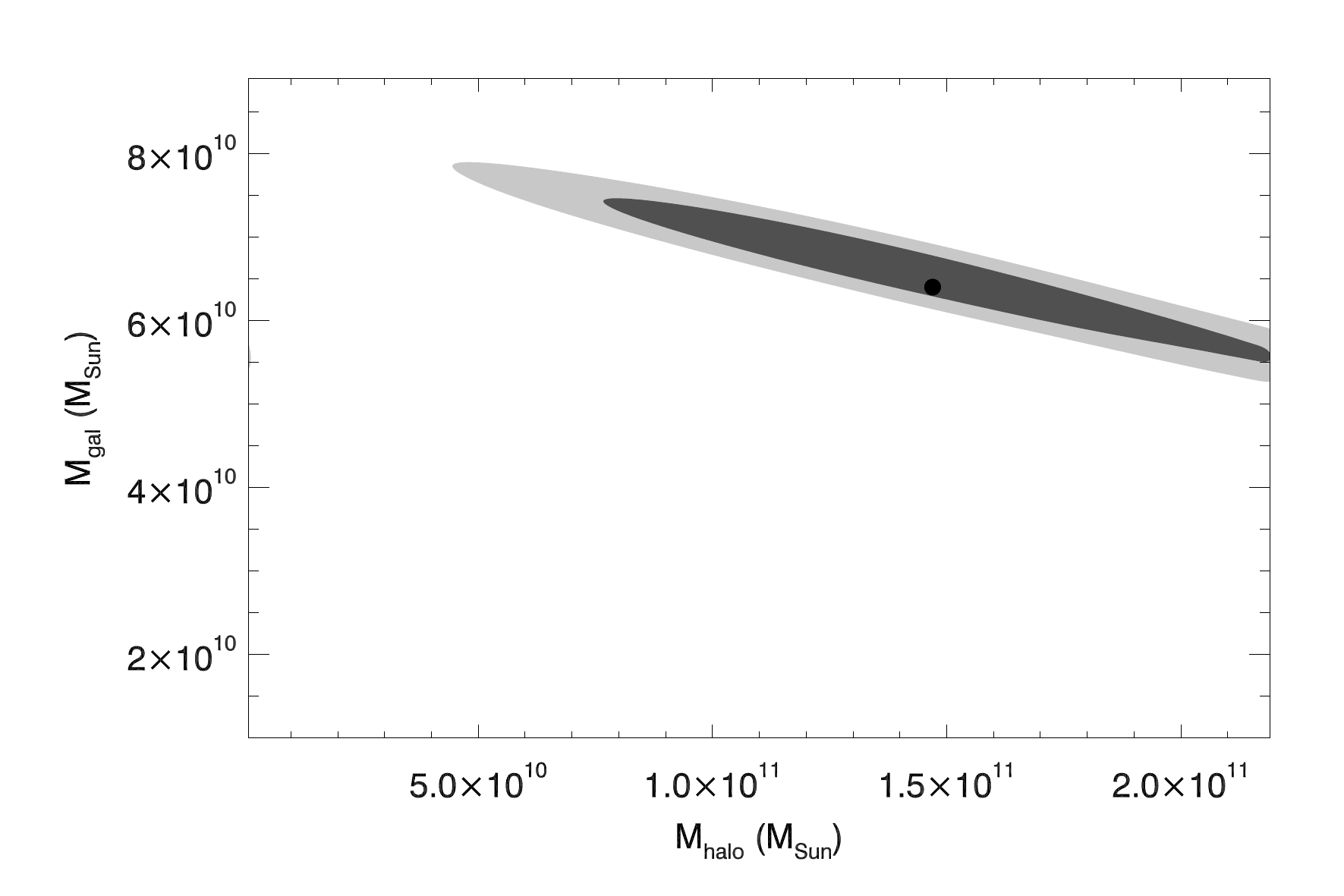}
      \caption{Marginalized probability in the ${\rm M}_{\rm halo}$--${\rm M}_{\rm gal}$ plane for the case $\alpha=1.4$. The best model is marked with a black dot and has a likelihood value of $ln(\mathcal{L}) = 0.1871$
              }
         \label{Fig_MhaloMgal_a1p4}
   \end{figure}

\section{Surface Brightness Constraints}\label{sect_James}

An alternative approach to fit the macro model is to fit the extended surface brightness emission of the lensed source and lens galaxy. Here, the lens model is constrained by hundreds of image-pixels containing the lens galaxy's and lensed source's emission in the HST images, which in principle offers more information to constrain the lens model than the twelve data points provided by the positional constraints. However, the modelling process now includes the effects of the HST Point Spread Function (PSF) and has to make assumptions regarding the source’s morphology. Furthermore, the low Einstein radius of iPTF16geu means there is significantly less extended source emission observed than is typically available in most strong lenses (e.g. \citet{Bolton2008}). 

We fit the surface brightness using the open-source lens modelling software \texttt{PyAutoLens}\footnote{\url{https://github.com/Jammy2211/PyAutoLens}} \citep{Nightingale2018,Nightingale2021}. We perform fits to the F814W HST image taken in November 2018 which therefore does not include the emission of the supernovae. 
All model fits use the nested sampling algorithm \texttt{dynesty} \citep{dynesty}. Fits assuming the following source models are compared: (i) a single Sersic light profile; (ii) two Sersic light profiles and; (iii) a pixelized source reconstruction \citep{Nightingale2015}. The single Sersic model has the highest Bayesian evidence, consistent with iPTF16geu's low Einstein Radius meaning that even after lensing magnification the source's detailed structure is not resolved. The macro model results presented therefore use the single Sersic source, however the models inferred for fits assuming these more complex source parameterizations are consistent with these results. Bayesian model comparison was also used to determine the lens light model, with two Sersic light profiles with centres aligned chosen. 

The first mass model fitted is an elliptical power-law (PowLaw) mass distribution of the form $\rho(r) \propto r^{-\gamma}$ \citep{Tessore2016} with an external shear. The Einstein mass inferred for this model is $1.79 \times 10^{10}$M$_{\rm \odot}$, consistent with the position-based constraints shown in figure 9 

For microlensing, the only relevant quantities are the $\kappa$ and $\gamma$ values at the locations of the multiple image positions.
Columns 1,3 in table \ref{table:KappaGamma} show the values of $\kappa$ and $\gamma$ at each multiple image location for the PowLaw model. 

\begin{table}
\tiny
\begin{tabular}{| c | c | c | c | c |} 

\hline
Image & $\kappa$ (PowLaw) & $\gamma$ (PowLaw) & $\kappa$ (stars+DM)& $\gamma$ (stars+DM)\\ 
\hline
1 & 0.700 & 0.460 & 0.791 & 0.261  \\[2pt]
2 & 0.570 & 0.390 & 0.606 & 0.339  \\[2pt]
3 & 0.555 & 0.353 & 0.543 & 0.336  \\[2pt]
4 & 0.652 & 0.439 & 0.698 & 0.352  \\[2pt]
\hline
\end{tabular}
\caption{The values of the convergence $\kappa$ and shear $\gamma$ at the four multiple image locations of iPTF16geu. Each value is that of the maximum likelihood model inferred via surface brightness modelling. Note how the parity of the images can not be recovered properly. This is however expected since the position of these images are not used as constraints.}
\label{table:KappaGamma}
\end{table}
We next perform model-fits assuming a stars plus dark matter lens model, following the approach described in \citet{Nightingale2019} (deflection angle calculations follow \citet{Anowar2019}). We assume two Sersic profiles for the stars, representing a bulge plus disk, and a spherical NFW dark matter halo profile \citep{Navarro1997} where the mass of the halo is set via the relation of \citet{Ludlow2016}. The Sersic's centres are forced to be aligned and the NFW profile's centre is a free parameter (these choices were informed via model comparison). This model infers an Einstein mass of $1.67 \times 10^{10}$M$_{\rm \odot}$, $6.7\%$ lower than that of the PowLaw model, but still consistent with it. 
Table \ref{table:KappaGamma} gives the values of $\kappa$ and $\gamma$ at the multiple images for the stars plus dark matter model in columns 3 and 4 respectively. 

By comparing with the values of $\kappa$ and $\gamma$ presented in figure \ref{Fig_Comparison_WilliamsMortsell}, we find good agreement at positions 1 and 2, but not so much at positions 3 and 4, indicating that different modelling techniques can result in substantial changes in the inferred values of $\kappa$ and $\gamma$ at the observed SN positions. 

Given the low resolution of the data, the extended surface brightness lens model does not necessarily infer a lens model that is more precise than the SN position based model, despite containing many more data points. The low Einstein radius of iPTF16geu means that only a small fraction of its lensed source's emission is resolved and there is insufficient detailed structure to constrain the lens model (contrast this to the three strong lenses fitted by \citet{Nightingale2019}, where distinct star forming clumps and spiral arm structures are resolved by the source reconstruction). 

\section{Expanding photosphere}\label{Sect_Photosphere}
\begin{figure} 
   \centering
   \includegraphics[width=9.0cm]{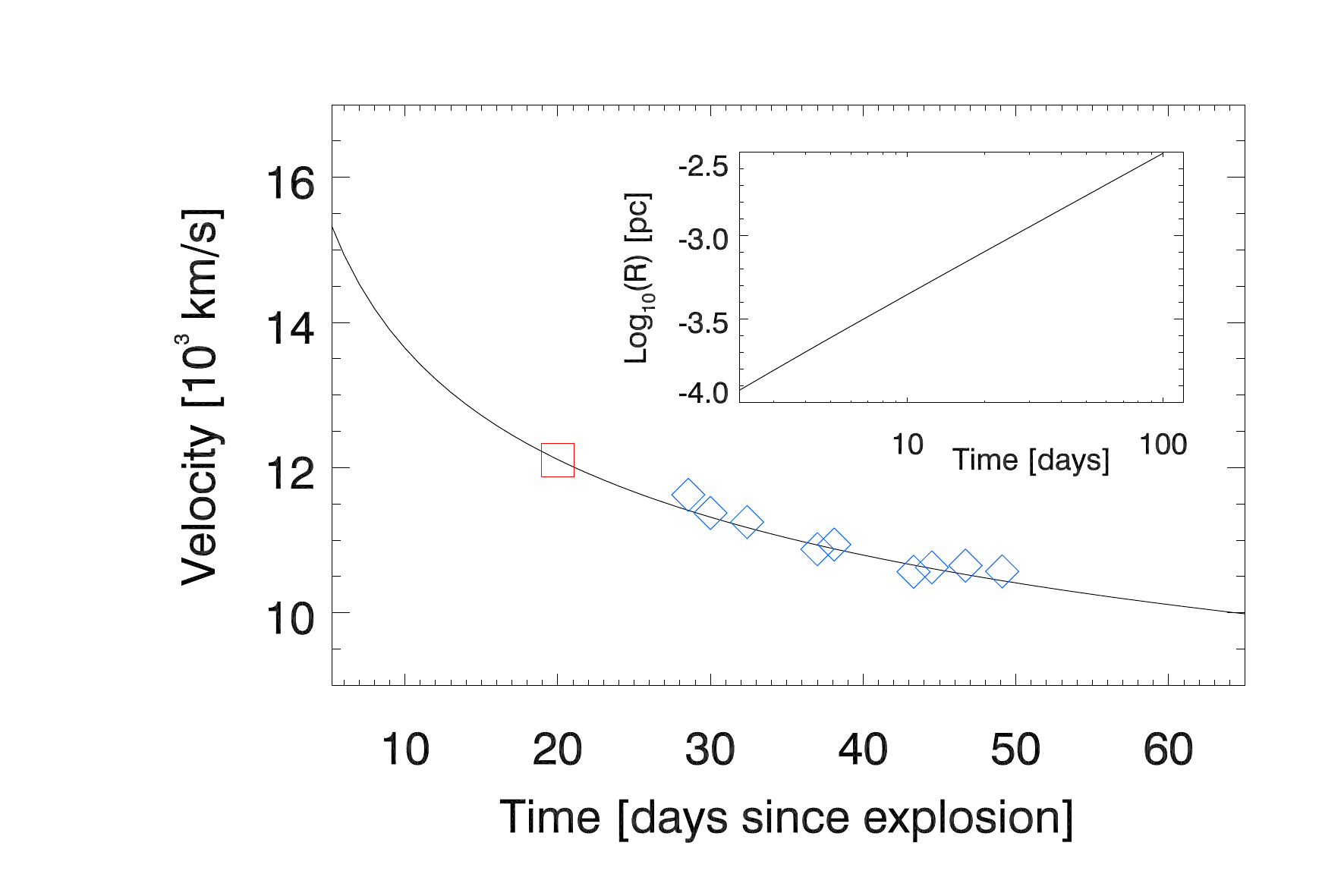}
      \caption{The blue diamond symbols are the estimated velocities from \cite{Johansson2021}. The red square shows the estimated velocity by the same authors at the time of maximum luminosity. The solid line is a model with the standard exponent of 0.22,  $v_{exp} \propto t^{-0.22}$ (see text). The small inset plot in the top right shows the radius of the photosphere for this model of the velocity expansion. 
              }
         \label{Fig_Photosphere}
   \end{figure}

To model the expansion velocity, we follow the exponential form in \cite{Piro2014}, 
\begin{equation}
    V_{exp}(t_d) = 1.8\times10^4 t_{d}^{-0.22} + 2.8\times10^3   {\rm km} {\rm s}^{-1}, 
    \label{Eq_Vexp}
\end{equation}
where $t_d$ is the time since explosion expressed in days and in the source frame of the SN. The constant velocity floor of $2.8\times10^3  {\rm km} {\rm s}^{-1}$ is added to reproduce the shallower shape of the velocity curve during the last days. In Fig.~\ref{Fig_Photosphere} we show the observed values from \cite{Johansson2021}, together with an extrapolated estimation at the time of maximum luminosity. The model in the above equation is shown as a solid line. The corresponding radius of a photosphere expanding with this velocity is shown in the smaller inset plot. The maximum in the light curve of iPTF16geu is observed $\approx 20$ days (or $\approx 15$ days in the SN rest frame) after the first observation. Assuming that first observation is within a day of the explosion time of the supernova, the photosphere radius at the time of maximum is $\approx 6.27\times10^{-4}$ pc. Observations of SN iPTF16geu span $\approx 95$ days in the observer frame. This corresponds to $\approx 65$ days in the SN rest frame. The last observation is $\approx 70$ days in the observer frame after the peak in the luminosity, or $\approx 50$ days in the SN rest frame. At this time, the photosphere radius is $\approx 2.16\times10^{-3}$ pc, based on the model for the expansion velocity in Eq.~\ref{Eq_Vexp} (see Fig.\ref{Fig_Photosphere}). 

\section{Software Citations}

This work uses the following software packages:

\begin{itemize}

\item
\href{https://github.com/astropy/astropy}{\texttt{Astropy}}
\citep{astropy1, astropy2}

\item
\href{https://bitbucket.org/bdiemer/colossus/src/master/}{\texttt{Colossus}}
\citep{colossus}

\item
\href{https://github.com/dfm/corner.py}{\texttt{corner.py}}
\citep{corner}

\item
\href{https://github.com/joshspeagle/dynesty}{\texttt{dynesty}}
\citep{dynesty}

\item
\href{https://github.com/matplotlib/matplotlib}{\texttt{matplotlib}}
\citep{matplotlib}

\item
\href{numba` https://github.com/numba/numba}{\texttt{numba}}
\citep{numba}

\item
\href{https://github.com/numpy/numpy}{\texttt{NumPy}}
\citep{numpy}

\item
\href{https://github.com/rhayes777/PyAutoFit}{\texttt{PyAutoFit}}
\citep{pyautofit}

\item
\href{https://github.com/Jammy2211/PyAutoLens}{\texttt{PyAutoLens}}
\citep{Nightingale2015, Nightingale2018, pyautolens}

\item
\href{https://github.com/equinor/pylops}{\texttt{PyLops}}
\citep{pylops}

\item
\href{https://github.com/AshKelly/pyquad}{\texttt{pyquad}}
\citep{pyquad}

\item
\href{https://www.python.org/}{\texttt{Python}}
\citep{python}

\item
\href{https://github.com/scikit-image/scikit-image}{\texttt{scikit-image}}
\citep{scikit-image}

\item
\href{https://github.com/scikit-learn/scikit-learn}{\texttt{scikit-learn}}
\citep{scikit-learn}

\item
\href{https://github.com/scipy/scipy}{\texttt{Scipy}}
\citep{scipy}

\end{itemize}

\end{document}